\documentclass[12pt,reqno]{amsart}%
\usepackage{graphics,graphicx,amsmath,amssymb,mathtools,epsfig,stmaryrd,color,flafter,xspace,lscape,threeparttable,pdfpages,csquotes,amsfonts,amsthm,geometry,array,booktabs,latexsym,caption,subfig,multirow,tabularx,rotating,fnpct}
\usepackage[font=small,labelfont=bf]{caption}
\usepackage{changepage,enumitem}
\usepackage{natbib}%

\usepackage[hidelinks]{hyperref}
\DeclareCaptionFont{tiny}{\tiny}
\everymath{\displaystyle}
\RequirePackage{setspace,indentfirst,epsfig,psfrag,ifthen,ifpdf}

\newtheorem{theorem}{Theorem}
\theoremstyle{plain}

\newtheorem{assumption}{Assumption}
\newtheorem{subassumption}{Assumption\normalfont}[assumption]

\newenvironment{assumption*}
{\ifnum\value{subassumption}=0 \stepcounter{assumption}\fi\subassumption}
{\endsubassumption}
\newenvironment{assumption+}[1]
{\renewcommand{\thesubassumption}{#1}\subassumption}
{\endsubassumption}
\theoremstyle{definition}
\newtheorem{assump}{Assumption}

\newtheorem{corollary}{Corollary}

\newtheorem{definition}{Definition}
\newtheorem*{definition*}{Definition}

\numberwithin{equation}{section}
\newcounter{steps}

\begin{document}
\title{An econometrician's guide to optimal transport
}
\author{Alfred Galichon and Marc Henry}
\address{NYU and Penn State}
\thanks{
This review benefited from discussions with Onil Boussim, Tim Christensen, Amine Fahli, Yanqin Fan, Christophe Gaillac, Florian Gunsilius, Xavier d'Haultfoeuille, Antoine Jacquet, Lixiong Li, Arnaud Maurel, Ju Hyun Oh, Brendan Pass, Guillaume Pouliot, Susanne Schennach, Hideyuki Tomiyama, and Yidi Zhao. 
Corresponding author: Marc Henry: \texttt{marc.henry@psu.edu}.}

\begin{abstract}

We propose an overview of optimal transport theory and its applications to econometric methodology. This review is specifically designed for practitioners, be they econometric theorists or applied econometricians. The review of applications of optimal transport to econometrics is organized around the particular aspects of the mathematical theory of optimal transport they rely on. 

\vskip20pt


\vskip10pt


\end{abstract}

\maketitle

\nocite{*}

\section*{Introduction}

Optimal transport traces its origins to 18th Century civil engineering. It draws its central ideas from several diverse areas of mathematics. It finds application in fields as diverse as non-Euclidean geometry, cosmology, fluid dynamics, traffic management, machine learning, economics and computational biology. There is a rapidly growing community of researchers who work {\em on} optimal transport as well as researchers who work {\em with} optimal transport. There is also a growing body of reference books, textbooks and surveys about various aspects of optimal transport theory and its applications. Among the latter, in economics alone, optimal transport has wide ranging applications, from family economics to mechanism design, with recent books and surveys to account for them. The objective of this new review is to give an account of optimal transport theory and methods specifically designed for applications in econometrics. The target audience includes all producers and consumers of econometric methods. Econometric theorists will find here new proof techniques, new computational tools, new ways to model phenomena and new ways to perform inference. Applied economists will find here many new econometric techniques applicable to a wide range of fields with step-by-step guides to implementation. 

We start with a brief overview of optimal transport designed mostly for readers who were previously unacquainted with the subject. We provide a gentle general introduction, which includes a short history of the subject, the basic formulations of the problem, a short description of the main aspects of the theory that are relevant to econometrics as well as its classical computational tools. We then devote the second sub-section to describing a collection of algorithms that are the main building blocks in most of the procedures presented in this review. We propose a simplex-based algorithm to derive optimal matchings, which are optimal transport plan between two discrete distributions. We derive the optimal transport map between two Gaussian distributions. We present the solution to optimal transport in~$\mathbb R$ and their relation with rearrangement inequalities and quantiles. We describe algorithms from computational geometry to compute optimal transport maps from a discrete distribution to a uniform in~$[0,1]^d$, and the iterated proportional fitting procedure to compute entropic relaxations of the optimal transport problem. Finally, we discuss methods to compute the Wasserstein distance between two distributions, including recent work on Wasserstein generative adversarial networks. 

We then proceed to the review of econometric papers, where the contribution relies in a significant way on optimal transport theory. The applications of optimal transport we review here range across most areas of econometrics, including causal inference, discrete choice, structural estimation, data combination, program evaluation and treatment assignment, risk and inequality measurement, nonparametric identification, mis-specification and robustness, distributional regression, and machine learning. We organize them not by area but according to the aspect of optimal transport theory that underlies the main innovation in each paper. We identify three main aspects of the theory as well as several extensions. The main aspects of the theory we identify are: First, the existence of optimal transport plans and Kantorovich duality theory; Second, the uniqueness of optimal transport maps and generalized monotonicity; Third, optimal transport distance between distributions. They correspond broadly, but not exactly with: First data combination applications; Second, multivariate quantiles and identification; Third, distributional robustness. For each aspect of the theory, we explain how it is useful in econometrics, and we review econometric innovations based on this aspect
.

Optimal transport has natural connections with the economics of matching. General applications to economic theory are beyond the scope of this review. However, an aspect of the economics of matching is directly related to econometric theory and practice. The problem of recovering transport costs from realized transport plans, called inverse optimal transport problem, is directly relevant to the identification and estimation of matching surpluses and complementarities in empirical matching theory. In section~\ref{sec:IOT}, we present the inverse optimal transport problem; we review its applications in empirical matching and its connections with parametric estimation of matching, Poisson regressions, gravity equations and metric learning.

For those readers who wish to work with optimal transport theory to develop new econometric methods, this review is an introduction, not a substitute for more advanced texts. Beyond the pioneering books of \citeauthor{rachev1998mass} [\citeyear{rachev1998mass,rachev1998mass-app}], which are treasure troves of useful results, but somewhat difficult to navigate, we recommend \citeauthor{villani2003topics} [\citeyear{villani2003topics,villani2008optimal}], \cite{santambrogio2015optimal}, \cite{galichon2016optimal}, \cite{peyre2019computational} and \cite{chewi2024statistical}. \cite{villani2003topics} is the book that ushered us both into the world of optimal transport. It's hard to overstate its elegance and influence. It provides a concise overview of the whole theory up to that point, suitable for any audience with a modicum of mathematical sophistication. It is, and is likely to remain, the best introduction to optimal transport theory. \cite{villani2008optimal} expands considerably on \cite{villani2003topics}. For the purposes of applications to econometrics, it is particularly valuable for the deep insights it provides into solutions to the Monge problem, i.e., existence, uniqueness and cyclical monotonicity of optimal transport maps. \cite{santambrogio2015optimal} is also a very valuable resource. Unlike \cite{villani2008optimal} which is primarily designed as a reference for researchers who wish to work on optimal transport and advance the theory, \cite{santambrogio2015optimal} targets researchers who wish to work with optimal transport, such as applied mathematicians and econometricians among others. \cite{galichon2016optimal} 
introduces optimal transport to economists as a unifying tool for modeling, identification, and computation, and sketches its core applications across matching, equilibrium analysis, and empirical economics; it is addressed to a general economics, rather than econometrics, audience, and it predates many of the examples we review. \cite{peyre2019computational} is a great resource on all computational aspects of optimal transport. Finally \cite{chewi2024statistical} contains an account of recent developments on the estimation of optimal transport plans and maps and optimal transport distances, which are particularly relevant to applications in econometrics. More recently,~\citet{Galichon2026DCM} focuses on discrete choice models in connection with multiple aspects of optimal transport. 


\newpage
\tableofcontents

\newpage

\section{A brief overview of optimal transport}

    \subsection{Historical vignette}
    The origin of optimal transport theory is always traced back to \cite{monge1781memoire}, with the original stylized formulation of the problem of moving specific piles of dirt to build specific structures at minimum cost. Its author was Gaspard Monge, influential scientist, engineer and politician during the French Revolution and Empire, and one of the three founding fathers of the \'Ecole polytechnique in Paris. The Russian mathematician Leonid Kantorovich rediscovered\footnote{Kantorovich made the connection with the Monge problem in~1948.} the problem in~1938 and proposed a probabilistic formulation with a duality theorem in \citeauthor{kantorovich1940} [\citeyear{kantorovich1940,kantorovich1942translocation}], which laid the foundations of linear programming methods, and for which he was awarded the 1975 Nobel Prize for economics, jointly with Tjalling Koopmans, “for their contributions to the theory of optimum allocation of resources”.\footnote{The American mathematician Frank Hitchcock independently formulated a special case of the optimal transport problem in \cite{hitchcock1941distribution}, together with an early version of the simplex algorithm to solve it.} He also used optimal transport to define a distance between probability measures, which, through the vagaries of misattribution, is now commonly known as Wasserstein distance. The solution to the original Monge problem, when the transportation cost is quadratic, was given by \citeauthor{brenier1987decomposition} [\citeyear{brenier1987decomposition,brenier1991polar}] and \cite{ruschendorf1990characterization} based on earlier work by \cite{sudakov1979geometric} and \cite{knott1984optimal} in particular. They show that solutions to the optimal transport problem with quadratic cost exist, are unique and are cyclically monotone (gradients of convex functions). \cite{mccann1995existence} extended this result by showing that two probability measures can always be mapped into each other with a cyclically monotone map, under minimal regularity and without the finite variance condition necessary to define quadratic optimal transport. Chapter~10 of \cite{villani2008optimal} gives comprehensive results on existence and uniqueness of solutions to optimal transport with more general costs based on work by \cite{mccann2001polar} and \cite{bernard2007optimal}. More recent developments that are relevant to econometric applications are displacement convexity in \cite{mccann1997convexity}, multi-marginal optimal transport in \cite{pass2015multi}, and weak optimal transport in \cite{gozlan2017kantorovich}. Optimal transport theory and its applications are now in an explosive phase of development, one of the latest twists being the massive interest in Wasserstein barycenters and Wasserstein Generative Adversarial Networks in machine learning and artificial intelligence.
    
    \subsection{Basic formulations of the optimal transport problem}\label{sec:basic}

    Optimal Transport is a coupling problem. It is the problem of finding a coupling of two probability measures to minimize an aggregate cost. Let~$\mu$ and~$\nu$ be probability measures on~$\mathcal X$ and~$\mathcal Y$ respectively. A coupling of~$(\mu,\nu)$ is the construction of a pair of random elements~$(X,Y)$ with probability distribution~$\pi$ on~$\mathcal X\times\mathcal Y$ such that~$X$ has distribution~$\mu$ and~$Y$ has distribution~$\nu$. The probability distribution~$\pi$ is then said to have marginals~$\mu$ and~$\nu$, which is equivalent to
    \begin{eqnarray*}
        \int_{\mathcal X\times\mathcal Y} (\varphi(x)+\psi(y))d\pi(x,y) & = & 
        \int_{\mathcal X}\varphi(x)d\mu(x)+\int_{\mathcal Y}\psi(y)d\nu(y),
    \end{eqnarray*}
    for all integrable functions~$\varphi$ and~$\psi$ on~$\mathcal X$ and~$\mathcal Y$ respectively. By extension, a distribution~$\pi$ with marginals~$\mu$ and~$\nu$ is also called a coupling of~$(\mu,\nu)$ and the collection of all such couplings is denoted~$\mathcal M(\mu,\nu)$. 
    
    The coupling is called deterministic if there is a measurable function~$T:\mathcal X\rightarrow\mathcal Y$ such that~$Y=T(X)$. The function~$T$ is then called a change of variables, which is equivalent to
    \begin{eqnarray}\label{eq:CofV}
        \int_{\mathcal Y} \psi(y)d\nu(y) & = & 
        \int_{\mathcal X}\psi(T(x))d\mu(x),
    \end{eqnarray}
    for all integrable function~$\psi$ on~$\mathcal Y$.

    The optimal transport problem minimizes an aggregate cost. Call~$c: \mathcal X\times\mathcal Y\rightarrow \mathbb R$ the cost function. Optimal transport has two main formulations.
    
    \subsubsection*{Kantorovich formulation} 
    The Kantorovich formulation of the optimal transport problem is that of finding a coupling~$(X,Y)$ of~$(\mu,\nu)$ with distribution~$\pi$ that minimizes the aggregate cost
    \begin{eqnarray*}
        \mathcal C^K(\mu,\nu) \; := \: \inf_{X\sim\mu, Y\sim\nu} 
        \mathbb E\;c\left( X,Y \right)
        &=&\int c(x,y) \; d\pi(x,y).
    \end{eqnarray*}
    Such a distribution~$\pi$ with marginals~$\mu$ and~$\nu$ is called an optimal transport plan from distribution~$\mu$ to distribution~$\nu$.
    \subsubsection*{Monge formulation} The Monge formulation of the optimal transport problem is that of finding a deterministic coupling~$(X,T_{\mu\rightarrow\nu}(X))$ of~$(\mu,\nu)$ that minimizes the aggregate cost
    \begin{eqnarray*}
        \mathcal C^M(\mu,\nu) 
        & = & \int c(x,T_{\mu\rightarrow\nu}(x)) \; d\mu(x).
    \end{eqnarray*}
    The map~$T_{\mu\rightarrow\nu}$ that defines such a deterministic coupling of~$(\mu,\nu)$ is called an optimal transport map from distribution~$\mu$ to distribution~$\nu$.
    
    \subsection{Main aspects of optimal transport theory}

    The main aspects of optimal transport theory that are relevant to econometrics are related to the following basic questions about optimal transport plans, optimal transport maps and the value of the optimal transport objective.
    \begin{enumerate}
        \item When do optimal transport plans and maps exist, and when are they unique?
        \item What properties can help characterize, compute and estimate optimal transport plans, maps and optimal value?
    \end{enumerate}
    The answers to these questions may naturally depend on the cost function~$c$ and the marginal distributions~$\mu$ and~$\nu$. Conversely, we may also ask:
    \begin{enumerate}
        \setcounter{enumi}{2}
        \item What can we learn about the relation between~$\mu$ and~$\nu$ from the value of the optimal transport objective~$\mathcal C(\mu,\nu)$?
    \end{enumerate}

    The answer to question~(1) is straightforward for optimal transport plans, which exist in great generality and are generically non unique. Optimal transport plans exist under the following condition on the cost function, which we will assume in the rest of this review.
    \begin{assumption}\label{ass:cost}
        The cost function~$c$ on~$\mathcal X\times\mathcal Y$ is lower semi-continuous and satisfies~$c(x,y)\geq a(x)+b(y)$ for two integrable real valued upper semi-continuous functions~$a$ on~$\mathcal X$ and~$b$ on~$\mathcal Y$.
    \end{assumption}
    Existence of an optimal transport plan holds because~$\mathcal M(\mu,\nu)$ is compact, and, as C\'edric Villani puts it, ``has probably been known since time immemorial'' \cite{villani2008optimal}. The answer to question~(1) for optimal transport maps is much more involved and is one of the most significant aspects of optimal transport theory. Existence of optimal transport maps is much trickier because the set of deterministic couplings is not compact. Proofs of existence and uniqueness of optimal transport maps for certain classes of transport cost functions and under certain regularity conditions on the probability distributions~$\mu$ and~$\nu$ rely on some of the answers to question~(2), so we will return to them later in this section.

    A first characterization of optimal transport plans and answer to question~(2) is the equivalence between cyclical monotonicity and optimality of a transport plan. Equivalence between cyclical monotonicity and optimality can be understood in the following way. Recall that a transport plan~$\pi$ is a joint distribution on~$\mathcal X\times\mathcal Y$. If a pair~$(x,y)$ is in the support of~$\pi$, then some mass is being transferred from~$x$ to~$y$. If~$\pi$ is an optimal transport plan, then the mass transferred from~$x$ to~$y$ cannot be more cheaply transferred via~$x\rightarrow y^\prime\rightarrow x^\prime\rightarrow y$, say. More generally, optimality implies the property of cyclical monotonicity of the support of~$\pi$.
    \begin{definition}[$c$-Cyclical monotonicity]\label{def:c-mon}
        A subset~$\Gamma$ of~$\mathcal X\times\mathcal Y$ is called $c$-cyclically monotone if for any integer~$K$ and any collection~$(x_1,y_1),$ $\ldots,$ $(x_K,y_K),$ in~$\Gamma$ and the convention~$y_{K+1}=y_1$, we have the inequality
        \begin{eqnarray*}
            \sum_{k=1}^Kc(x_k,y_k) & \leq & \sum_{k=1}^Kc(x_k,y_{k+1}).
        \end{eqnarray*}
    \end{definition}
    Conversely, it can be shown that cyclical monotonicity of the support implies optimality of the transport plan under assumption~\ref{ass:cost}. Equivalence of cyclical monotonicity and optimality, true for real-values and lower semicontinuous cost functions is proved in \cite{schachermayer2009characterization}. See equivalence of (a) and (b) in Theorem 5.10(ii) of \cite{villani2008optimal}.

    Cyclical monotonicity also relates to the central answer to question~(2) and a major aspect of optimal transport theory, namely Kantorovich duality, which also holds in great generality.
    The dual Kantorovich problem is that of maximizing loading and unloading costs
    \begin{eqnarray*}
        \tilde{\mathcal C}(\mu,\nu) & := & \sup_{(\varphi,\psi)\in\Phi_c} \left( \int_{\mathcal X}\varphi(x)d\mu(x) + \int_{\mathcal Y}\psi(y)d\nu(y) \right)
    \end{eqnarray*}
    under the constraint that loading cost~$\varphi(x)$ at~$x$ and unloading cost~$\psi(y)$ at~$y$ are continuous and bounded and do not exceed the total transportation cost~$c(x,y)$. Hence the supremum in the dual Kantorovich problem is taken over the set
    \begin{eqnarray*}
        \Phi_c& := & \left\{ (\varphi,\psi)\in\mathcal C_b(\mathcal X)\times\mathcal C_b(\mathcal Y): \varphi(x)+\psi(y)\leq c(x,y) \mbox{ for all }(x,y)\in\mathcal X\times\mathcal Y\right\}.
    \end{eqnarray*}
    Since all transport plans are couplings of~$(\mu,\nu)$, integrating the constraint yields
    \begin{eqnarray*}
        \int_{\mathcal X}\varphi(x)d\mu(x) + \int_{\mathcal Y}\psi(y)d\nu(y) & = & \int_{\mathcal X\times\mathcal Y}\left( \varphi(x) + \psi(y) \right) d\pi(x,y)\\
        & \leq & \int_{\mathcal X\times\mathcal Y} c(x,y) d\pi(x,y).
    \end{eqnarray*}
    Hence the Kantorovich dual~$\tilde{\mathcal C}(\mu,\nu)$ is no larger than the primal~$\mathcal C(\mu,\nu)$ (weak duality). It turns out they are equal (strong duality) under very general conditions, which is the content of the Kantorovich duality theorem.
    \begin{theorem}[Kantorovich duality]\label{thm:dual}
        Under assumption~\ref{ass:cost}, $\tilde{\mathcal C}(\mu,\nu)=\mathcal C(\mu,\nu)$, and there is an optimal transport plan that attains~$\mathcal C(\mu,\nu)$. If in addition~$\mathcal C(\mu,\nu)$ is finite and~$c(x,y)\leq c_{\mathcal X}(x)+c_{\mathcal Y}(y)$ for integrable functions~$c_{\mathcal X}$ and~$c_{\mathcal Y}$, then there is a pair~$(\varphi,\psi)$, called transport potentials, that attains the dual~$\tilde{\mathcal C}(\mu,\nu)$.
    \end{theorem}
    The dual constraint can be rewritten~$\psi(y)\leq c(x,y)-\varphi(x)$. Hence, function~$\psi$ can be taken equal to
    \begin{eqnarray*}
        \varphi^c(y) & := & \inf_x\left\{c(x,y)-\varphi(x)\right\},
    \end{eqnarray*}
    which is called the $c$-conjugate of~$\varphi$. Similarly, the dual objective can be further improved if we replace~$\varphi$ with
    \begin{eqnarray*}
        \varphi^{cc}(x) & := & \inf_y\left\{c(x,y)-\varphi^c(y)\right\},
    \end{eqnarray*}
    which is the double conjugate of~$\varphi$. If we iterate the procedure, the pair~$(\varphi^{cc},\varphi^c)$ stays unchanged. Hence, for a pair~$(\varphi,\psi)$ to achieve the dual optimal value,~$\psi=\varphi^c$ and~$\varphi$ must be $c$-concave in the sense of the following definition, due to \cite{ruschendorf1990characterization}.
    \begin{definition}\label{def:c-concave}
        A function~$\varphi$ is called $c$-concave if it is equal to its double conjugate~$\varphi^{cc}$.
    \end{definition}
    Then, if~$\varphi$ is indeed $c$-concave, and~$(\varphi,\varphi^c)$ achieves the dual optimal value, the set where the constraints bind
    \begin{eqnarray}\label{eq:subdiff}
        \partial_c\varphi & := & \{ (x,y)\in\mathcal X\times\mathcal Y: \varphi(x)+\varphi^c(y)=c(x,y)\}
    \end{eqnarray}
    is the cyclically monotonic support of an optimal transport plan, i.e., of a solution to the primal Kantorovich optimal transport problem. It is called the $c$-subdifferential of~$\varphi$, which explains the notation. The terminology will become clear when we consider the special case of quadratic cost later in this subsection. With definition~\ref{def:c-concave} and~(\ref{eq:subdiff}), we can state the second part of the Kantorovich duality, which gives necessary and sufficient conditions for optimality.
    \begin{theorem}\label{theorem:dual-iff}
        Under the conditions of theorem~\ref{thm:dual}, if in addition~$c$, $c_{\mathcal X}$ and~$c_{\mathcal Y}$ are continuous, then there is a closed $c$-cyclically monotone set~$\Gamma\subset \mathcal X\times\mathcal Y$ such that for any~$\pi\in\mathcal M(\mu,\nu)$ and any $c$-concave function~$\varphi$,
        \begin{enumerate}
            \item $\pi$ is optimal for the Kantorovich problem if and only if it is supported on~$\Gamma$;
            \item $(\varphi,\varphi^c)$ is optimal in the dual Kantorovich problem if and only if~$\Gamma\subseteq\partial_c\varphi$.
        \end{enumerate}
    \end{theorem}

    This brings us back to the second part of question~(1): when do optimal transport maps~$T$ exist? In other words, when is there an optimal coupling~$\pi$ that is deterministic? Heuristically, for this to happen, the support of an optimal transport plan must be concentrated on a curve. Since the support of an optimal transport plan is~$\partial_c\varphi$ for some $c$-concave function~$\varphi$, for it to be concentrated on a curve, the set~$\partial_c\varphi(x):=\{ y\in\mathcal Y: \varphi(x)+\varphi^c(y)=c(x,y)\}$ must be reduced to a single~$y$ for almost all~$x$. Since the pair~$(\varphi,\varphi^c)$ satisfies the dual constraint, for any~$x^\prime$, we have~$\varphi(x^\prime)+\varphi^c(y)\leq c(x^\prime,y)$. Eliminating~$\varphi^c(y)$ yields~$\varphi(x^\prime) - \varphi(x) \leq c(x^\prime,y) - c(x,y).$
    Since this is true for all~$x^\prime$ approaching~$x$, assuming suitable differentiability, we obtain
    \begin{eqnarray}\label{eq:T}
        \nabla\varphi(x) & = & \nabla_xc(x,y).
    \end{eqnarray}
    The latter has a unique solution~$y$, as desired, if~$\nabla_xc(x,y)$ is injective:
    \begin{assumption}(Twist condition)\label{ass:twist}
        
        On its domain of definition, if~$x,y,y^\prime$ satisfy~$\nabla_xc(x,y^\prime)=\nabla_xc(x,y)$, then~$y^\prime=y$.
    \end{assumption}
    Assumption~\ref{ass:twist} rules out, for instance, additive costs of the form $c(x,y)=a(x)+b(y)$, which would lead to multiple optimal couplings.  In order for equation~(\ref{eq:T}) to make sense, we need differentiability of~$\varphi$. All we know about~$\varphi$ is that it is $c$-concave, as part of dual optimal pair. Hence, a sufficient condition is the ($\mu$-almost sure) differentiability  of all $c$-concave functions, which brings us to formulate: 
    \begin{theorem}[Existence and uniqueness of optimal transport maps]
    \label{thm:T}
    \,\vskip1pt\,
        Suppose the following conditions hold:
        \begin{enumerate}
            \item The optimal value~$\mathcal C(\mu,\nu)$ of the optimal transport problem is finite;
            \item The cost~$c$ is differentiable, bounded below, and satisfies assumption~\ref{ass:twist};
            \item Any $c$-concave function is differentiable~$\mu$-almost surely;
            \item The space~$\mathcal X$ is a closed subset of~$\mathbb R^n$ and~$\mu$ is absolutely continuous with respect to Lebesgue measure.
        \end{enumerate}
        Then there is a unique optimal coupling $\pi$ of~$(\mu,\nu)$. It is deterministic and there is a $c$-concave function~$\varphi$ (the transport potential) that satisfies~(\ref{eq:T}) for $\pi$-almost all~$(x,y)$.
    \end{theorem}
    The conditions of theorem~\ref{thm:T} are satisfied\footnote{Note that theorem~\ref{thm:T} does not apply to distance costs~$c(x,y)=d(x,y)$.} in the special case, where the cost function takes the form~$c(x,y)=h(x-y)$ with~$h$ strictly convex. From~(\ref{eq:T}), the optimal transport map then takes the special form~$y=T(x)=x-\left(\nabla h\right)^{-1}\left(\nabla\varphi(x)\right)$. The even more special case of quadratic cost~$c(x,y)=\|x-y\|^2/2$ is particularly important both historically and in terms of applications. In that case, the optimal transport map takes the form~$y=T(x)=x-\nabla\varphi(x)$, which is the gradient of the convex function~$\vartheta (x) := x^\top x/2-\varphi(x)$.
    \begin{theorem}[Brenier-McCann]\label{thm:BMcC}
        Let~$\mathcal X$ and~$\mathcal Y$ be closed subsets of~$\mathbb R^d$. Let~$\mu$ be absolutely continuous with respect to Lebesgue measure. 
        \begin{enumerate}
            \item There exists a deterministic coupling~$(X,\nabla\vartheta(X))$ of~$(\mu,\nu)$, where~$\vartheta$ is convex, and~$\nabla\vartheta$ is~$\mu$ almost surely unique;
            \item If in addition~${\textstyle\int_{\mathcal X}}x^2d\mu(x)<\infty$ and~${\textstyle\int_{\mathcal Y}}y^2d\nu(y)<\infty$, then~$\nabla\vartheta$ is the optimal transport map from~$\mu$ to~$\nu$ with quadratic cost function~$c(x,y)=\|x-y\|^2/2$.
        \end{enumerate}   
    \end{theorem}
    The first part of theorem~\ref{thm:BMcC} says that for any two probability distributions~$\mu$ and~$\nu$ on~$\mathbb R^d$, if~$\mu$ is absolutely continuous, there is a unique gradient of a convex function that pushes~$\mu$ to~$\nu$ in the sense of the change of variables formula~(\ref{eq:CofV}). When the optimal transport problem from~$\mu$ to~$\nu$ is well defined, the second part of theorem~\ref{thm:BMcC} states that this gradient of a convex function is the unique optimal transport map from~$\mu$ to~$\nu$ from theorem~\ref{thm:T}. This relates to a classical result in convex analysis, which states that gradients of convex functions are characterized by cyclical monotonicity of their graph\footnote{With strictly convex costs in~$\mathbb R^n$, $\nabla T$ is diagonalizable with nonnegative eigenvalues. See the remark on the bottom of page~278 of \cite{villani2008optimal}.}. See for instance section~24 of \cite{rockafellar1970convex}.

    Turning now to question~(3), we examine the aspects of optimal transport theory relating to measuring discrepancy between probability distributions. The optimal transport cost between two probability distributions can be used to measure the discrepancy between the two. More precisely, optimal transport can be used to define a family of distances on the space of probability distributions called Wasserstein distances.
    \begin{definition}[Wasserstein distance]\label{def:W}
        The Wasserstein distance of order~$p\in[1,\infty)$ between two probability distributions~$\mu$ on~$\mathcal X\subseteq \mathbb R^d$ and~$\nu$ on~$\mathcal Y\subseteq \mathbb R^d$ is defined by
        \begin{eqnarray*}
            W_p(\mu,\nu) & = & \left( \inf_{\pi\in\mathcal M(\mu,\nu)} \int_{\mathcal X\times\mathcal Y}\|x-y\|^p \; d\pi(x,y)\right)^{1/p}
        \end{eqnarray*}
    \end{definition}
    
    Definition~\ref{def:W} is a theorem-definition, in that it implies that the optimal transport functional~$W_p$ satisfies the properties of a distance, i.e., it is non negative, it equals zero if and only if the two probability distributions are identical, and it satisfies the triangle inequality. In case~$p=1$, which corresponds to cost function~$c(x,y)=\|x-y\|$, the $c$-concave functions (definition~\ref{def:c-concave}) are the 1-Lipschitz functions. Hence, the Kantorovich dual takes the special form
    \begin{eqnarray}\label{eq:Rubinstein}
        W_1(\mu,\nu) & = & \sup_{\varphi\in\mbox{Lip}_1}\left( \int_{\mathcal X}\varphi(x)\;d\mu(x)-\int_{\mathcal Y}\varphi(y)\;d\nu(y) \right),
    \end{eqnarray}
    which is called Kantorovich-Rubinstein dual. Wasserstein distance~$W_1$ was historically named Kantorovich-Rubinstein distance, and is now commonly called Earth Mover's Distance in computer science. The latter name refers to the fact that Wasserstein distances, by definition of optimal transport, define the distance with the amount of mass that needs to be shifted to move from one distribution to another. This is in sharp contrast with traditional distances between absolutely continuous probability distributions that aggregate the difference between densities at each point. 

    Under the conditions of theorem~\ref{thm:T}, there is a unique optimal transport map~$T$ and a deterministic coupling~$(X,T(X))$ of~$(\mu,\nu)$ that solves the optimal transport problem of~$\mu$ to~$\nu$. The conditions of theorem~\ref{thm:T} are satisfied in particular if~$\mu$ is absolutely continuous and the cost function is~$c(x,y)=\|x-y\|^p$, $p>1$. Hence, the Wasserstein distance has the following expression:
    \begin{eqnarray}\label{eq:W}
        W_p(\mu,\nu) & = & \left( \int_{\mathcal X} \| x-T(x) \|^p \; d\mu(x) \right)^{1/p}.
    \end{eqnarray}
    
    Wasserstein distances~$W_p$ induce a notion of convergence of probability distributions.
    \begin{theorem}[Wasserstein convergence]\label{thm:W}
        A sequence~$(\mu_k)_{k\in\mathbb N}$ of probability distributions converges to a probability distribution~$\mu$ in $p$-Wasserstein distance, namely, we have~$W_p(\mu_k,\mu)\rightarrow0$ as~$k\rightarrow\infty$ if and only if the following hold:
        \begin{enumerate}
            \item $\int\zeta(x)d\mu_k(x)\rightarrow\int\zeta(x)d\mu(x)$ for all continuous and bounded functions~$\zeta$;
            \item $\int\|x\|^pd\mu_k(x)\rightarrow\int\|x\|^pd\mu(x)<\infty$.
        \end{enumerate}
    \end{theorem}
    A first observation from theorem~\ref{thm:W} is that convergence in~$W_p$ implies convergence in~$W_q$ when~$p>q$, and convergence in~$W_1$ is the weakest. The second observation is that convergence in~$W_p$ is stronger than traditional convergence in distribution, since it also requires convergence of moments of order~$p$. Another feature of Wasserstein distances is that they satisfy the Kantorovich duality theorem, which provides many computational and technical advantages. The space of probability distributions also inherits geometric features from the base space~$\mathcal X$. A first way to see this is to compare Wasserstein distance with the total variation distance~$\mbox{TV}(\mu,\nu)=\sup_{A\subset\mathcal X}|\mu(A)-\nu(A)|$. It can be shown by Kantorovich duality that~$\mbox{TV}(\mu,\nu)$ is the optimal cost of transporting~$\mu$ to~$\nu$ with cost function~$c(x,y)=1\{x\ne y\}$. Hence, none of the geometry of~$\mathcal X$ is preserved, since the indicator~$1\{x\ne y\}$ says nothing about how far apart~$x$ and~$y$ are. In contrast, the Wasserstein distance~$W_p(\delta_x,\delta_y)$ between the two Dirac masses at~$x$ and~$y$ is equal to the distance between~$x$ and~$y$ themselves. More generally, the Wasserstein distances endow the space of probability distributions with a rich geometry, that involves notions of geodesics, interpolation, convexity, and barycenters.


    Among geometric concepts, interpolation and barycenters are the most directly relevant to econometric applications. Most of the relevant theory on barycenters is with respect to the $2$-Wasserstein distance~$W_2$ on~$\mathcal W_2:=\{\mu:{\textstyle\int_{\mathcal X}}\|x\|^2\;d\mu(x)<\infty\}$. A barycenter of~$K$ probability distributions~$\mu_1,\ldots,\mu_K$ on~$\mathcal X_1\subset\mathbb R^d,\ldots,\mathcal X_K\subset\mathbb R^d$ respectively, is defined, in analogy with the Euclidean case, as the minimizer
    \begin{eqnarray}\label{eq:barycenter}
        \min_\mu\sum_{k=1}^K\lambda_kW_2^2(\mu_k,\mu),
    \end{eqnarray}
    given a collection~$\lambda_1,\ldots,\lambda_K$ of non negative weights that sum to~$1$. Let~$x$ be a point in~$\mathcal X_1\times\ldots\times\mathcal X_K$. In analogy with the fact that the Euclidean barycenter~$\Sigma_{k=1}^K\lambda_kx_k$ of~$(x_1,\ldots,x_K)$ satisfies
    \begin{eqnarray*}
        \sum_{k=1}^K\lambda_k\|x_k-\sum_{j=1}^K\lambda_jx_j\|^2
        & = & \inf_y\sum_{k=1}^K\lambda_k\left\|x_k-y\right\|^2,
    \end{eqnarray*}
    it can be shown that the Wasserstein barycenter is obtained from the following optimization problem:
    \begin{eqnarray}\label{eq:multi-bary}
        \inf_{\pi\in\mathcal M(\mu_1,\ldots,\mu_K)}\int
        \sum_{k=1}^K\lambda_k\|x_k-\sum_{j=1}^K\lambda_jx_j\|^2
        \;d\pi(x_1,\ldots,x_K).
    \end{eqnarray}
    Problem~(\ref{eq:multi-bary}) is called a multi-marginal optimal transport problem with cost function~$c(x_1,\ldots,x_K)=\Sigma_{k=1}^K\|x_k-\Sigma_{j=1}^K\lambda_jx_j\|^2$. In the particular case of problem~(\ref{eq:multi-bary}), by theorem~\ref{thm:quadratic multi}, there is a unique solution, which yields a useful characterization of the Wasserstein barycenter.

    \subsection{Extensions of optimal transport}
    This section briefly describes the main variants of the duality of optimal transport. First, entropic optimal transport is the case, where the primal Kantorovich formulation contains an entropic regularization that penalizes departures of the optimal transport plan from the independent coupling. Second, unbalanced and weak optimal transport relax the sharp constraint on the marginals. Finally, multi-marginal optimal transport extends optimal transport to the problem of finding a coupling (or {\em teaming}) of more than two marginals, while minimizing an aggregate cost.

        \subsubsection{Entropic optimal transport}\label{sec:entropy}

        Entropic optimal transport is a regularization of the Kantorovich formulation of the optimal transport problem that was originally proposed for computational purposes. We discuss this in section~\ref{sec:sink}. However, some theoretical features of entropic optimal transport are also useful in econometric applications. We describe the basic formulation and duality in what follows.

        Let~$H(\pi\vert \pi^\prime)$ be the relative entropy between probability distributions~$\pi$ and~$\pi^\prime$, defined as follows.
        \begin{eqnarray}\label{eq:entropy}
            H(\pi\vert \pi^\prime) & := & \left\{ 
            \begin{array}{ll}
                 \int \ln\left( \frac{d\pi}{d\pi^\prime}\right) \; d\pi &  \mbox{if } \pi \ll \pi^\prime \\ \\
                +\infty & \mbox{otherwise}.
            \end{array}\right.
        \end{eqnarray}
        Entropic optimal transport solves the following problem:
        \begin{eqnarray}\label{eq:EOT}
            \mbox{EOT}(\mu,\nu;\varepsilon) & := & \inf_{\pi\in\mathcal M(\mu,\nu)} \left\{ \int_{\mathcal X\times\mathcal Y}c(x,y)\;d\pi(x,y) + \varepsilon H(\pi\vert \mu\otimes\nu) \right\}.
        \end{eqnarray}
        Entropic optimal transport penalizes joint distributions that are far from the independent coupling~$\mu\otimes\nu$ of~$(\mu,\nu)$. It's a convex problem, so it provides a computationally convenient approximation of~$\mathcal C(\mu,\nu)$ when~$\varepsilon$ is small. The dual version of EOT is also useful in econometric applications.
        \begin{theorem}\label{thm:EOT}
            Under assumption~\ref{ass:cost}, the minimum in~(\ref{eq:EOT}) is attained and equals
            \begin{eqnarray*}
                \sup_{\varphi,\psi} \left\{ \int_{\mathcal X}\varphi(x)d\mu+\int_{\mathcal Y}\psi(y)d\nu -\int_{\mathcal X\times\mathcal Y}\varepsilon e^{\frac{\varphi(x)+\psi(y)-c(x,y)}{\varepsilon}}d\mu(x)d\nu(y)\right\},
            \end{eqnarray*}
            where the supremum is taken over all pairs of continuous functions~$(\varphi,\psi)$.
        \end{theorem}
        Optimal pairs satisfy the first order optimality conditions
        \begin{eqnarray}\label{eq:sink}
            \left\{ \begin{array}{ccc}
                \varphi(x) & = & -\varepsilon \ln \int e^{\frac{\psi(y)-c(x,y)}{\varepsilon}}d\nu(y), \\
                \psi(y) & = & - \varepsilon \ln \int e^{\frac{\varphi(x)-c(x,y)}{\varepsilon}}d\mu(x).
            \end{array}\right.
        \end{eqnarray}
        Moreover, if~$\pi$ is an entropic optimal transport plan, i.e., solves the primal~(\ref{eq:EOT}) and~$(\varphi,\psi)$ is a pair of entropic optimal transport potentials, i.e., solve the dual in theorem~\ref{thm:EOT}, then they are related by
        \begin{eqnarray*}
            \int_{\mathcal X\times\mathcal Y}\zeta(x,y)\;d\pi(x,y) & = & \int_{\mathcal X\times\mathcal Y}\zeta(x,y)\;e^{\frac{\varphi(x)+\psi(y)-c(x,y)}{\varepsilon}}\;d\mu\;d\nu,
        \end{eqnarray*}
        for all integrable functions~$\zeta$. This can be written~$d\pi=e^{\frac{\varphi+\psi-c}{\varepsilon}}d(\mu\otimes\nu)$ in short.

        \subsubsection{Partial and unbalanced optimal transport}

        Partial and unbalanced optimal transport are variants of optimal transport motivated by situations in which the origin and destination marginals have different masses, or the particular application prescribes only part of the mass be transported. It can also accommodate applications where there is some flexibility in the specification of the marginals. In the most common version, unbalanced optimal transport solves the following problem:
        \begin{eqnarray}\label{eq:UOT}
            \mbox{UOT}(\mu,\nu;\varepsilon) & := & \inf_{\pi} \left\{ \int_{\mathcal X\times\mathcal Y}c(x,y)\;d\pi(x,y) + \varepsilon H(\pi_x\vert \mu)  + \eta H(\pi_y\vert \nu)\right\},\hskip20pt
        \end{eqnarray}
        where~$H$ denotes relative entropy, defined in~(\ref{eq:entropy}).
        Unbalanced optimal transport penalizes marginals that are far from~$\mu$ and~$\nu$. The dual version of UOT is also useful in econometric applications.
        \begin{theorem}\label{thm:UOT}
            Under assumption~\ref{ass:cost}, the minimum in~(\ref{eq:UOT}) is attained and equals
            \begin{eqnarray*}
            \mbox{UOT}(\mu ,\nu ;\varepsilon ) &=&\sup_{\varphi ,\psi }\left\{ - \varepsilon  \int_{%
            \mathcal{X}}\exp (-\frac{\varphi (x)}{\varepsilon}-1)\,d\mu -\eta \int_{\mathcal{Y}}\exp (-\frac{\psi (y)}{\eta }-1)\,d\nu \right\}  \\
            &&\hskip50pt\mbox{ s.t. }\varphi (x)+\psi (y)\leq c(x,y) \\
            &=&\sup_{\psi }\left\{ -\varepsilon \int_{\mathcal{X}}\exp (-\frac{\psi ^{c}(x)}{%
            \varepsilon}-1)\,d\mu -\eta \int_{\mathcal{Y}}\exp (-\frac{\psi (y)}{%
            \eta }-1)\,\,d\nu \right\} ,
            \end{eqnarray*}
            where~$\psi^c$ is the~$c$-conjugate of~$\psi$, and the supremum is over continuous functions~$\psi$.
        \end{theorem}

        \subsubsection{Weak optimal transport}\label{sec:weak}

        The basic Kantorovich formulation of the optimal transport problem from section~\ref{sec:basic} can be rewritten:
        \begin{eqnarray*}
            \mathcal C(\mu,\nu) & = & \inf_{\pi\in\mathcal M(\mu,\nu)}\int_{\mathcal X\times\mathcal Y}c(x,y)\;d\pi(x,y) \\
            & = & \inf_{\pi\in\mathcal M(\mu,\nu)}\int_{\mathcal X} \left[ \int_{\mathcal Y} c(x,y) \; d\pi_x(y) \right] \;d\mu(x).
        \end{eqnarray*}
        The term in brackets in the previous display is a function of both~$x$ and the conditional distribution~$\pi_x$ of~$Y$ conditional on~$X=x$ when the joint distribution of~$(X,Y)$ is~$\pi$. Weak optimal transport generalizes this to more general functions~$\mathcal F(x,\pi_x)$ of~$x$ and~$\pi_x$.
        \begin{eqnarray*}
            \mbox{WOT}(\mu,\nu) & = & \inf_{\pi\in\mathcal M(\mu,\nu)}\int_{\mathcal X} \mathcal F(x,\pi_x) \;d\mu(x).
        \end{eqnarray*}
        In addition to traditional optimal transport, many extensions are  special cases of weak optimal transport.
        \begin{enumerate}
            \item Martingale optimal transport. With cost function~$\mathcal F(x,\pi_x):={\textstyle\int} c(x,y)d\pi_x$ if~${\textstyle\int} yd\pi_x(y)=x$ and~$+\infty$ otherwise, weak optimal transport is equivalent to optimal transport with the constraint that~$(X,Y)$ with distribution~$\pi$ is a martingale. 
            \item Entropic optimal transport. If~$\mathcal F(x,\pi_x)= {\textstyle\int} c(x,y)d\pi_x+\varepsilon H(\pi_x | \nu)  $, we obtain the entropic regularization in~\eqref{eq:EOT}.
        \end{enumerate}
        Duality theory for weak optimal transport closely follows Kantorovich duality.
        \begin{theorem}[Duality of weak optimal transport]\label{thm:WOT}
            Let~$\mathcal F(x,\pi_x)$ be lower semi-continuous, bounded below and convex in its second argument. Then~WOT$(\mu,\nu)$ admits a minimizer, and is equal to the dual
            \begin{eqnarray*}
                \sup_{(\varphi,\psi)} \left( \int_{\mathcal X}\varphi(x)d\mu(x)+\int_{\mathcal Y}\psi(y)d\nu(y) \right)
                \mbox{ s.t. } \varphi(x)+\int_{\mathcal Y}\psi(y)d\pi_x(y)\leq\mathcal F(x,\pi_x),
            \end{eqnarray*}
            where the constraint should be satisfied for all Markov kernel $\pi_x$.
            As in the case of traditional optimal transport, if $\pi$ is an optimal primal solution and if the pair~$(\varphi,\psi)$ is an optimal dual solution, then  complementary slackness holds:~$\varphi(x)+{\textstyle\int_{\mathcal Y}}\psi(y)d\pi_x(y)=\mathcal F(x,\pi_x)$, $\mu$-almost surely.
        \end{theorem}

        \subsubsection{Multi-marginal optimal transport}
    
    The optimal transport problem has a natural extension with more than two prescribed marginals~$\mu_1,\ldots,\mu_K$ on~$\mathcal X_1,\ldots,\mathcal X_K$. The Kantorovich formulation is the following.
    \begin{eqnarray}\label{eq:multi}
        \inf_{\pi\in\mathcal M(\mu_1,\ldots,\mu_K)}
        \int_{\mathcal X_1\times\ldots\times\mathcal X_K}c(x_1,\ldots,x_K)
        \;d\pi(x_1,\ldots,x_K).
    \end{eqnarray}
    Considered as a problem of shifting mass~$\mu_1$ on~$\mathcal X_1$ to~$\mathcal X_2\times\ldots\times\mathcal X_K$, it differs from a standard optimal transport problem in two ways. First, the origin and destination dimensions are different. Second, the probability distribution on~$\mathcal X_2\times\ldots\times\mathcal X_K$ is not fully specified, only its marginals~$\mu_2,\ldots,\mu_K$. The Kantorovich duality extends straightforwardly, with dual
    \begin{eqnarray}\label{eq:dual multi}
        \sup_{\varphi_1,\ldots,\varphi_K} \sum_{k=1}^K\int_{\mathcal X_k}\varphi_k(x_k)\; d\mu_k(x_k) \; \mbox{ subject to } \; \sum_{k=1}^K\varphi_k(x_k) \;\leq \;c(x_1,\ldots,x_K),
    \end{eqnarray}
    where the supremum is taken over continuous and bounded functions satisfying the dual constraints. The theory on existence and uniqueness of optimal transport maps, on the other hand, is more involved. Up to relabeling of the variables, the Monge version of the problem takes the following form.
    \begin{eqnarray}\label{eq:Monge multi}
        \inf_{F_2,\ldots,F_K} \int_{\mathcal X_1} c(x_1,F_2(x_1),\ldots,F_K(x_1)) \;d\mu_1(x_1),
    \end{eqnarray}
    where the infimum is taken over functions~$F_k$ that push~$\mu_1$ forward to~$\mu_k$ in the sense of~(\ref{eq:CofV}).
    Two cases are well understood: the case with quadratic cost, and the case with marginals on~$\mathbb R$. In the first case, the cost function is taken to be:
    \begin{eqnarray}\label{eq:MQC}
        c(x_1,\ldots,x_K) & = & \sum_{k<l} \frac{|x_k-x_l|^2}{2}.
    \end{eqnarray}
    The following theorem shows uniqueness of the optimal transport maps~$(F_2,\ldots,F_K)$.
    \begin{theorem}\label{thm:quadratic multi}
        Let~$\mu_1,\ldots,\mu_K$ be absolutely continuous probability distributions on~$\mathbb R^d$. If the cost function is given by~(\ref{eq:MQC}), then problems~(\ref{eq:multi}), (\ref{eq:dual multi}) and~(\ref{eq:Monge multi}) admit solutions, and have equal values. In addition, the Monge version~(\ref{eq:Monge multi}) has a $\mu_1$-almost surely unique solution~$F_k(x_1)=\nabla \phi_k^\ast(\nabla \phi_1(x_1))$ with~$\phi_k(x_k)=|x_k|^2/2-\varphi_k(x_k)$,
        where~$\varphi_1,\ldots,\varphi_K$ are (convex) solutions of the dual~(\ref{eq:dual multi}) and~$\phi^\ast$ is the convex conjugate of~$\phi$ (see notation and preliminaries).
    \end{theorem}

    The second case, which is well understood is the case with univariate marginals with submodular transport cost. We first define submodularity.
    \begin{definition}
        Let~$e_i$ be the characteristic vector of coordinate~$i$ in~$\mathbb R^K$. A function~$c:\mathbb R^K\rightarrow\mathbb R$ is called submodular if for all~$x\in\mathbb R^K$ and all~$t,s>0$, we have
        \begin{eqnarray*}
            c(x+te_i+se_j)+c(x) & \leq & c(x+te_i)+c(x+se_j).
        \end{eqnarray*}
        If~$c$ is twice continuously differentiable, this is equivalent to~$\partial^2 c/\partial x_i\partial x_j\leq0$ for all pairs of distinct~$(i,j)$. The function is called strictly submodular if the inequalities are strict.
    \end{definition}
    \begin{theorem}\label{thm:multi sub}
        Let~$\mu_1,\ldots,\mu_K$ be absolutely continuous probability distributions on strictly compact subsets~$\mathcal X_k$ of~$\mathbb R$. Assume the cost function is continuous, bounded and strictly submodular. Then problems~(\ref{eq:multi}), (\ref{eq:dual multi}) and~(\ref{eq:Monge multi}) admit solutions, and have equal values. In addition, the Monge version~(\ref{eq:Monge multi}) has a $\mu_1$-almost surely unique solution~$(F_2,\ldots,F_K)$, where each~$F_k$ is non-decreasing.
    \end{theorem}
    Theorem~\ref{thm:multi sub} tells us that the Monge multi-marginal optimal transport problem admits the deterministic coupling~$(X_1,F_2(X_1),\ldots,F_K(X_1))$ as solution. This coupling is obtained by monotone rearrangements of each component of the random vector. As a corollary, we have the following rearrangement inequality:
    \begin{eqnarray*}
        \int c(x_1,\ldots,x_K) \; d\pi(x_1,\ldots,x_K) & \geq &
        \int_0^1 c(Q_{\mu_1}(t),\ldots,Q_{\mu_K}(t)) \; dt
    \end{eqnarray*}
    for any joint distribution~$\pi$ with marginals~$(\mu_1,\ldots,\mu_K)$, where~$Q_\mu$ is the quantile function of probability distribution~$\mu$.

    \subsection{Important special cases}

    Important special cases include optimal transport on the real line, between Gaussians and between discrete distributions.

        \subsubsection{Optimal transport in~$\mathbb R$}\label{sec:scalar OT}
        
        The case$\mathcal X\subseteq\mathcal R$ and~$\mathcal Y\subseteq\mathbb R$ yields rich connections between optimal transport theory and the theory of monotone rearrangements of probability distributions. As a result, it yields closed form solutions for optimal transport maps and Wasserstein distances. 
    
    The class of cost functions relevant to econometric applications is the class of submodular costs. For convenience, we give a useful equivalent definition of submodularity in the case of~$\mathbb R^2$:
    \begin{definition}
        A function~$c:\mathbb R^2\rightarrow\mathbb R$ is called submodular if~$x^\prime\leq x$ and~$y^\prime\leq y$ imply
        \begin{eqnarray*}
            c(x,y)+c(x^\prime,y^\prime) & \leq & c(x,y^\prime) + c(x^\prime,y).
        \end{eqnarray*}
        In case~$c$ is twice continuously differentiable, it is equivalent to~$\partial^2c(x,y)/\partial x\partial y\leq0$. The function is called strictly submodular if the inequalities are strict.
    \end{definition}
    The class of (strictly) submodular functions includes functions of the form~$c(x,y)=h(x-y)$, with~$h$ (strictly) convex, or~$h(x+y)$ with~$h$ (strictly) concave. This includes the case~$c(x,y)=|x-y|^p,$ for~$p\geq1$, which is relevant to Wasserstein distances in particular. The general theory simplifies greatly in case of submodular costs in~$\mathbb R^2$ because $c$-cyclical monotonicity of a subset~$\Gamma$ of~$\mathbb R^2$ simplifies to monotonicity. A subset~$\Gamma$ of~$\mathbb R^2$ is said to be monotone if~$(x,y)\in\Gamma$ and~$(x^\prime,y^\prime)\in\Gamma$ imply~$(x-x^\prime)(y-y^\prime)\geq0$. Therefore, optimal transport plans for submodular costs in~$\mathbb R^2$ have monotone supports. Heuristically, this means that mass is transported from the lowest~$x$ to the lowest~$y$, then the second lowest~$x$ to the second lowest~$y$, and so on from left to right. Hence, an optimal transport plan is a monotone rearrangement of mass~$\mu$ into mass~$\nu$. This means that an optimal coupling~$(X,Y)$ of~$(\mu,\nu)$ is such that~$X$ and~$Y$ are comonotone random variables.
    \begin{definition}\label{def:comonotone-1}
        Two random variables~$X$ and~$Y$ are called comonotone if they can be written~$X=Q_X(U)$ and~$Y=Q_Y(U)$ for some~$U$ uniformly distributed on~$[0,1]$, where~$Q_X$ and~$Q_Y$ are the quantile functions of~$X$ and~$Y$ respectively.
    \end{definition}
    By the probability integral transform, it is always possible to write~$X=Q_X(U)$ and~$Y=Q_Y(V)$, for a pair~$(U,V)$ of uniform random variables. The random variables~$X$ and~$Y$ are comonotone, when~$U=V$, so that they are ordered in the same way. Optimality of the comonotone coupling can be formalized with the following result.
    \begin{theorem}\label{thm:CSS}
        Under assumption~\ref{ass:cost} and submodularity of the cost function~$c$, the primal optimal transport problem has the closed form:
        \begin{eqnarray*}
            \mathcal C(\mu,\nu) & = & \int c(Q_\mu(t),Q_\nu(t)) \;dt,
        \end{eqnarray*}
        where~$Q_\mu$ and~$Q_\nu$ are the quantile functions associated with~$\mu$ and~$\nu$ respectively.
    \end{theorem}
    The Cambanis-Simons-Stout monotone rearrangement inequalities are a direct corollary of theorem~\ref{thm:CSS}. So is the closed form solution for Wasserstein distances in~$\mathbb R$:
    \begin{corollary}
        The~$p$-Wasserstein distance between two probability distributions~$\mu$ and~$\nu$ on~$\mathbb R$, when it is defined, is equal to
        \begin{eqnarray*}
            W_p(\mu,\nu) & = & \left( \int | Q_\mu(t) - Q_\nu(t) |^p \; dt \right)^{1/p}.
        \end{eqnarray*}
    \end{corollary}
    When~$\mu$ is absolutely continuous with respect to Lebesgue measure, the comonotonic coupling~$(X,Y):=(Q_\mu(U),Q_\nu(U))$ is also equal to~$(X,Q_\nu(F_\mu(X)))$, where~$F_\mu$ is the cumulative distribution function of~$\mu$. Therefore, the monotone non decreasing map~$T:=Q_\nu\circ F_\mu$ is the unique optimal transport map. Notice that the optimal transport plan and the optimal transport map are independent of the submodular cost function in the definition of the optimal transport problem.
    
        \subsubsection{Gaussian optimal transport}

    Beyond the scalar case described in the previous section, closed form solutions for optimal transport maps are rare. One such closed form solution is the optimal transport map between Gaussian distributions with quadratic transportation cost~$c(x,y)=\|x-y\|^2$. Let~$\mu$ and~$\nu$ be Gaussian probability distributions on~$\mathbb R^d$, with means~$m_1$ and~$m_2$ and covariance matrices~$\Sigma_1$ and~$\Sigma_2$ respectively. If we can find a deterministic coupling~$(X,T(X))$ of~$(\mu,\nu)$, such that~$T$ is the gradient of a convex function, we know from theorem~\ref{thm:BMcC} that such a coupling is unique and that~$T$ is the optimal transport map. 

    Let~$b$ be a vector in~$\mathbb R^d$ and~$A$ a~$d\times d$ matrix. Since affine maps of the form~$x\mapsto b+Ax$, preserve Gaussianity, it is natural to look for~$T$ among affine maps. If~$A$ is symmetric and positive semidefinite, then~$T:x\mapsto b+Ax$ is the gradient of the convex function~$x\mapsto x^\top b+x^\top Ax/2$. There remains to find~$A$ and~$b$ such that~$A$ is symmetric positive semidefinite and~$(X,b+AX)$ is a coupling of~$(\mu,\nu)$.

    If~$A$ is symmetric positive semidefinite and~$X$ has Gaussian distribution~$\mu$ with mean~$m_1$ and covariance~$\Sigma_1$, then~$b+AX$ has Gaussian distribution with mean~$b+Am_1$ and covariance~$A\Sigma_1 A$. Hence, $(X,b+AX)$ is a coupling of~$(\mu,\nu)$ if~$m_2=b+Am_1$ and~$\Sigma_2=A\Sigma_1A$. A symmetric positive semidefinite solution of~$\Sigma_2=A\Sigma_1A$ can be found by noticing that
    \begin{eqnarray*}
        \begin{array}{lllll}
            \left(\Sigma_1^\frac{1}{2}A\Sigma_1^\frac{1}{2}\right)^2 & = & \Sigma_1^\frac{1}{2}A\Sigma_1A\Sigma_1^\frac{1}{2} & = & \Sigma_1^\frac{1}{2}\Sigma_2\Sigma_1^\frac{1}{2}.
        \end{array}
    \end{eqnarray*}
    From the latter, we obtain~$A$ and therefore the optimal transport map~$T$ is
     \begin{eqnarray*}
        T(x) & = & m_2+\Sigma_1^{-\frac{1}{2}}\left(\Sigma_1^\frac{1}{2}\Sigma_2\Sigma_1^\frac{1}{2}\right)^\frac{1}{2}\Sigma_1^{-\frac{1}{2}}(x-m_1).
    \end{eqnarray*}

    Using expression~(\ref{eq:W}) for the Wasserstein distance, we have the closed form
    \begin{eqnarray}
    W_2(\mu,\nu) & = & \left( \|m_1-m_2\|^2 + \mbox{tr}\left( \Sigma_1+\Sigma_2-2\left(\Sigma_1^\frac{1}{2}\Sigma_2\Sigma_1^\frac{1}{2}\right)^{\frac{1}{2}}\right) \right)^{1/2}.    
    \end{eqnarray}
    
        \subsubsection{Optimal transport with binary cost functions}

    The case, where the cost function~$c(x,y):=1\{(x,y)\in \Gamma\}$ is binary is very useful in probability theory and its applications. We will show later that it has several uses in econometrics, particularly in data combination and partial identification of structural models. 

    With cost function~$c(x,y)=1\{(x,y)\in \Gamma\}$, the Kantorovich dual of the optimal transport problem
    \begin{eqnarray*}
        \inf_{\pi\in\mathcal M(\mu,\nu)} \int 1\{(x,y)\in\Gamma\} \;d\pi(x,y) & = & \inf_{\pi\in\mathcal M(\mu,\nu)}\pi(\Gamma)
    \end{eqnarray*}
    takes the form
    \begin{eqnarray*}
        \sup_{\varphi,\psi} \int \varphi(x)\;d\mu(x)+\int \psi(y)\;d\nu(y) \quad \mbox{ s.t. }\quad\varphi(x)+\psi(y)\leq 1\{(x,y)\in \Gamma\}.
    \end{eqnarray*}
    It can be shown that the value of the dual is unchanged if the potential functions~$\varphi$ and~$\psi$ are restricted to be indicator functions themselves, specifically if~$\varphi=1\{x\in A\}$ and~$\psi(y)=-1\{y\in B\}$, and the supremum is taken over pairs of sets~$(A,B)$. The dual constraint then becomes
    \begin{eqnarray*}
        1\{x\in A\} - 1\{y\in B\} & \leq & 1\{(x,y)\in\Gamma\},
    \end{eqnarray*}
    which implies
    \begin{eqnarray*}
        1\{y\in B\} & \geq & \sup_x [1\{x\in A\}-1\{(x,y)\in\Gamma\}].
    \end{eqnarray*}
    The right-hand side of the latter expression is equal to~$1$ if and only if~$y$ is in the set
    \begin{eqnarray}\label{eq:binary costs}
        A^\Gamma:=\{y:\exists x\in A, (x,y)\notin\Gamma\}.
    \end{eqnarray} 
    Therefore, the value of the dual is unchanged if the potential functions~$\varphi$ and~$\psi$ are restricted to~$\varphi=1\{x\in A\}$ and~$\psi(y)=-1\{y\in A^\Gamma\}$, and the supremum is taken over sets~$A$.

    We therefore have the following result (theorem~2.27 in \cite{villani2003topics}):
    \begin{theorem}\label{thm:binary costs}
        Let~$\Gamma$ be a non-empty open subset of~$\mathcal X\times\mathcal Y$. Then,
        \begin{eqnarray*}
            \inf_{\pi\in\mathcal M(\mu,\nu)}\pi(\Gamma) & = & \sup_{A} \left[ \mu(A)-\nu\left(A^\Gamma\right)\right],
        \end{eqnarray*}
        where the supremum is taken over closed subsets of~$\mathcal X$ and~$A^\Gamma$ is defined in~(\ref{eq:binary costs}).
    \end{theorem}
    Taking~$\Gamma:=\{(x,y):d(x,y)>\varepsilon\}$ for some distance~$d$ and some~$\varepsilon\geq0$, a direct application of theorem~\ref{thm:binary costs} is one of Volker Strassen's many celebrated theorems, namely the characterization of the L\'evy-Prokhorov distance between probability distributions (see corollary~1.28 and remark~1.29 in \cite{villani2003topics} for details).

        \subsubsection{Discrete optimal transport}\label{sec:sink}

    We now investigate the formulations of the optimal transport problem and aspects of the theory when the transported distributions have finite support. In that case, the optimal transport problem is also called the optimal assignment problem. We now have finite spaces~$\mathcal X=\{1,\ldots,N\}$ and~$\mathcal Y=\{1,\ldots,M\}$. The distributions~$\mu$ and~$\nu$ on~$\mathcal X$ and~$\mathcal Y$ are now characterized by probability mass vectors~$(\mu_1,\ldots,\mu_N)$ and~$(\nu_1,\ldots,\nu_M)$. A coupling of~$(\mu,\nu)$ is a probability mass vector~$\pi=(\pi_{xy})$ on~$\mathcal X\times\mathcal Y$ with marginals~$\mu$, and~$\nu$, which is now equivalent to the adding up constraints
    \begin{eqnarray}\label{eq:addup}
        \begin{array}{lll}
            \sum_{y=1}^{M}\pi_{xy} = \mu_x & \mbox{and } & \sum_{x=1}^{N}\pi_{xy} = \nu_y.
        \end{array}
    \end{eqnarray}
    In the finite case, a coupling~$\pi$ of~$(\mu,\nu)$ is deterministic if~$M=N$, $\mu_x=\nu_y=1/N$ and~$\pi$ is the matrix of a permutation of~$\{1,\ldots,N\}$. Each~$x\in\mathcal X$ is assigned to exactly one~$y\in\mathcal Y$. The discrete version of the Kantorovich formulation of the optimal transport problem is that of finding a coupling~$\pi$ of~$(\mu,\nu)$, i.e., a distribution on~$\mathcal X\times\mathcal Y$ that satisfies~(\ref{eq:addup}), that achieves
    \begin{eqnarray}\label{eq:discrete primal}
        \mathcal C(\mu,\nu) & = & \min_{\pi}\sum_{x=1}^N\sum_{y=1}^M\pi_{xy}c(x,y).
    \end{eqnarray}
    The discrete version of the Monge formulation of the optimal transport problem is that of finding a deterministic coupling, i.e., a permutation~$\sigma$ of~$\{1,\ldots,N\}$, that achieves
    \begin{eqnarray}\label{eq:discrete Monge}
        \min_{\sigma}\sum_{x=1}^Nc(x,\sigma(x)).
    \end{eqnarray}
    The latter is also called the pure assignment problem. Kantorovich duality in case of discrete distributions is a special instance of the duality of linear programming. Once specialized to distributions with finite support, theorem~\ref{thm:dual} states equality of the primal problem~(\ref{eq:discrete primal}) under the adding up constraints~(\ref{eq:addup}) with the dual
    \begin{eqnarray*}\label{eq:discrete dual}
        \tilde C(\mu,\nu) & = & \max_{\varphi,\psi}\left( \sum_{x=1}^N\mu_x\varphi_x+\sum_{y=1}^M\nu_y\psi_y \right)
        \mbox{ subject to }\varphi_x+\psi_y\leq c(x,y).
    \end{eqnarray*}
    Both the primal and the dual have solutions~$\pi$ and~$(\varphi,\psi)$ respectively. Finally, in the language of linear programming, the concentration of a solution~$\pi$ to the primal on the cyclically monotone set~$\partial_x\varphi$ of~(\ref{eq:subdiff}), where dual constraints are binding, i.e.,
    \begin{eqnarray}\label{eq:CS}
        \pi_{xy} > 0 & \Rightarrow & \varphi_x+\psi_y=c(x,y)
    \end{eqnarray}
    is called complementary slackness. By theorem~\ref{theorem:dual-iff}, if~$\pi\in\mathcal M(\mu,\nu)$ and~$\varphi$ is $c$-concave, then the plan~$\pi$ and the pair of potentials~$(\varphi,\varphi^c)$ are optimal transport solutions if and only if they are complementary in the sense of~(\ref{eq:CS}).
    
    When~$M=N$ and~$\mu_x=\nu_y=1/N$, the Kantorovich form of the discrete optimal transport problem has a deterministic coupling as a solution. This coupling is concentrated on the graph of the map~$y=\sigma(x)$, where~$\sigma$ is a permutation of~$\{1,\ldots,N\}$. The Wasserstein metric also specializes easily to discrete distributions in case~$N=M$. Letting~$d(x,y)$ be a distance on~$\{1,\ldots,N\}$, and letting the cost function be~$c(x,y)=d(x,y)^p$ for some~$p\geq1$, then~$W_p(\mu,\nu)=(\mathcal C(\mu,\nu))^{1/p}$ defines a distance on the set of probability measures on~$\{1,\ldots,N\}$.
    
    \subsection{Computation of optimal transport}\label{sec:comp}

        \subsubsection{Computation of discrete optimal transport}\label{sec:simplex}

        The discrete optimal transport problem in its Monge formulation is also called the optimal assignment problem. The classical algorithms to solve the assignment problem are the Hungarian algorithm of \citeauthor{kuhn1955hungarian} [\citeyear{kuhn1955hungarian,kuhn1956variants}], \cite{munkres1957algorithms}, \cite{edmonds1972theoretical} and the auction algorithm of \cite{bertsekas1979distributed}. All standard computing packages contain efficient implementations of these algorithms. Their generic complexity is~$O(n^3)$. 
        
        However, econometric applications rely mostly on the Kantorovich formulation of the optimal transport problem. Extensions of the Hungarian and the auction algorithms exist for the Kantorovich formulation, but variants of the network simplex algorithm \cite{hitchcock1941distribution}, and shortest path algorithms, \cite{tomizawa1971some}, \cite{jonker1987shortest}, are often preferable. An even more favored alternative is entropy regularization. The discrete Kantorovich optimal transport problem is replaced with the discrete version of the entropic optimal transport with a suitably small~$\varepsilon$. The problem is solved using the iterated proportional fitting (IPFP), or Sinkhorn algorithm of \cite{deming1940least}, \cite{sinkhorn1964relationship}, and \cite{ruschendorf1995convergence}. The latter iterates between~$\varphi$ and~$\psi$ to solve~(\ref{eq:sink}).

        In what follows, we describe a specific form of the network simplex algorithm for the Kantorovich formulation of the optimal transport problem. We also propose a Python implementation of the algorithm for illustration purposes. However, for best results, we recommend using optimized code. For instance, the Python optimal transport library POT, see \cite{flamary2021pot}, implements a variant of the network simplex algorithm, while the Julia optimal transport library OptimalTransport.jl implements a variant of the shortest path algorithm of \cite{jonker1987shortest}. Both also implement the IPFP algorithm for entropy regularized optimal transport.

        \subsubsection*{An illustrative network simplex algorithm}

        Recall the notation from section~\ref{sec:sink}. The origin set is~$\mathcal X:=\{x_1,\ldots,x_M\}$ and the destination set is~$\mathcal Y:=\{y_1,\ldots,y_N\}$. The marginal probability mass vectors are~$\mu:=(\mu_i:=\mu_{x_i})_{i=1}^M$ and~$\nu:=(\nu_j:=\nu_{y_j})_{j=1}^N$ respectively. We look for a joint probability~$\pi:=(\pi_{ij}:=\pi_{x_iy_j})_{ij}$ with marginals~$\mu$ and~$\nu$ that minimizes transport cost~$C$ with elements~$C_{ij}:=c(x_i,y_j)$
        .
        Consider the bipartite graph~$G$ with~$M+N$ nodes in~$\mathcal X\cup\mathcal Y$ and~$MN$ edges in~$\mathcal X\times\mathcal Y$. The network simplex algorithm finds an optimal~$\pi$ by finding optimal weights on the edges of graph~$G$. 

        \begin{enumerate}
            \item First initialize~$\pi$ with an element of~$\mathcal M(\mu,\nu)$ using the {\em North-West Corner rule}: Initialize a row capacity~$r\leftarrow\mu_1$ and a column capacity~$c\leftarrow\nu_1$. Assign to edge~$(i,j)$ the maximum possible weight~$\pi_{ij}\leftarrow\min(r,c)$ and update row capacity~$r\leftarrow r-\pi_{ij}$ and column capacity~$c\leftarrow c-\pi_{ij}$. If~$r=0$ and~$i\leq M$, update~$i\leftarrow i+1$ and adjust~$r\leftarrow \mu_i$. If~$c=0$ and~$j\leq N$, update~$j\leftarrow j+1$ and adjust~$c\leftarrow \nu_j$. Repeat until~$i>M$ or~$j>N$. Assign weight zero to any edge~$(i,j)$ not visited by the algorithm. The North-West corner rule produces a set~$E^\pi$ of exactly~$M+N-1$ edges. By construction, the graph~$G^\pi$ with set of nodes~$\mathcal X\cup\mathcal Y$ and edges~$E^\pi$ has  a single connected component and no cycles.
            \item Take plan~$\pi$ obtained with the North-West corner rule, and look for complementary dual variables~$\varphi=(\varphi_i:=\varphi_{x_i})_i$ and~$\psi=(\psi_j:=\psi_{y_j})_j$. First, set~$\varphi_i^1=0$ for~$i=1$ and set~$\varphi_i^1=\psi_j^1=-\infty$ for all other~$i$'s and all~$j's$. Then propagate along each connected component by setting~$\psi_j^{t+1}:=C_{ij}-\varphi_i^t$ and~$\varphi_i^{t+1}:=C_{ij}-\psi_j^t$ at each step~$t\geq1$ and each active edge, i.e., each pair~$(x_i,y_j)\in E^\pi$.
            \item If no inactive pair~$(i,j)\notin E^\pi$ violates the dual constraint~$\varphi_i+\psi_j\leq C_{ij}$, the dual pair~$(\varphi,\psi)$ is feasible, and therefore~$\pi$ and~$(\varphi,\psi)$ are optimal by theorem~\ref{theorem:dual-iff}. 
            Otherwise, take a pair~$(i,j)\notin E^\pi$ such that~$\varphi_i+\psi_j> C_{ij}$ and add it to~$E^\pi$.
            $E^\pi\cup(i,j)$ now has a cycle, which we denote
            \begin{eqnarray*}
            (i_1=i,j^\prime_1=j), (j^\prime_1,i_2), (i_2,j^\prime_2),\ldots,(i_l,j^\prime_l),(j^\prime_l,i_{l+1}:=i).
            \end{eqnarray*}
            The cycle above, called alternating path, starts with the newly added edge~$(i,j)$. Update~$\pi$ by adding~$+\epsilon>0$ to each edge from~$\mathcal X$ to~$\mathcal Y$ and~$-\epsilon$ to each edge from~$\mathcal Y$ to~$\mathcal X$. Increase~$\epsilon$ until one edge weight hits zero. Hence the latter edge is no longer active, and is replaced by~$(i,j)$ in the set~$E^\pi$ of edge with positive weight. With this updated plan~$\pi$, return to step~(2) and repeat until the dual compatible variables are feasible.
        \end{enumerate}    

        \subsubsection{Power diagrams and computational geometry}\label{sec:power}

        The computation of optimal transport maps often involves a discrete marginal, typically an empirical distribution, and a continuous marginal, often the uniform on~$[0,1]^d$. The computation of the optimal transport map in such contexts is called the semi-discrete optimal transport problem. Let random vector~$X$ have absolutely continuous distribution~$\mu$ on a closed and compact~$\mathcal X$, and~$Y$ have discrete distribution~$\nu$ with probability mass $((y_1,q_1),\ldots,(y_K,q_K))$. The mass points~$y_1,\ldots,y_K$ are in~$\mathbb R^d$, and the weights~$q_1,\ldots,q_K$ are in~$[0,1]$ and sum to~$1$. The Kantorovich dual of the optimal transport problem from~$\mu$ to~$\nu$ is
        \begin{eqnarray*}
            \sup_{\varphi,\psi} \left\{ \int_{\mathcal X}\varphi(x)\,d\mu(x)+\sum_{j=1}^K\psi_j\,q_j \right\} \; \mbox{ s.t. } \varphi(x)+\psi_j\leq c(x,y_j), \mbox{ all }x\in\mathcal X,j\leq K.
        \end{eqnarray*}
        The $c$-conjugate of the vector~$\psi\in\mathbb R^K$ is the 
        function~$\psi^c(x)=\min_{1\leq j\leq K}\{c(x,y_j)-\psi_j\}.$
        The Kantorovich semi-dual given by
        \begin{eqnarray*}
            \sup_{\psi\in\mathbb R^K} \left\{ \int_{\mathcal X}\psi^c(x)\,d\mu(x)+\sum_{j=1}^K\psi_j\,q_j \right\}
        \end{eqnarray*}
        is therefore a finite dimensional optimization problem. For each~$j\leq K$, on the set~$\mathcal L_j(\psi)$, called {\em Laguerre cell} and defined by
        \begin{eqnarray*}
            \mathcal L_j(\psi) & := & \{x\in\mathcal X: \forall k\ne j, c(x,y_k)-\psi_k\geq c(x,y_j)-\psi_j\},
        \end{eqnarray*}
        the $c$-conjugate of~$\psi$ is equal to~$c(x,y_j)-\psi_j$. Hence, the Kantorovich semi-dual can be rewritten
        \begin{eqnarray*}
            \sup_{\psi\in\mathbb R^K} \sum_{j=1}^K\left(\int_{\mathcal L_j(\psi)}(c(x,y_j)-\psi_j)\,d\mu(x)+\psi_j\,q_j \right),
        \end{eqnarray*}
        which is a convex program with optimizer~$\hat\psi$. The first order conditions are~$q_j=\mu(\mathcal L_j(\hat\psi))$ for all~$j$ and the optimal transport map is the piecewise constant map, which takes values~$y_j$ on each Laguerre cell~$\mathcal L_j(\hat\psi)$. The partition of~$\mathcal X$ into Laguerre cells is called a power diagram, and the original algorithm is due to \cite{aurenhammer1998minkowski}. A variant of this algorithm is implemented in the power diagram component of the Julia DelauneyTriangulation.jl library, see \cite{VandenHeuvel2024DelaunayTriangulation}.
        

        \subsubsection{Inverse optimal transport problem: Recovering transport cost}

        In many applications in economics, such as matching markets and trade gravity equations, the computational problem is reversed: instead of computing the optimal transport plan given marginals and cost, one is concerned with recovering transport costs from the optimal transport plan. Here we address the issue of computing the optimal transport cost given perfect knowledge of the optimal transport plan. 
        Here we give details of the algorithm proposed in \cite{carlier2023sista}. 
        
        Assume that the transport plan solves the discrete version of entropic optimal transport with transport cost~$c(x,y)$:
        \begin{eqnarray*}
            \min_{\pi\in\mathcal M(\mu,\nu)}\sum_{xy}c(x,y)\pi_{xy}+\varepsilon\pi_{xy}\ln\pi_{xy},
        \end{eqnarray*}
        with dual (where we have slightly redefined the dual variables in order to drop the $-1$ inside the exponential) 
        \begin{eqnarray*}
            \max_{\varphi,\psi} \left( \sum_x\varphi_x\mu_x+\sum_y\psi_y\nu_y-\varepsilon\sum_{xy}\exp\frac{\varphi_x+\psi_y-c(x,y) }{\varepsilon}\right).
        \end{eqnarray*}
        Optimal~$\pi$ and~$(\varphi,\psi)$ are related by
        \begin{eqnarray*}
            \pi_{xy} & = & \exp\frac{\varphi_x+\psi_y-c(x,y) }{\varepsilon}.
        \end{eqnarray*}
        Assume that the true transport cost belongs to a parametric family~$c(x,y;\beta)$ with finite dimensional parameter~$\beta$. The object of the procedure is to recover~$\beta$ from the knowledge of~$\pi$ and the optimality conditions. Hence we are looking for~$\beta$ such that
        \begin{eqnarray*}
            \pi_{xy}^\beta & := & \exp\frac{\varphi_x+\psi_y-c(x,y;\beta)}{\varepsilon}
        \end{eqnarray*}
        satisfies the marginal constraints 
        \begin{eqnarray}\label{eq:marges}
            \sum_{y}\pi_{xy}^\beta = \mu_x \; \mbox{ and } \; \sum_{x}\pi_{xy}^\beta = \nu_y,
        \end{eqnarray}
        and moment constraints
        \begin{eqnarray}\label{eq:moments}
            \sum_{xy}\nabla_\beta c(x,y;\beta)\;\pi_{xy}^\beta & = & \sum_{xy}\nabla_\beta c(x,y;\beta)\;\pi_{xy}.
        \end{eqnarray}
        Conditions~(\ref{eq:marges}-\ref{eq:moments}) can be seen as the first order conditions of a Poisson maximum likelihood, i.e., the maximization of
        \begin{eqnarray*}
            F(\varphi,\psi,\beta) & := & \sum_{xy} \pi_{xy}(\varphi_x+\psi_y-c(x,y;\beta)) -\sum_{xy} \exp(\varphi_x+\psi_y-c(x,y;\beta)) .
        \end{eqnarray*}
        The algorithm builds on the Sinkhorn algorithm in the following way. Choose step size~$\rho$ and initial values~$\beta^0,\varphi^0$ and~$\psi^0$. The algorithm alternates Sinkhorn updates on the potentials~$\varphi$ and~$\psi$ (to enforce the marginals): 
        \begin{eqnarray*}
            \varphi_x^{t+1} & := & \varepsilon \ln \frac{\mu_x}{\sum_y\exp\frac{\psi_y^t-c(x,y;\beta^t)}{\varepsilon}}, \\
            \psi_y^{t+1} & := & \varepsilon \ln \frac{\nu_y}{\sum_x\exp\frac{\varphi_x^t-c(x,y;\beta^t)}{\varepsilon}},
        \end{eqnarray*}
        with a gradient descent to update~$\beta$:
        \begin{eqnarray*}
            \beta^{t+1} & := & \beta^t-\rho\nabla_\beta F(\varphi^{t+1},\psi^{t+1},\beta^t).
        \end{eqnarray*}
        \cite{carlier2023sista} add a LASSO regularization and show convergence of the proximal version of the algorithm for a class of sparse parametric cost functions.

        \subsubsection{Wasserstein Generative Adversarial Networks}\label{sec:WGAN}
        
        The $1$-Wasserstein distance is increasingly used in machine learning algorithms. Conversely, deep neural nets can be used to compute the~$1$-Wasserstein distance between two probability distributions. The starting point is always the Kantorovich-Rubinstein dual~(\ref{eq:Rubinstein}) of~$W_1(\mu,\nu)$. 
        
        The idea of generative adversarial networks is the competition between a {\em generator} who tries to mimic samples from the true data generating process, and a {\em critic}, who tries to discriminate between the generator's data and data from the true data generating process. The generator uses data~$(Z_1,\ldots,Z_n)$ sampled from a prior~$P$, and transforms them through a parametric function~$g_\theta$ to mimic true data. The critic tries to discriminate between the generator data~$(g_\theta(Z_1),\ldots,g_\theta(Z_n))$ drawn from~$\nu_\theta$ (the push-forward of~$P$ by~$g_\theta$) and true data~$(X_1,\ldots,X_n)$ drawn from the true data generating process~$\mu$. In Wasserstein generative adversarial networks (hereafter WGAN) of \cite{arjovsky2017wasserstein}, the critic uses a parametric family of Lipschitz functions~$\varphi_\eta$ and the criterion
        \begin{eqnarray}\label{eq:WGAN}
            \mathcal D(\eta,\theta) & := & \sum_{i=1}^n\varphi_\eta(X_i)-\sum_{i=1}^n\varphi_\eta(g_\theta(Z_i)).
        \end{eqnarray}
        The critic chooses~$\eta$ to maximize~(\ref{eq:WGAN}) and therefore maximize discrimination. The generator then chooses parameter~$\theta$ to make discrimination as difficult as possible, hence solving
        \begin{eqnarray*}
            \min_\theta\max_\eta \mathcal D(\eta,\theta) & \approx & \min_\theta W_1(\mu,\nu_\theta).
        \end{eqnarray*}
        As indicated in the display above, the inner maximization is an approximation of the $1$-Wasserstein distance between two fixed probability distributions~$\mu$ and~$\nu_\theta$. Parameter values~$\eta^\ast$ that maximize~$\mathcal D(\eta,\theta)$ in~(\ref{eq:WGAN}) and hence approximate~$W_1(\mu,\nu_\theta)$ are typically computed using root mean square propagator (RMSProp). Efficient implementations of RMSProp can be found in the machine learning libraries such as PyTorch in Python and Optimizers.jl in Julia. One aspect of the algorithm that remains unsatisfactory is the way to impose the Lipschitz constraint on the family of neural nets~$\varphi_\eta$ to conform with the Kantorovich-Rubinstein dual~(\ref{eq:Rubinstein}). The current preferred solution is to add a gradient penalty to the critic's objective~(\ref{eq:WGAN}). This term directly penalizes the norm of the derivative of the potential~$\varphi_\eta$:
        \begin{eqnarray*}
            \lambda\frac{1}{N}\sum_{i=1}^N\left(\max\left\{ 0,\left\| \nabla\varphi_\eta\left(\hat X_i\right)\right\|-1\right\}\right)^2,
        \end{eqnarray*}
        where the~$\hat X_i=\varepsilon_i X_i+(1-\varepsilon_i)g_\theta(Z_i)$ are random convex combinations of real and generated observations.
    \subsection{Estimation of optimal transport}
    
    The primitives in optimal transport problems are the marginal distributions to be coupled and the transport cost. In many applications of optimal transport, at least one of the marginals must be estimated from data. In this section, we survey the theory that addresses the question of how the estimation of (one of) the marginals affects:
    \begin{enumerate}
        \item optimal transport plans,
        \item optimal values of transport problems, and Wasserstein distances,
        \item optimal transport maps?
    \end{enumerate}
    We address each question in turn. In all this section, unless otherwise specified, $\mu_n$ denotes the empirical distribution based on a size~$n$ i.i.d. sample of observations~$(X_1,\ldots,X_n)$ with distribution~$\mu$. Most of the results in this section extend to more general estimators of the marginals, denoted~$\hat\mu_n$, as long as they converge in distribution with a suitable rate to the true marginals.

    \subsubsection{Stability of optimal transport plans}

    Under very general conditions, optimal transport plans between estimated marginals converge to an optimal transport plan between the true marginals.

    \begin{theorem}[Stability of optimal transport]
    \label{thm:stability}
        Let~$(c_n)_{n\in\mathbb N}$ be a sequence of functions that converge uniformly to a continuous function~$c$ on~$\mathcal X\times\mathcal Y$. Let~$(\mu_n)_{n\in\mathbb N}$ (resp. $(\nu_n)_{n\in\mathbb N}$) be a sequence of probability distributions on~$\mathcal X$ (resp. $\mathcal Y$) that converge in distribution to~$\mu$ (resp. $\nu$). Finally, for each~$n$, let~$\pi_n$ be an optimal transport plan between~$\mu_n$ and~$\nu_n$ such that
        \begin{eqnarray*}
            \int_{\mathcal X\times\mathcal Y} c_n(x,y)\;d\pi_n(x,y) & < & \infty.
        \end{eqnarray*}
        Then there is a subsequence of~$(\pi_n)_{n\in\mathbb N}$ that converges to an optimal transport plan between~$\mu$ and~$\nu$ with transport cost~$c$.
    \end{theorem}

    In order to quantify the rate of convergence of optimal transport, we need to specify particular classes of transport costs. Most of the results in the literature relate to Wasserstein distances~$W_p$.

    \subsubsection{Estimation of Wasserstein distances}\label{sec:W estimation}

    We consider the rate of convergence of Wasserstein distance~$W_p(\mu_n,\mu)$ to zero. Rates of convergence of~$W_p(\mu_n,\nu_m)$ to~$W_p(\mu,\nu)$ follow from the triangle inequality. First consider~$W_1$. Let~$X$ be a random vector with distribution~$\mu$ and~$\mu_n$ be the empirical distribution relative to the i.i.d. sample~$(X_1,\ldots,X_n)$. By the Kantorovich-Rubinstein duality~(\ref{eq:Rubinstein}),
    \begin{eqnarray*}\label{eq:Empirical Rubinstein}
    W_1(\mu_n,\mu) & = & \sup_{\varphi\in\mbox{\tiny Lip}_1} \frac{1}{n}\sum_{i=1}^n \left( \varphi(X_i)  - \mathbb E \varphi(X_i) \right).
    \end{eqnarray*}
    The right-hand side is the supremum over the class of Lipschitz functions of the empirical process~$\varphi\mapsto\Sigma_{i=1}^n (\varphi(X_i)  - \mathbb E \varphi(X_i))/n$. Hence, asymptotic distributions and rates of convergence can be obtained with empirical process methods. Asymptotic bounds on the rate of convergence of~$\mathbb EW_1(\mu_n,\mu)$ obtained using empirical process theory are the following:
    \begin{eqnarray*}
        \mathbb EW_1(\mu_n,\mu) & \lesssim & \left\{ \begin{array}{cc}
            \frac{1}{\sqrt n} & \mbox{ if }d=1; \\ \\
            \frac{\ln n}{\sqrt n} & \mbox{ if }d=2; \\ \\
            n^{-\frac{1}{d}} & \mbox{ if }d\geq3.
        \end{array}
        \right.
    \end{eqnarray*}
    The rates above are the best attainable rates of convergence, except when~$d=2$, where they are off by a~$\sqrt{\ln n}$ factor. The rate~$n^{-1/d}$ for~$d\geq3$ is an instance of the curse of dimensionality. The rate of convergence slows exponentially with the dimension of the space~$d$. This is related to the fact that the supremum of the empirical process in~$W_1$ is taken over a very large class of functions.

    Some ways to circumvent the curse of dimensionality are the following. In each case, the curse of dimensionality is circumvented in the sense that the rate of convergence is~$n^{-1/2}$ irrespective of dimension~$d$.
    \begin{enumerate}
        \item Smooth Wasserstein Distances: Better rates can be obtained with smoothness conditions on the measures~$\mu$ and~$\nu$. Hence, the smoothed Wasserstein distance is defined as the Wasserstein distance between smoothed versions of the measures. Let~$\mu^\sigma$ be the convolution of~$\mu$ with a normal~$N(0,\sigma^2 I)$, for some (small)~$\sigma>0$. The smoothed $p$-Wasserstein distance~$W_p^\sigma(\mu,\nu)$ between~$\mu$ and~$\nu$ is defined as the $p$-Wasserstein distance~$W_p(\mu^\sigma,\nu^\sigma)$ between the smoothed versions of~$\mu$ and~$\nu$.
        \item Sliced Wasserstein Distances: The curse of dimensionality can also be circumvented by projections into a single dimension. Let~$\alpha$ be a direction, i.e., an element of the unit sphere~$\mathbb S^{d-1}$ in~$\mathbb R^d$. Call~$\mbox{Proj}_\alpha:x\mapsto \mbox{Proj}_\alpha(x)=x^\top\alpha$, and~$\mu_\alpha$ the push-forward of~$\mu$ by~$\mbox{Proj}_\alpha$ in the sense of the change of variables formula~(\ref{eq:CofV}). The one dimensional Wasserstein distances~$W_p(\mu_\alpha,\nu_\alpha)$ can be aggregated over the uniform distribution~$\rho$ on the set of directions~$\mathbb S^{d-1}$ to yield the sliced Wasserstein distance
        \begin{eqnarray*}
            \mbox{SW}_p(\mu,\nu) & := & \left( \int W_p(\mu_\alpha,\nu_\alpha)^p\;d\rho(\alpha)\right)^\frac{1}{p}.
        \end{eqnarray*}
        As a Wasserstein distance between probability distributions~$\mu_\alpha$ and~$\nu_\alpha$ on~$\mathbb R$, the integrand has the closed form solution. Define the directional cumulative distribution function
        \begin{eqnarray*}
            G_\mu(s;\alpha) & := & \int 1\{x^\top \alpha\leq s\} d\mu(x).
        \end{eqnarray*}
        Then
        \begin{eqnarray*}
            \mbox{SW}_p(\mu,\nu) & = & \left( \int \int_0^1\left|G_\mu^{-1}(u;\alpha)-G_\nu^{-1}(u;\alpha)\right|^pdu\;d\rho(\alpha)\right)^\frac{1}{p}.
        \end{eqnarray*}
        The sliced Wasserstein distance defines a metric over the set of probability measures for each~$p\geq1$.
        \item Maximum Mean Discrepancy: Since the curse of dimensionality is due to the size of the function class~$\mbox{Lip}_1$, another way to circumvent it is to take the supremum in~(\ref{eq:Rubinstein}) over a smaller class of functions. Maximum Mean Discrepancy~$\mbox{MMD}(\mu,\nu)$ between~$\mu$ and~$\nu$ is defined in this way, where the chosen space of functions is the unit ball of a reproducing kernel Hilbert space.
        \item Entropic regularization: \cite{mena2019statistical} derive a central limit theorem for the entropy regularized Wasserstein distance, i.e., $EOT(\mu,\nu,\varepsilon)$ of section~\ref{sec:entropy} for quadratic transport cost. Let~$\mu_n$ be the empirical distribution associated with an i.i.d. sample of size~$n$ from~$\mu$. Assume~$\mu$ is subgaussian, i.e., $\mathbb E_\mu\exp(\|X\|^2/(2d))\leq 2$. Then the following CLT holds:
        \begin{eqnarray*}
            \sqrt n\left( \mbox{EOT}(\mu_n,\nu;\varepsilon)-\mathbb E[\mbox{EOT}(\mu,\nu;\varepsilon)]\right) & \rightsquigarrow_d & N(0,\mbox{Var}_\mu(\varphi(X))),
        \end{eqnarray*}
        where~$\varphi$ is the transport potential. \cite{franguridi2025inference} extend the result to more general cost functions.
    \end{enumerate}


    \subsubsection{Estimation of optimal transport maps}\label{sec:Testimation}

    Existing theory on the estimation of optimal transport maps concerns mostly the case of optimal transport with at least one absolutely continuous marginal and quadratic cost. We will concentrate on this case here. As we have seen in theorem~\ref{thm:BMcC}, the optimal transport map from~$\mu$ to~$\nu$ with transport cost~$c(x,y)=\|x-y\|^2$ is~$T=x-\nabla\varphi(x)$, where~$\varphi$ is a $c$-concave function that achieves the dual. Equivalently, the optimal transport map is~$T=\nabla\vartheta(x)$, where~$\vartheta$ is a convex function that achieves
    \begin{eqnarray*}
        \vartheta & \in & \mbox{arg}\min_{\tilde\vartheta} \int\tilde\vartheta(x)\;d\mu(x)
        + \int\tilde\vartheta^\ast(y)\;d\nu(y),
    \end{eqnarray*}
    where~$\tilde\vartheta^\ast$ is the convex conjugate of~$\tilde\vartheta$. This form is called semi-dual, because the dual constraints are incorporated in the dual objective using the fact that any optimal dual pair must be of the form~$(\vartheta,\vartheta^\ast)$. The optimal transport map can be estimated as the gradient~$\hat T:=\nabla\hat\vartheta$ of a solution~$\hat\vartheta$ to
    \begin{eqnarray}\label{eq:Map estimator}
        \hat\vartheta & \in & \mbox{arg}\min_{\tilde\vartheta\in\mathcal F} \int\tilde\vartheta(x)\;d\mu_n(x)
        + \int\tilde\vartheta^\ast(y)\;d\nu_n(y),
    \end{eqnarray}
    with the usual normalization~$\vartheta(0)=0$.
    Computational issues are taken up in section~\ref{sec:comp}. To derive statistical properties of this estimator of the optimal transport map, we need to specify the class~$\mathcal F$ of convex functions over which the minimum in~(\ref{eq:Map estimator}) is taken. We also need to specify the smoothness of marginals~$\mu$ and~$\nu$. 
    There exists classes~$\mathcal F$ of convex functions such that the solution of~(\ref{eq:Map estimator}) achieves the following rates
    \begin{eqnarray*}
        \mathbb E\|\nabla\hat\vartheta-\nabla\vartheta\|_{L^2(\mu)}^2 & \lesssim & \left\{ \begin{array}{cc}
            n^{-1} & \mbox{ if }d=1; \\ \\
            (\ln n)^2 \; n^{-1} & \mbox{ if }d=2; \\ \\
            (\ln n)^2 \; n^{-\frac{2s}{2s-2+d}} & \mbox{ if }d\geq3.
        \end{array}
        \right.
    \end{eqnarray*}
    In the display above, $s$ is the number of bounded derivatives of~$\nabla\vartheta$.
    Up to logarithm factors, the rates above are minimax rates for estimators of $s$-smooth optimal transport maps between marginals~$\mu$ and~$\nu$ with densities bounded away from~$0$ and~$\infty$ on a compact support. When~$\nabla\hat\vartheta$ minimizes~(\ref{eq:Map estimator}) where~$\mu_n$ and~$\nu_n$ are replaced with Gaussian kernel estimators with bandwidth~$h_n$, the following type of pointwise central limit theorem
    \begin{eqnarray*}
        \sqrt{nh_n^{d-2}}\left(\nabla\hat\vartheta(x)-\nabla\vartheta(x)\right) & \rightsquigarrow & N(0,\Sigma(x)),
    \end{eqnarray*}
    can be shown on the torus~$\mathbb R^d/ \mathbb Z^d$ (to avoid boundary issues) with~$d\geq3$.
   

    
    \subsection{Guide to further reading}

    The most complete account of the history of research on optimal transport theory up to that point is given in chapter~3 of \cite{villani2008optimal}. The history of some more recent developments can be found in the preface of \cite{santambrogio2015optimal}. Each chapter of \cite{villani2008optimal} also contains very detailed bibliographical notes that precisely trace the history of ideas and contributions in each aspect of the theory. The basic formulations and main mathematical questions relating to optimal transport are elegantly presented in the introduction of \cite{villani2003topics}. Theorem~\ref{thm:dual} (the principal Kantorovich duality theorem) is taken from Theorem~1.3 page~19 of \cite{villani2003topics} and theorem~5.10 page~58 of \cite{villani2008optimal}.

    Chapter~4 of \cite{villani2008optimal} gives a proof of existence of optimal transport plans, Theorem~4.1. Chapter~5 of \cite{villani2008optimal} gives a very thorough introduction to $c$-cyclical monotonicity, $c$-concavity, and their relation to optimality and the dual Kantorovich problem.  The relation between $c$-cyclical monotonicity and duality is also detailed in section~1.6 of \cite{santambrogio2015optimal}. Chapter~5 of \cite{villani2008optimal} also presents as complete a picture of Kantorovich duality as needed for applications to econometrics. Theorem~\ref{theorem:dual-iff} is taken from theorem~5.10 of \cite{villani2008optimal}. Theorem~\ref{thm:stability} is taken from theorem~5.20 page 77 of \cite{villani2008optimal}.

    Chapter~10 of \cite{villani2008optimal} gives the best account of the theory underlying solutions to the Monge problem, namely existence of optimal transport maps, and sufficient conditions on the cost function and the marginal distributions. Theorem~\ref{thm:T} is implied by theorem~10.28 page~243 of \cite{villani2008optimal}. Section~1.3 of \cite{santambrogio2015optimal} is a good resource to understand the results in the special case, where the cost function has the form~$c(x,y)=h(x-y)$, with~$h$ strictly convex. Theorem~\ref{thm:CSS} can be found as theorem~2.9 page~63 of \cite{santambrogio2015optimal} for instance. This includes the quadratic cost, which is particularly relevant to econometric applications. Theorem~\ref{thm:BMcC} is a combination of theorem~9.4 page~209 and theorem~10.42, corollary~10.44 and particular case~10.45 on page~256 of \cite{villani2008optimal}. See also chapter~3 of \cite{villani2003topics}.

    Sections~5.1 and~5.2 of \cite{santambrogio2015optimal} give a thorough treatment of Wasserstein distances between probability distributions on~$\mathbb R^d$. Theorem~\ref{thm:W} is theorem~5.9 page 184 of \cite{santambrogio2015optimal}. Chapter~6 of \cite{villani2008optimal} considers the more general case of distributions on a Polish space, which can be useful in applications with distributions over the space of distributions, for example. Section~1.2 of \cite{chewi2024statistical} is also a very good introduction. Section~7.1 of \cite{chewi2024statistical} is a very accessible introduction to geodesics and interpolation in the Wasserstein metric. Chapter~5 of \cite{villani2003topics} introduces displacement convexity. Chapters~16 and~17 of \cite{villani2008optimal} are much more thorough and include more recent material. The original paper \cite{agueh2011barycenters} is still the best reference for the theory of Wasserstein barycenters.

    \cite{nutz2021introduction} gives a very complete account of the mathematics of entropic optimal transport. Theorem~\ref{thm:EOT} is derived from theorem~7 page~35 of \cite{nutz2021introduction}. Theorem~\ref{thm:UOT} is a special case of theorem~1.2.19 page~28 of \cite{chizat2017unbalanced}. \cite{pass2015multi} is an in-depth survey of the theory of multi-marginal optimal transport up to that point. Theorem~\ref{thm:quadratic multi} is taken from theorem~2.1 page~27 of \cite{gangbo1998optimal} and theorem~\ref{thm:multi sub} is taken from theorem~4.1 page~527 of \cite{carlier2003class}. \cite{gozlan2017kantorovich} introduce weak optimal transport\footnote{See also \cite{chone2023weak}.}, and \cite{veraguas2018existence} prove the duality theorem presented in section~\ref{sec:weak}. More precisely, theorem~\ref{thm:WOT} is taken from theorem~3.1 page~203 of \cite{veraguas2018existence}.

    Chapter~3 of \cite{galichon2016optimal} gives a concise presentation of discrete optimal transport duality and a simple constructive proof of the existence of a deterministic solution to the discrete Kantorovich optimal transport problem. Section~2.4 of \cite{peyre2019computational} gives a proof of the triangle inequality for the discrete Wasserstein distance which helps understand the proof for continuous distributions in theorem~7.3 of \cite{villani2003topics}.

    The classical algorithms to compute optimal assignments are presented and reviewed in chapter~3 of \cite{peyre2019computational}: The network simplex algorithm in section~3.5, the Hungarian algorithm in section~3.6 and the auction algorithm in section~3.7.

    \cite{chewi2024statistical} is an excellent recent resource on the estimation of optimal transport. Chapter~2 of \cite{chewi2024statistical} covers the estimation of Wasserstein distances (mostly~$W_1$), rates of convergence and ways of circumventing the curse of dimensionality. The semi-dual estimator of optimal transport maps was introduced in \cite{chernozhukov2017monge}, with a proof of uniform convergence to the population map. Chapter~3 of \cite{chewi2024statistical} derives rates of convergence for the semi-dual estimator of optimal transport maps. The minimax rates of section~\ref{sec:Testimation} are derived in \cite{hutter2021minimax}. The central limit theorem for optimal transport maps can be found in \cite{manole2023central}.

\newpage


\section{Optimal transport as a tool: Econometric methodology}

In this second part of our review, we examine applications of optimal transport methods in econometrics around three major aspects of the theory: first, the duality of optimal transport and its entropic, unbalanced, weak and multi-marginal extensions; second, the uniqueness and cyclical monotonicity of optimal transport maps; third, the metric on the space of probability distributions based on optimal transport. Broadly, the corresponding categories of econometric applications are the following: for the first aspect, partial identification and data combination problems; for the second one, multivariate quantiles and their applications; and for the third one, distributional robustness.

\subsection{Existence of optimal transport plans and Kantorovich duality}

This section reviews econometric applications that involve direct use of optimal transport formulation and the duality of optimal transport for computational advantage in a variety of problems or  for the characterization of identified sets and sharp bounds in partially identified problems, specifically in incomplete models and broadly defined data combination problems. These applications mostly rely on the convenience of the simplex algorithm, which relies on theorem~\ref{theorem:dual-iff}, on rearrangement inequalities of theorem~\ref{thm:CSS}, and on the Kantorovich duality theorem~\ref{thm:dual} and its extensions to entropic~\ref{thm:EOT}, unbalanced~\ref{thm:UOT}, weak~\ref{thm:WOT}, and multi-marginal~\ref{thm:multi sub} optimal transport. Econometric applications often involve conditioning on exogenous covariates, which, as we shall see, can be handled in a variety of ways. 

In a variety of different applications, optimal transport is used to formulate the problem and unlock computational advantages. This includes applications to discrete games with multiple equilibria, in \cite{galichon2011set}, \cite{henry2015combinatorial}, \cite{gu2023dual}, to discrete choice models in \cite{galichon2022cupid}, \cite{chiong2016duality}, \cite{shi2018estimating}, \cite{bonnet2022yogurts}, to treatment assignment under budget constraints in \cite{sunada2025optimal}. Optimal transport duality is used to transform an infinite dimensional optimization problem to a finite dimensional one, as in the partial identification of incomplete models in \citeauthor{galichon2006inference} (\citeyear{galichon2006inference,galichon2009test,galichon2011set}), to transform optimization of expectations with respect to the joint distribution to expectations with respect to the marginals only, in the references above, as in the estimation of distributional treatment effects in \cite{ji2023model}. Monotone rearrangement inequalities of theorem~\ref{thm:CSS} are applied to the characterization of treatment effect bounds in \cite{fan2009partial}, \cite{fan2010sharp}, \cite{fan2014identifying}, and in data combination problems as in \cite{gechter2024generalizing,fan2024multidimensional,d2024linear,meango2025combining}.

In the rest of this section, we review several classes of applications: treatment effects, data combination, incomplete models, and discrete choice and matching estimation. We explain in detail how optimal transport is applied in each and we provide algorithms.

    \subsubsection{Optimal treatment assignment} Optimal transport methods have been used in solving optimal assignment problems, but there is no established literature yet. \cite{kitagawa2025who} consider the problem of optimally matching one population with another to minimize a matching cost (the opposite of a matching surplus). They estimate the entropy regularized optimal assignment based on estimated cost and marginals. \cite{sunada2025optimal} consider a utilitarian planner seeking to maximize utility~$w(x,t)$ from giving treatment~$t\in\{0,1\}$ to individuals with characteristics~$x$. Only a proportion~$p$ of the total population can be treated. What is the propensity score~$\mu(t\vert x)$ that maximizes this constrained optimization problem
    \begin{eqnarray*}
        \max_{\mu(\cdot\vert\cdot)}\int ( w(x,1)\mu(1\vert x)+w(x,0)\mu(0\vert x) )dF_X(x)
    \end{eqnarray*}
    subject to~$\int \mu(1\vert x) dF_X(x)=p$?
    Since the treatment assignment must be Bernoulli random variable with probability of success~$p$ (distribution~$F_T$), the maximum above solves the optimal transport problem
    \begin{eqnarray*}
        \max_{\pi\in\mathcal M(F_X,F_T)}\int w(x,t)d\pi(x,t).
    \end{eqnarray*}
    Here the optimal transport formulation is used for computational convenience.
              
    \subsubsection{Treatment effects}

    A large class of applications of optimal transport duality concerns treatment effects. The treatment effects model we consider has the following main ingredients. A sample of variables~$(Y,D,X)$ is observed. The variable~$D$ is a binary treatment indicator. Unobserved potential outcomes~$(Y_0,Y_1)$ determine the observed outcome through the relation~$Y=Y_0(1-D)+Y_1D$. Outcomes~$Y$ may be scalar or multivariate, depending on the application. Finally, $X$ denotes a vector of exogenous covariates. Assume selection on observables~$(Y_0,Y_1)\perp\!\!\!\!\perp D\vert X$ or any other condition that guarantees identification of the joint distributions of~$(Y_0,X)$ and~$(Y_1,X)$. In all cases reviewed below, $(Y_i,D_i,X_i)_{i=1}^n$ is an i.i.d. sample of observations. We denote~$P_{Y_0,X}$ and~$P_{Y_1,X}$ the joint distributions of~$(Y_0,X)$ and~$(Y_1,X)$ respectively, and~$P_{Y_0,X;n}$ and~$P_{Y_1,X;n}$ their empirical counterparts based on the sample. Similarly, we denote~$P_{Y_0\vert X}$ and~$P_{Y_1\vert X}$ the conditional distributions of~$Y_0\vert X$ and~$Y_1\vert X$ respectively, and~$P_{Y_0\vert X;n}$ and~$P_{Y_1\vert X;n}$ their empirical counterparts based on the sample.

    Optimal transport is applied to this framework in two very different ways. First, optimal transport is applied to covariate matching procedures in conditional average treatment effects estimation in \cite{gunsilius2021matching}, \cite{dunipace2021optimal}, and \cite{charpentier2023optimal}, and optimal experimental design, such as site selection for external validity in \cite{bouyamourn2025where}
    . Second, \cite{fan2009partial}, \cite{fan2010sharp}, \cite{fan2014identifying}, \cite{fan2017partial}, \cite{ji2023model}, \cite{ober2023estimating}, \cite{kaji2023assessing}, \cite{lin2025estimation}, \cite{cote2026worthwhile} and \cite{fan2024multidimensional} apply optimal transport methods to the (partial) identification and estimation of functionals of the joint distribution of~$(Y_0,Y_1,X)$. This is an instance of data combination problem, since~$Y_0$ and~$Y_1$ are never observed for the same individual. A variety of applications to other types of data combination problems will be reviewed in the next subsection.


    \cite{gunsilius2021matching} propose an alternative to propensity score matching for the estimation of average treatment effects in a setting with selection on observables. The idea is to match each treated observation with an average of control observations weighted by the similarity in their covariates to the treated individual. We have 
    \begin{eqnarray}\label{eq:covariate match}
        \begin{array}{lll}
             \mathbb E[Y_1] & = & \mathbb E[\mathbb E[Y_1\vert D=1,X]\mathbb P(D=1\vert X)+\mathbb E[Y_1\vert D=0,X]\mathbb P(D=0\vert X)].
        \end{array}
    \end{eqnarray}

    
    The idea of the covariate matching procedure is to replace the term~$\mathbb E[Y_1\vert D=0,X]$ by a weighted average of controls. Let~$\mu_0$ and~$\mu_1$ be the covariate distributions of control and treated units respectively, and let~$\mu_{0;n}$ and~$\mu_{1;n}$ be their empirical counterparts based on the sample. Let~$\hat\pi$ be the coupling of~$\mu_{0;n}$ and~$\mu_{1;n}$ that solves the unbalanced optimal transport problem~(\ref{eq:UOT}). As an optimal transport solution, this matching of covariates does not require the support of covariates to be the same for treated and control units. In addition, the unbalanced optimal transport problem discards both control and treated units that don't have sufficiently close covariate matches in the opposite group. \cite{gunsilius2021matching} show that it reduces estimation bias. With the unbalanced optimal transport solution~$\hat\pi$, we can construct the weights as the conditional probability distribution~$\hat\pi_{1\vert 0}(x_1\vert x_0)$ of treated group covariates given the value of the control group covariate. Each treated unit~$i$ with covariate~$X_i$ is matched with control unit outcomes~$Y_k$
    with a weight~$\hat\pi_{1\vert 0}(X_k\vert X_i)$. Finally, $\mathbb EY_1$ is estimated with the following empirical version of~(\ref{eq:covariate match}):
    \begin{eqnarray*}
        \frac{1}{n}\sum_{i=1}^n \left( Y_i\;1\{D_i=1\}+\sum_{k\ne i}Y_k\;1\{D_k=0\}\;\hat\pi_{1\vert 0}(X_k\vert X_i)\right).
    \end{eqnarray*}

    We turn now to the second major application of optimal transport to treatment effects, which is partial identification and estimation of functionals of the joint distribution of potential outcomes. When outcomes are scalar, and the function~$h$ is submodular or supermodular, \cite{fan2014identifying} propose to apply monotone rearrangement inequalities of theorem~\ref{thm:CSS} to derive closed form sharp bounds for~$\mathbb E[h(Y_0,Y_1)]$. These closed form solutions can be extended to conditional sharp bounds on~$\mathbb E[h(Y_0,Y_1)\vert X=x]$ based on conditional versions of the monotone rearrangement inequalities of theorem~\ref{thm:CSS}. \cite{kaji2023assessing} extend monotone rearrangement inequalities to derive sharp bounds on subgroup treatment effects. They also derive sharp bounds on the subgroup proportion of winners (those who benefit from the treatment). The latter can be derived using optimal transport duality with zero-one costs functions as we show below. Their bounds analysis is motivated by the policy relevance of the average effect of a treatment or the sign of the treatment effect for the section of the population with the lowest outcomes before treatment. The subgroup treatment effects are defined as~$\mathbb E[Y_1-Y_0\vert a<U_0<b]$. In the latter expression, $U_0$ is the rank of an individual in the untreated outcome distribution. The latter is defined by~$Y_0=Q_0(U_0)$, where~$Q_0$ is the quantile function of~$Y_0$. The sharp upper bound on~$\mathbb E[Y_1-Y_0\vert a<U_0<b]$ is obtained in the rearrangement where individuals ranked from rank~$a$ to rank~$b$ in the untreated distribution are ranked from rank~$1-b+a$ to rank~$1$ in the treated distribution. Similarly, the sharp lower bound is obtained in the rearrangement where individuals ranked from rank~$a$ to rank~$b$ in the untreated distribution are ranked from rank~$0$ to rank~$b-a$ in the treated distribution. The sharp bounds are therefore given by
    \begin{eqnarray*}
        \left[\frac{1}{b-a}\int_a^b[Q_1(u-a)-Q_0(u)]du \, , \, \frac{1}{b-a}\int_a^b[Q_1(1-u+a)-Q_0(u)]du \right].
    \end{eqnarray*}
    The subgroup proportion of winners from theorem~3 in \cite{kaji2023assessing}, can be obtained with an application of theorem~\ref{thm:binary costs} on Kantorovich duality with binary transport cost functions. The subgroup proportion of winners is defined as~$\mathbb P(Y_1>Y_0\vert a< U_0 <b)$ and can be obtained by dividing~$\mathbb P(Y_1>Y_0, a< U_0 <b)$ by~$b-a$. By construction, $U_0$ is the uniform random variable on~$[0,1]$ such that~$Y_0=Q_0(U_0)$. Define~$U_1$ similarly, so that~$Y_1=Q_1(U_1)$ and define~$V_1:=F_0(Q_1(U_1))$, so that by definition~$Y_1>Y_0$ if and only if~$V_1>U_0$. The lower bound on~$\mathbb P(Y_1>Y_0, a< U_0 <b)$ is equal to the solution of the binary cost optimal transport problem
    \begin{eqnarray*}
        \inf_{\pi\in\mathcal M(U_0,V_1)} \pi((U_0,V_1)\in\Gamma), \mbox{ where }\Gamma:=\{(u,v):u<v,\;a<u<b\}.
    \end{eqnarray*} 
    Since~$\mathbb P(Y_1\leq Y_0, a< U_0 <b)=b-a-\mathbb P(Y_1>Y_0, a< U_0 <b)$, the upper bound can be obtained analogously. By theorem~\ref{thm:binary costs}, the lower bound~$\inf_\pi \pi((U_0,V_1)\in\Gamma)$ is equal to its dual~$\sup\left[P_{U_0}(A)-P_{V_1}\left(A^\Gamma\right)\right]$, where~$A^\Gamma:=\{v:\exists u\in A,(u,v)\notin\Gamma\}$. Hence, when~$A\not\subseteq(a,b)$, $v$ is always in~$A^\Gamma$, and~$A$ therefore does not contribute to the dual. Hence, take the supremum over~$A\subseteq(a,b)$. Note that there exists~$u\in A, u\geq v$ if and only if~$v=\bar a:=\sup A$. Hence~$A^\Gamma=[0,\bar a]$. The dual value is non smaller if~$A$ is restricted to take the form~$(a,\bar a]$ (in the terminology of \cite{galichon2011set}, the sets~$A=(a,\bar a]$ form a {\em core determining class}). Finally, since~$V_1=F_0(Q_1(U_1))$, with~$U_1$ uniform on~$[0,1]$, its cdf is~$[F_0\circ Q_1]^{-1}=F_1\circ Q_0$. Hence, the dual is equal to
    \begin{eqnarray*}
        \sup_{\bar a}\left[P_{U_0}((a,\bar a])-P_{V_1}\left([0,\bar a]\right)\right] & = & \sup_{\bar a}\left[\bar a-a-F_1(Q_0(\bar a))\right].
    \end{eqnarray*}
    Dividing by~$b-a$ and adding the non-negativity constraint, we obtain the lower bound in theorem~3 of \cite{kaji2023assessing}. 

    \cite{cote2026worthwhile} consider the probability that the treatment effect exceeds a policy-relevant threshold under monotone treatment response. Building on the directional optimal transport of \cite{nutz2022directional}, which restricts feasible couplings to satisfy~$(Y_1\geq Y_0)$, they derive sharp lower and upper bounds on $(\mathbb{P}(Y_1-Y_0\geq\delta))$ from the marginal distributions of the potential outcomes and provide explicit couplings attaining both bounds. Their analysis thus combines a binary transport cost with an economically meaningful support restriction on the admissible couplings.
    
    Monotone rearrangement inequalities rely on scalar potential outcomes. More generally, one of the major advantages of optimal transport methods is the ability to handle multivariate outcomes. \cite{ji2023model}, \cite{fan2024multidimensional} and \cite{lin2025estimation} apply optimal transport methods to derive sharp bounds for functionals of the joint distribution of possibly multivariate potential outcomes. \cite{fan2024multidimensional} and \cite{lin2025estimation} rely on the primal formulation of the bounds and the \cite{ruschendorf1991bounds} method of conditioning: The lower bound on~$\mathbb E[h(Y_0,Y_1,X)]$ is obtained by taking the minimum over joint distributions for~$(Y_0,Y_1,X)$ with given multivariate marginals for~$(Y_0,X)$ and~$(Y_1,X)$. This situation is different from the standard coupling problem because of the overlapping marginals: $X$ is common to both multivariate marginals. The method of conditioning in \cite{ruschendorf1991bounds} consists in writing
    \begin{eqnarray*}
        \min_{\pi\in\mathcal M(P_{Y_0,X},P_{Y_1,X})} \mathbb E_\pi[h(Y_0,Y_1,X)]
        & = & \int \left[ \min_{\pi\in\mathcal M(P_{Y_0\vert X=x},P_{Y_1\vert X=x})} \mathbb E_\pi[h(Y_0,Y_1,x)] \right] \;dP_X,
    \end{eqnarray*}
    to remove the overlapping marginals. \cite{lin2025tightening} propose an alternative approach to remove the overlapping marginals. They relax the problem to \begin{eqnarray*}
        \min_\pi \mathbb E[h(Y_0,Y_1,X_0)+\eta\|X_0-X_1\|^2]
    \end{eqnarray*}
    under the constraint~$\pi_{Y_0,X_0}\sim P_{Y_0,X}$ and~$\pi_{Y_1,X_1}\sim P_{Y_1,X}$. This relaxation is similar to the approach to conditioning in \cite{li2022finite}.
    \cite{ji2023model} apply a conditional version of Kantorovich duality to derive and estimate sharp bounds on~$\mathbb E[h(Y_0,Y_1,X)]$. For instance, the lower bound is equal to
    \begin{eqnarray}\label{eq:Ji}
        \begin{array}{lllll}
            \theta_L & := & \min_{\pi\in\mathcal M(P_{0,X},P_{1,X})}\mathbb Eh(Y_0,Y_1,X)
            & = & \max_{\varphi_{0,x},\varphi_{1,x}} \mathbb E[\varphi_{0,X}(Y_0)]+\mathbb E[\varphi_{1,X}(Y_1)] \\ \\
            &&& \geq & \mathbb E[\varphi_{0,X}(Y_0)]+\mathbb E[\varphi_{1,X}(Y_1)]
        \end{array}
    \end{eqnarray}
    where the last inequality holds for any~$\varphi_{0,x},\varphi_{1,x}$ such that
    \begin{eqnarray}\label{eq:Ji2}
        \varphi_{0,x}(y_0) + \varphi_{1,x}(y_1) \leq h(y_0,y_1,x),
    \end{eqnarray}
    for all~$x,y_0,y_1$. For each~$x$, the dual functions solve 
    \begin{eqnarray}\label{eq:Ji3}
        (\varphi_{0,x},\varphi_{1,x}) & = & \mbox{arg}\max_{\varphi_{0,x},\varphi_{1,x}} \mathbb E[\varphi_{0,x}(Y_0)\vert X=x]+\mathbb E[\varphi_{1,x}(Y_1)\vert X=x].
    \end{eqnarray}
    The dual formulation~(\ref{eq:Ji}) has two main benefits. First, the minimization problem over the joint distribution is turned into a problem involving only the marginals. Second, the inequality in~(\ref{eq:Ji}) is valid for any choice of function pair~$(\varphi_{0,x},\varphi_{1,x})$ that satisfies the dual constraint~(\ref{eq:Ji2}). Therefore the lower bound~$\mathbb E[\varphi_{0,X}(Y_0)]+\mathbb E[\varphi_{1,X}(Y_1)]$ is valid (if potentially conservative) even if~$(\varphi_{0,x},\varphi_{1,x})$ is not a pair of dual solutions, i.e., it doesn't solve~(\ref{eq:Ji3}). Using this dual representation, \cite{ji2023model} propose the following procedure to conduct inference on the lower bound~$\theta_L$ (and symmetrically for the upper bound).
    \begin{enumerate}
        \item Divide the data sample into two disjoint subsets~$\mathcal D_1$ and~$\mathcal D_2$.
        \item Step~1 using~$\mathcal D_1$:
        \begin{enumerate}
            \item Compute estimators~$\hat P_{Y_0\vert X}$ and~$\hat P_{Y_1\vert X}$ for the conditional distributions of potential outcomes using a machine learning algorithm, a regularized quantile regression or a distributional regression.
            \item Compute a solution pair~$(\hat\varphi_{0,x},\hat\varphi_{1,x})$ to~(\ref{eq:Ji3}), where the expectations are taken with respect to~$\hat P_{Y_0\vert X}$ and~$\hat P_{Y_1\vert X}$. This can be performed using the simplex algorithm of section~\ref{sec:simplex} with a well chosen discretization.
        \end{enumerate}
        \item Given an estimator~$\hat p(x)$ of the propensity score, compute an inverse probability weighting estimator~$\hat\theta_L$ of~$\theta_L$ based on~$\mathcal D_2$:
        \begin{eqnarray*}
            \frac{1}{|\mathcal D_2|}\sum_{i\in\mathcal D_2}\left( \frac{\hat\varphi_{0,X_i}(Y_i)(1-D_i)}{1-\hat p(X_i)} + \frac{\hat\varphi_{1,X_i}(Y_i)D_i}{\hat p(X_i)} \right).
        \end{eqnarray*}
    \end{enumerate}


    \subsubsection{Data combination}

    The treatment effect problems described in the previous subsection is a particular instance of data combination, since both potential outcomes are never observed for the same individual. A variety of other data combination problems can also be addressed with optimal transport methods. Various instances of the short and long regression problem of \cite{cross2002regressions} can be solved with the monotone rearrangement inequalities of theorem~\ref{thm:CSS}, as in \cite{fan2014identifying}, \cite{gechter2024generalizing}, \cite{d2024linear} and \cite{meango2025combining}, and with Kantorovich duality, as in \cite{meango2025combining}. \cite{fan2024multidimensional} provide a unified conditional optimal transport framework to address these and more general data combination problems. \cite{d2025partially} derive sharp bounds in partially linear models with data combination using weak optimal transport.

    Two recent applications of the short and long regression framework of \cite{cross2002regressions} directly use optimal transport methods. The first, \cite{gechter2024generalizing}, considers external validity of the results of randomized experiments. Population~$e$ is treated at random, so~$F^e_{Y_0}$ and~$F^e_{Y_1}$ are identified. In population~$a$ (for alternative), no one is treated, so~$F^a_{Y_0}$ is identified. Call~$f^e_0$ and~$f^a_0$ the identified densities of~$Y_0$ in the experimental and alternative populations respectively. What remains to be identified for the treatment effect in the alternate population is~$\mathbb E^a[Y_1]$. Under the population similarity assumption~$\mathbb E^a[Y_1\vert Y_0]=\mathbb E^e[Y_1\vert Y_0]$, we have
    \begin{eqnarray*}
        \begin{array}{lllllll}
           \mathbb E^a[Y_1]  & = & \mathbb E^a [\mathbb E^a[Y_1\vert Y_0]] & =
             & \mathbb E^a [\mathbb E^e[Y_1\vert Y_0]] & = & \mathbb E^e \left[Y_1\frac{f^a_0(Y_0)}{f^e_0(Y_0)}\right].
        \end{array}
    \end{eqnarray*}
    The monotone rearrangement theorem~\ref{thm:CSS} then provides sharp bounds on~$\mathbb E^a[Y_1]$ as desired: the Fréchet upper bound is attained when the two terms in the product, namely $Y_1$ and $f^a_0(Y_0) /  f^e_0(Y_0) $, are comonotone, and the lower bound, when they are anti-monotone. 
    
    In the second application of optimal transport to short and long regressions, \cite{meango2025combining} propose to use stated preferences from surveys to identify revealed preferences from actual choices. Actual binary choices~$D\in\{0,1\}$ are observed together with an endogenous driver of choice~$X$ in the revealed preference data set. A different data set contains~$X$ and the variable~$P$, which is the stated expected probability of choosing~$D=1$. The quantity of interest is the counterfactual (or potential) choice~$D(x)$ when the driver of choice is externally set to~$x$. The main identifying assumption is that stated preferences reveal the unobserved heterogeneity relevant to choices, i.e., $D(x)\perp\!\!\!\!\perp X\vert P$. Under this assumption,
    \begin{eqnarray*}
        \begin{array}{lllll}
           \mu(x) := \mathbb E[D(x)]  &  = & \int \mathbb E[D(x)\vert P=p]\;dF_{P}(p) & = & \int \mathbb E[D\vert X=x, P=p]\;dF_{P}(p).
        \end{array}
    \end{eqnarray*}
    Using the same strategy as in the external validity example, we can write the previous expression
    \begin{eqnarray*}
        \mu(x) & = & \mathbb E\left[ D\frac{f_P(P)}{f_{P\vert X=x}(P\vert X=x)}\;\vert\; X=x\right],
    \end{eqnarray*}
    and use the monotone rearrangement theorem to derive sharp bounds. However, the latter involve the quantile function of a ratio of density functions, which is not conducive to inference. \cite{meango2025combining} therefore provide an alternative characterization of the bounds based on a direct application of Kantorovich duality.

    \cite{fan2025partial} develop a general framework that includes many data combination problems. They analyze the (partial) identification of finite dimensional~$\theta$ in the moment equality model~$\mathbb Em(Y_0,Y_1,X;\theta)=0$, where~$m$ is a vector of~$d_m$ moment functions. The distributions of~$(Y_0,X)$ and~$(Y_1,X)$ are identified in different data sets, but the joint distribution of~$(Y_0,Y_1,X)$ is not. While they use the potential outcomes notation~$(Y_0,Y_1)$ of treatment effects, their framework is more general. The identified set~$\Theta_I$ is defined as the set of parameter values~$\theta$ such that~$\mathbb E_\pi m(Y_0,Y_1,X;\theta)=0$ for some joint probability distribution~$\pi$, which is compatible with the overlapping marginals~$P_{Y_0,X}$ and~$P_{Y_1,X}$. By the \cite{ruschendorf1991bounds} method of conditioning, the identified set is characterized by the existence of a joint probability distribution~$\pi$ with marginals~$P_{Y_0\vert X=x}$ and~$P_{Y_1\vert X=x}$ such that
    \begin{eqnarray}\label{eq:moment}
        \int\left(\int m(y_0,y_1,x;\theta)d\pi(y_0,y_1\vert X=x) \right) dP_X(x)=0.
    \end{eqnarray} 
    The latter is a problem of existence of a joint probability distribution. It can be transformed into an optimal transport problem, as in \cite{galichon2006inference} and \cite{ekeland2010optimal} (see section~\ref{sec:incomplete} below). See the related result in theorem~1 of \cite{franguridi2025inference}. Heuristically, the existence of a joint probability~$\pi$ that makes~${\textstyle\int} m\;d\pi$ equal to zero is equivalent to~$0$ being larger than~$\inf_\pi{\textstyle\int} \lambda^\top m\;d\pi$ and smaller than~$\sup_\pi{\textstyle\int} \lambda^\top m\;d\pi$ for any direction~$\lambda$ in the unit sphere~$\mathcal S^{d_m}$ ($d_m$ is the dimension of the vector~$m$ of moment functions). Hence, the identified set~$\Theta_I$ is also characterized by
    \begin{eqnarray}\label{eq:moment2}
        \int\left(\inf_{\pi\in\mathcal M(P_{Y_0\vert X=},P_{Y_1\vert X=x})}\int \lambda^\top m(y_0,y_1,x;\theta)\;d\pi(y_0,y_1\vert x) \right) dP_X(x) & \leq & 0,
    \end{eqnarray} 
    for all~$\lambda\in\mathcal S^{d_m}$. The special case of the framework in \cite{fan2024multidimensional}, where the function~$m(y_0,y_1,x;\theta)$ is linear in~$\theta$ yields a useful characterization of the identified set~$\Theta_I$ in a variety of empirically relevant settings. We illustrate the procedure and the characterization of the identified set~$\Theta_I$ in the special case of the linear projection model. Let~$X=(X_p,X_{np})$, where covariates in~$X_p$ appear in the projection and~$X_{np}$ collects additional covariates that do not appear in the projection. Consider the model with scalar~$Y_1$ satisfying
    \begin{eqnarray*}
        Y_1 = (Y_0^\top,X_p^\top) \vartheta +\varepsilon & \mbox{ with } & \mathbb E[\varepsilon(Y_0^\top,X_p^\top)]=0.
    \end{eqnarray*}
    The parameter of interest is~$\vartheta$. The model can be rewritten in the general framework of \cite{fan2024multidimensional} with
    \begin{eqnarray*}
        m(y_0,y_1,x;\theta) = \theta-y_1 y_0 & \mbox{ and } & \vartheta = \left( \begin{array}{ll}
           \mathbb E[Y_0Y_0^\top]  & \mathbb E[Y_0X_p^\top] \\
           \mathbb E[X_pY_0^\top]  & \mathbb E[X_pX_p^\top]          
        \end{array}\right)^{-1} \left(\begin{array}{c}
           \theta  \\ \mathbb E[X_pY_1] 
        \end{array}\right).
    \end{eqnarray*}
    In this case, characterization~(\ref{eq:moment2}) can immediately be rewritten
    \begin{eqnarray}\label{eq:projection}
        \lambda^\top\theta & \leq & \int\left(\sup_{\pi\in\mathcal M(P_{Y_0\vert X=x},P_{Y_1\vert X=x})}\int y_1\left(\lambda^\top y_0\right)\;d\pi(y_0,y_1\vert x) \right) dP_X(x),
    \end{eqnarray}
    for all~$\lambda\in\mathcal S^{d_m}$ (here~$d_m$ is equal to the dimension of~$\theta$, which is also the dimension of~$Y_0$). When~$Y_0$ is scalar, the monotone rearrangement theorem~\ref{thm:CSS} applies directly to the term in brackets in~(\ref{eq:projection}), which is equal to
    \begin{eqnarray}\label{eq:Maurel}
        \int_0^1F^{-1}_{Y_1\vert X=x}(u)F^{-1}_{\lambda^\top Y_0\vert X=x}(u)\;du.
    \end{eqnarray}
    When~$Y_0$ is multivariate, it can be shown that the knowledge of the distribution of~$Y_0$ gives no additional information relative to the knowledge of the distribution of~$\lambda^\top Y_0$, so that bound~(\ref{eq:Maurel}) is still sharp. See the discussion below theorem~1 in \cite{d2024linear}.
    Bound~(\ref{eq:Maurel}) leads directly to the characterization of the identified set for the parameter of interest~$\vartheta$ in theorems~1 and~2 of \cite{d2024linear} and proposition~4.2 in \cite{fan2024multidimensional}. The monotone rearrangement theorem~\ref{thm:CSS} applies in~(\ref{eq:projection}) because~$Y_1$ is scalar. When~$Y_1=(Y_{1l},Y_{1r}^\top)^\top$ is multivariate, as in the case of the linear projection model
    \begin{eqnarray*}
        Y_{1l} = (Y_0^\top,Y_{1r}^\top,X_p^\top) \vartheta +\varepsilon & \mbox{ with } & \mathbb E[\varepsilon(Y_0^\top,Y_{1r}^\top,X_p^\top)]=0,
    \end{eqnarray*}
    the optimal transport characterization of the identified set still provides computational advantages. An important feature of all these bounds in linear projection models is that conditioning on the full vector~$(X_p,X_{np})$ of covariates yields tighter bounds than conditioning only on~$X_p$. Hence, covariates that are irrelevant in the point identified linear projection are no longer irrelevant in case of data combination.
    
    \cite{d2025partially} consider the partially linear regression model
    \begin{eqnarray}\label{eq:partially}
        \mathbb E[Y\vert X_c,X_{nc}]=f(X_c)+\beta^\top X_{nc}.
    \end{eqnarray}
    In the display above, $Y$ is the scalar outcome, and~$X_c$, $X_{nc}$ are vectors of covariates. The unknown function~$f$ and finite dimensional parameter~$\beta$ are the objects of inference. Both~$Y$ and~$X_c$ are observed in one data set, and~$X_{nc}$ and~$X_c$ are observed in another data set. Hence, the distributions of~$(Y,X_c)$ and $(X_{nc},X_c)$ are identified, but the joint distribution of~$(Y,X_{nc},X_c)$ is not. The identification procedure in \cite{d2025partially} follows the steps:
    \begin{enumerate} 
        \item Integrate with respect to~$X_c$ yields:
        \begin{eqnarray*}
            \mathbb E[Y\vert X_c=x]=f(x)+\beta^\top \mathbb E[X_{nc}\vert X_c=x].
        \end{eqnarray*}
        This identifies the value of~$f$ for each fixed value of~$\beta$.
        \item As in \cite{robinson1988root}, take residuals
        \begin{eqnarray*}
            Y^x & := & Y - \mathbb E[Y\vert X_c=x],\\
            X_{nc}^x & := & X_{nc} - \mathbb E[X_{nc}\vert X_c=x],
        \end{eqnarray*} 
        to transform the partially linear model into a linear one:
        \begin{eqnarray*}
            \mathbb E[Y^x\vert X_{nc}^x]=\beta^\top X_{nc}^x.
        \end{eqnarray*}
        \item Characterize the identified set for~$\beta$ using optimal transport.
        \item Add constraints on~$f$ to tighten the identified set. For instance, impose monotonicity and/or convexity of
        \begin{eqnarray*}
            f(x) = \mathbb E[Y\vert X_c=x] - \beta^\top \mathbb E[X_{nc}\vert X_c=x].
        \end{eqnarray*}
    \end{enumerate}
    The crucial step is the characterization of the identified set for~$\beta$ in point~3 of the previous list. Start with the case~$X$ is scalar. Assume~$Y$ has mean zero without loss of generality. The random variables~$Y$ and~$X$ can be coupled in any way that preserves the marginal distribution and the martingale constraint~$\mathbb E[Y\vert X]=\beta X$. The first observation is that we cannot expect much informativeness on~$\beta$ with only marginal information. In particular,~$X\perp\!\!\!\!\perp Y$ rationalizes the data. However, the model is not devoid of empirical content, and the bounds can be informative, when tightened with additional restrictions. With the notation~$X^\beta:= \beta X$, the identified set is the set of~$\beta$'s such that~$\mathbb E[\widetilde Y\vert \widetilde{X^\beta}]=\widetilde{X^\beta}$ for some~$\widetilde Y$ distributed like $Y$ and some~$\widetilde{X^\beta}$ distributed like~$X^\beta$. The marginal distributions are known for $Y$ and $X^\beta$. But nothing is known of their joint distribution, except the martingale constraint. This is a classic problem of coupling under constraints, which is equivalent to Lorenz dominance (see for instance chapter~17.C of \cite{marshall1979inequalities}):
    \begin{eqnarray*}
        \int_\alpha^1F_Y^{-1}(u)\,du\geq\int_\alpha^1F_{X^\beta}^{-1}(u)\,du \mbox{ for all }\alpha\in[0,1].
    \end{eqnarray*}
    Of course, things are not that simple when~$X$ is not scalar, because~$\mathbb E[Y\vert X]$ is no longer the same as~$\mathbb E[Y\vert \beta^\top X]$. There remains to show that the set of~$\beta$'s such that~$\mathbb E[\widetilde Y\vert \widetilde{X^\beta}]=\widetilde{X^\beta}$ for some~$\widetilde Y$ distributed like $Y$ and some~$\widetilde{X^\beta}$ distributed like~$X^\beta$ is the same as the set of~$\beta$'s such that~$\mathbb E[\widetilde Y\vert \widetilde{X}]=\beta^\top\widetilde{X}$ for some~$\widetilde Y$ distributed like $Y$ and some~$\widetilde{X}$ distributed like~$X$. \cite{d2025partially} show this using the duality of weak optimal transport.\footnote{Beyond the scope of this review, weak optimal transport is also used to characterize labor market equilibrium in \cite{chone2021matching} and \cite{paty2022algorithms}.}\footnote{In related work, \cite{d2021rationalizing} characterize rational expectations with the existence of a martingale coupling.}
    

    \subsubsection{Incomplete models}\label{sec:incomplete}

    The first class of applications of optimal transport methods concerns inference in parametric incomplete structural models considered in \citeauthor{galichon2006inference} (\citeyear{galichon2006inference,galichon2009test,galichon2011set}). The vector of variables of interest~$(Y,X,U)$ satisfies support constraint $(Y,X,U) \in \Gamma(\theta)$ almost surely, and~$U$ has fixed and known distribution~$Q_U$.\footnote{The distribution~$Q_U$ may also depend on an unknown finite dimensional parameter.} Both vectors or variables~$Y$ and~$X$ are observed, in the sense that available data consists in a sample~$((Y_1,X_1),\ldots,(Y_n,X_n))$. Variables in vector~$U$ are unobserved. Variables in vector~$X$ are exogenous\footnote{The distribution of~$U$ may also depend on~$X$ as long as the conditional distribution~$Q_{U\vert X}$ is known up to a finite dimensional parameter vector.} (in the sense that~$U\perp X$), and there are no restrictions on the process generating~$(X_1,\ldots,X_n)$. All endogenous variables are subsumed in vector~$Y$. 
    The model is incomplete in that multiple values of endogenous variables may be consistent with a single value of exogenous and unobserved variables. This can be seen in the fact that
    the set $\{ y: (y,x,u)\in\Gamma(\theta) \}$ may not be a singleton for all~$(u,x)$. This corresponds to the fact that the model fails to produce a unique prediction. Recent empirical examples of such models include discrete choice with unobserved heterogeneity in consideration sets in \cite{barseghyan2021heterogeneous} and market structure and competition in airline markets in \cite{ciliberto2021market}.

    The object of inference is the finite dimensional parameter~$\theta\in\Theta$. The identified set for~$\theta$ (also known as sharp identified region) is the set~$\Theta_I$ of all~$\theta$ such that the joint distribution~$P_{Y,X}$ of the data is one of the data generating processes predicted by the model under~$\theta$. As \citeauthor{galichon2006inference} (\citeyear{galichon2006inference}) and \cite{ekeland2010optimal} point out, this is equivalent to the existence of joint distribution with marginals~$P_{Y,X}$ and~$Q_U$ with support contained in~$\Gamma(\theta)$, which in turn is equivalent to~$0$ being the value of the optimal transport problem
    \begin{eqnarray*}
        \min_{\pi\in\mathcal M(P_{Y\vert X=x},Q_U)} \; \pi(\{(y,x,u)\notin\Gamma(\theta)\}\vert X=x).
    \end{eqnarray*}
    This can be equivalently written
    \begin{eqnarray}\label{eq:GHprimal}
        \min_{\pi\in\mathcal M(P_{Y\vert X=x},Q_U)} \int 1{\{(y,x,u)\notin\Gamma(\theta)\}} \; d\pi(y,u\vert X=x) & \leq & 0 \; \mbox{ for all }x.
    \end{eqnarray}
    Since the cost function is an indicator function, we can directly apply theorem~\ref{thm:binary costs} to conclude that~(\ref{eq:GHprimal}) is equivalent to
    \begin{eqnarray*}
        \sup_{A} \;
        \left[ P_{Y\vert X=x}(A\vert X=x)-Q_U\left(A^\Gamma\right) \right] & \leq & 0 \; \mbox{ for all }x, 
    \end{eqnarray*}
    where~$A^\Gamma=\{u\in\mathcal U:\,\exists y\in A,\,(y,x,u)\in\Gamma(\theta)\}$, and where the supremum is over all measurable subsets of the set of outcomes~$y$. This gives a characterization of the identified set with a collection of conditional moment inequalities, called {\em Strassen-Artstein inequalities}. The same characterization of the identified set was obtained by \cite{beresteanu2011sharp} using random set theory and the Artstein theorem (corollary~1.4.11 page~83 of \cite{molchanov2005theory}).
    When the set~$\mathcal Y$ of outcomes is finite, such as the set of strategy profiles in finite action finite player games, this duality is an instance of transformation of an infinite dimensional into a finite dimensional problem. \cite{ekeland2010optimal} propose an extension of the Kantorovich duality to characterize the identified set in incomplete models, where the probability distribution of the latent variable is not specified, but is known to satisfy a set of moment constraints. \cite{schennach2014entropic} and \cite{li2026identification} offer alternative approaches to the same general question based on different optimization problems.

    \cite{galichon2011set} show that the identified set for~$\theta$ can be characterized by the solution of a finite optimal transport problem when the set of outcomes is discrete. Let~$\mathcal Y:=\{y_1,\ldots,y_L\}$ be the finite set of outcomes. Then the set of predicted outcomes (set of equilibria in a finite game) when~$u$ ranges over its (usually continuous) domain is also finite. Call it~$\mathcal U^\ast(x;\theta):=\{u^\ast_1(x;\theta),\ldots,u^\ast_K(x;\theta)\}$.
    Let~$P_{Y\vert X=x}:=(p(y_1\vert x),\ldots,p(y_L\vert x))$ be the probability mass function of~$Y$, and let~$Q_{U^\ast\vert X=x;\theta}:=(q(u^\ast_1\vert x;\theta),\ldots,q(u^\ast_K\vert x;\theta))$, where for each~$k$, 
    \begin{eqnarray*}
        q(u_k^\ast\vert x;\theta) & = & Q_U(\{u:\, y\in u_k^\ast \iff (y,x,u)\in\Gamma(\theta)\}).
    \end{eqnarray*}
    The characterization~(\ref{eq:GHprimal}) of the identified set for~$\theta$ can now be rewritten as the finite optimal transport problem
    \begin{eqnarray*}
        \min_\pi
        \sum_{k=1}^K\sum_{l=1}^L\;\pi_{kl} \; 1_{\{y_l\notin u_k^\ast\}} & \leq & 0,
    \end{eqnarray*}
    where the minimization is over~$\pi$ satisfying the margin constraints
    \begin{eqnarray*}
        \begin{array}{lllllll}
            \sum_{l=1}^L\;\pi_{kl}  & = & q(u_k^\ast\vert x;\theta) & \mbox{ and }
             & \sum_{k=1}^K\;\pi_{kl} & = & p(y_l\vert x),
        \end{array}
    \end{eqnarray*}
    for each~$k\leq K$ and~$l\leq L$.
    
    \cite{li2022finite} also apply optimal transport method to derive a finite sample valid inference method for parameter vector~$\theta$ or a lower dimensional transformation of~$\theta$ such as a subvector. They therefore consider the problem of testing~$H_0(S):\Theta_I\cap S\ne\varnothing$, for some region~$S$ of the parameter space. In the most common case, where a component, say~$\theta_1$, of the vector of structural parameters, with true value~$\theta_{01}$, is of interest, $S:=\{\theta\in\Theta: \theta_1=\theta_{10}\}$, and a confidence region for~$\theta_1$ is obtained by inverting the test, i.e., including all values~$\theta_{10}$ such the test fails to reject at the chosen significance level. 

    The test statistic in \cite{li2022finite} is inspired by the characterization~(\ref{eq:GHprimal}) of the identified set. However, the conditioning over covariates~$X$ is handled by replacing the indicator transport cost function~$1_{\{(y,x,u)\notin\Gamma(\theta)\}}$ with a discrepancy~$\delta$ between~$(u,x)$ and~$(y,x^\prime)$ which is non negative, lower semi-continuous and equal to zero if and only if~$x=x^\prime$ and~$(y,x,u)\in\Gamma(\theta)$.
    The idea of the discrepancy is to simultaneously penalize departures from the model structure and bad covariate matches. \cite{li2022finite} recommend to construct the discrepancy as follows:
    \begin{eqnarray*}
        &&\delta((u,x),(y,x^\prime);\theta) \;:=\\
        &&\hskip30pt \inf_{u^\prime:\;(y,x^\prime,u^\prime)\in\Gamma(\theta)}
        \left( 
        \begin{array}{cc}
            (u-u^\prime)^\top & (x-x^\prime)^\top
        \end{array}\right)
        \left( 
        \begin{array}{cc}
            \Sigma_U & 0 \\
            0 & \hat\Sigma_X
        \end{array}\right)
        \left( 
        \begin{array}{c}
            u-u^\prime \\
            x-x^\prime
        \end{array}\right).
    \end{eqnarray*}
    In the expression above,~$\Sigma_U$ is the (known) covariance matrix of the random vector~$U$ with distribution~$Q_U$ and~$\hat\Sigma_X$ the empirical covariance matrix of the sample~$(X_1,\ldots,X_n)$. This relaxation approach to conditional optimal transport is similar to the one later proposed in \cite{lin2025tightening}.

    The test statistic in \cite{li2022finite} is based on a discrete optimal transport problem based on~(\ref{eq:GHprimal}) with the indicator replaced with the discrepancy~$\delta$. Then the value of the optimal transport problem, which is a function of~$\theta$, is profiled by taking the infimum over all~$\theta\in S$. The inner optimal transport problem is discrete. When the support of outcomes~$y$ is finite, this can be achieved with the \cite{galichon2011set} strategy described above. More generally, it is achieved with a low discrepancy sequence~$(\tilde u_1,\ldots,\tilde u_n)$ that approximates the distribution~$Q_U$. Call~$\Pi_n$ the set of non negative matrices such that $\Sigma_i\pi_{ij}=\Sigma_j\pi_{ij}=1/n$,  for all $i,j \leq n.$ The test statistic is defined as
    \begin{eqnarray}\label{eq:profile}
    T_n(S) & := & \inf_{\theta\in S} \; \min_{\pi\in\Pi_n} \; \sum_{i,j=1}^n\pi_{ij} \; \delta((\tilde u_i,X_i),(Y_j,X_j);\theta).
    \end{eqnarray}
    Replace in~(\ref{eq:profile}) the sample~$(Y_1,\ldots,Y_n)$ with an arbitrary sequence~$\tilde y:=(\tilde y_1,\ldots,\tilde y_n)$ of elements of the outcome space and call the resulting statistic~$T_n(\tilde y;S)$. Call~$\tilde U:=(\tilde U_1,\ldots,\tilde U_n)$ a sample of independent draws from~$Q_U$. Critical values are obtained as quantiles of the statistic
    \begin{eqnarray}\label{eq:critical profile}
        \tilde T_n(S) & := & \sup_{\theta^\prime\in S} \; \sup_{\forall j,\;(\tilde y_j,X_j,\tilde U_j)\in\Gamma(\theta')} \; T_n(\tilde y;S).
    \end{eqnarray}
    In the construction of~(\ref{eq:critical profile}), the null hypothesis is enforced because~$(\tilde y_j,X_j,\tilde U_j)\in\Gamma(\theta')$ for some~$\theta^\prime$ in~$S$. The test has exact validity, because the supremum over~$\theta^\prime$ and~$\tilde y$ means that~$\tilde T_n(S)$ achieves the worst case (largest in first order stochastic dominance) distribution for~$T_n(S)$ under the null. In practice, the critical values are obtained numerically using a large number of replications of~$\tilde T_n(S)$ each with an independent sample~$\tilde U:=(\tilde U_1,\ldots,\tilde U_n)$.

    
    \subsubsection{Discrete choice}

    \cite{galichon2022cupid}, \cite{chiong2016duality}, and \cite{bonnet2022yogurts} formulate the estimation of discrete choice models as an optimal transport problem, which provides a computationally attractive solution to the demand inversion problem. Let agent~$i$ choose option~$y$ among a finite set~$\mathcal Y$. Agent~$i$ chooses~$y$ to maximize utility 
    \begin{eqnarray*}
        u_{iy}(x) & = & \delta_y(x)+\varepsilon_{iy}.
    \end{eqnarray*}
    In the display above, the mean utility~$\delta_y$ of option~$y$ depends on observable characteristics~$X=x$ of the collection of choices~$\mathcal Y$, and~$\varepsilon_{iy}$ is the random utility component. As in \cite{berry1993automobile}, the conditional distribution of the random utility shocks~$\varepsilon$ given~$X=x$ is known and equal to~$P_{\varepsilon\vert X=x}$. Market shares are known and represented by a probability mass function~$q_{Y\vert X=x}$ over~$\mathcal Y$ conditional on~$X=x$. The demand inversion problem is the problem of finding the set of mean utility vectors~$\delta(x)=(\delta_y(x))_{y\in\mathcal Y}$ compatible with market shares~$q_{Y\vert X=x}$. Hence, the object of interest is the set
    \begin{eqnarray*}
        D(q_{Y\vert X=x}) & := & \{\delta(x): \exists Y\sim q_{Y\vert X=x}, Y\mbox{ maximizes }\delta_y(x)+c(y,\varepsilon_{i}\vert x)\}.
    \end{eqnarray*}
    We let~$c(y,\varepsilon_i\vert x_i)=\varepsilon_{iy}$, and define~$G(\delta(x)):=\mathbb {\textstyle\int}[\max_y(\delta_y(x)+c(y,\varepsilon\vert x))]dP_{\varepsilon\vert X=x}(\varepsilon\vert X=x)$. By the envelope theorem~$\delta(x)\in D(q_{Y\vert X=x})$ if and only if~$q_{Y\vert X=x}$ is in the subdifferential~$\partial G(\delta(x))$. By Theorem~23.5 page~218 of \cite{rockafellar1970convex}, the latter is equivalent to~$\delta(x)\in\partial G^\ast(q_{Y\vert X=x})$. Hence, by definition of the convex conjugate~$G^\ast$,
    \begin{eqnarray*}
        \begin{array}{lllll}
            D(q_{Y\vert X=x}) & = & \partial G^\ast(q_{Y\vert X=x}) & = & \mbox{arg}\min_\delta\left(G(\delta(x))-\delta(x)^\top q_{Y\vert X=x}\right).
        \end{array} 
    \end{eqnarray*}
    Therefore~$\delta(x)=-\psi(x)\in D(q_{Y\vert X=x})$ if and only if it minimizes
    \begin{eqnarray*}
            \min_\psi \int[\max_y(c(y,\varepsilon\vert x)-\psi_y(x))]\;dP_{\varepsilon\vert X=x}(\varepsilon\vert X=x)+\sum_yq_{Y\vert X=x}(y\vert X=x)\psi_y(x),
    \end{eqnarray*}
    which is the dual of the optimal transport problem with cost~$-c(y,\varepsilon\vert x)$. In this optimal transport problem, the transport potential~$\psi(x)$ is equal to minus the mean utility~$\delta(x)$. Hence, this optimal transport is a parametric conditional optimal transport problem, where the parametric model is imposed on the potential. \cite{galichon2022cupid} estimate matching surplus in matching markets using a parameterization of the potential in an entropic optimal transport formulation of the problem.
        
        
\subsection{Uniqueness and cyclical monotonicity of optimal transport maps}

This section reviews  econometric applications of optimal transport maps. The aspects of optimal transport theory involved here relate to cyclical monotonicity of optimal transport plans and maps (definition~\ref{def:c-mon} and theorem~\ref{theorem:dual-iff}), the uniqueness of optimal transport maps (theorem~\ref{thm:T}), and optimal transport theory with quadratic costs (theorem~\ref{thm:BMcC}). Cyclical monotonicity provides a multivariate notion of monotonicity. This allows to define multivariate monotone rearrangements in \cite{ekeland2012comonotonic} and multivariate comonotonicity in \cite{ekeland2012comonotonic} and \cite{puccetti2010multivariate}. Multivariate rearrangements are used to define  vector quantile functions in \cite{ekeland2012comonotonic}, \cite{galichon2012dual}, and \cite{chernozhukov2017monge} (see also \cite{hallin2021distribution}) and vector quantile regressions in \cite{carlier2016vector}. In turn, vector quantile functions are used in econometrics to identify nonlinear simultaneous equations in \cite{chernozhukov2021identification} and \cite{gunsilius2023condition}, define robust risk measures in \cite{ekeland2012comonotonic} and \cite{ruschendorf2012law}, copulas between multivariate marginals in \cite{fan2023vector}, multidimensional inequality measures in \cite{fan2022lorenz} and \cite{hallin2025multiple}, distributional difference-in-differences in \cite{torous2024optimal}, principal component analysis in \cite{gunsilius2023independent} and rank based distribution-free nonparametric inference in \cite{deb2023multivariate}.

As explained in section~\ref{sec:scalar OT}, the theory of optimal transport in~$\mathbb R$ is closely related to the theory of monotone rearrangements. Consider the problem of transporting the uniform distribution~$\mu$ on~$[0,1]$ to a given distribution~$\nu$ with quadratic cost~$c(t,y)=|t-y|^2$ for~$t\in[0,1]$ and~$y\in\mathcal Y\subseteq\mathbb R$. Let~$t\mapsto Q_\nu(t)$ be the quantile function of~$\nu$. The cost function is submodular, hence by theorem~\ref{thm:CSS}, the value of the optimal transport problem is
\begin{eqnarray*}
    \mathcal C(\mu,\nu) & = & \int_0^1\left|t-Q_\nu(t)\right|^2\;dt,
\end{eqnarray*}
and the optimal transport map from the uniform~$\mu$ to~$\nu$ is simply~$t\mapsto Q_\nu(t)$, i.e., the quantile function of~$\nu$. The optimal coupling is the comonotone coupling~$(T,Q_\nu(T))$, where~$T$ is uniformly distributed and~$Q_\nu(T)$ is the monotone rearrangement of~$Y\sim\nu$ that minimizes the transport cost from~$\mu$ to~$\nu$.

Consider now the case of a distribution~$\nu$ in~$\mathbb R^d$. By theorem~\ref{thm:BMcC}, we know that there is a unique map~$Q_\nu$ on~$[0,1]^d$ that minimizes the optimal transport problem 
\begin{eqnarray*}
    \min \int_{u\in [0,1]^d}\left\| u-Q_\nu(u)\right\|^2 \;du,
\end{eqnarray*}
which is the cost of transporting the uniform distribution~$\mu$ on~$[0,1]^d$ to~$\nu$ with quadratic cost function~$c(u,y):=\|u-y\|^2$. As an optimal transport map, $Q_\nu$ satisfies a multivariate notion of monotonicity, i.e., $c$-cyclical monotonicity of definition~\ref{def:c-mon}. With quadratic cost~$c(u,y)=\|u-y\|^2$, $c$-cyclical monotonicity boils down to the traditional cyclical monotonicity of \cite{rockafellar1966characterization}, which requires
\begin{eqnarray*}
    \sum_{i=1}^K (Q_\nu(u_i)-Q_\nu(u_{i+1}))^\top\left( u_i-u_{i+1} \right) & \geq & 0
\end{eqnarray*}
for any~$K$, and any collection of vectors~$(u_1,\ldots,u_K)$, setting~$u_{K+1}=u_1$. Cyclical monotonicity characterizes maps that are the gradient of a convex function, which is another way to define multivariate monotonicity (since the derivatives of convex functions are the non-decreasing functions in~$\mathbb R$). Hence, optimal transport maps can be interpreted as multivariate monotone rearrangements. Optimal transport maps from the uniform on~$[0,1]^d$ can therefore define vector quantiles. If~$\nu$ does not have finite second moments, then the unique gradient of a convex function~$Q_\nu$ that transports mass from~$\mu$ to~$\nu$ can still define the vector quantile, even though the optimal transport problem with quadratic cost is no longer well defined. \cite{chernozhukov2017monge} therefore define vector quantiles in the following way.

\begin{definition}[Vector quantile]
    \label{def:VQ}
    A vector quantile~$Q_\nu$ associated with distribution~$\nu$ on~$\mathbb R^d$ is a map~$Q_\nu:[0,1]^d\rightarrow\mathbb R^d$ with the following properties.
    \begin{enumerate}
        \item The map~$Q_\nu$ is cyclically monotone, i.e., the gradient of a convex function.
        \item For any uniformly distributed random variable~$U$ on~$[0,1]^d$, the random vector~$Q_\nu(U)$ has distribution~$\nu$.
    \end{enumerate}
    As shown in theorem~\ref{thm:BMcC}, there exists a map that satisfies~(1) and~(2) and it is unique in the sense that two such maps are equal almost everywhere.
\end{definition}

When~$d=1$, $Q_\nu$ is the traditional quantile function (automatically satisfied when~(1) and~(2) hold), which explains why this notion is proposed as an extension of the traditional notion of quantile. When~$d\geq 1$, and if, in addition, $\nu$ is absolutely continuous, $Q_\nu$ is invertible and~$Q_\nu^{-1}(Y)$ is uniformly distributed on~$[0,1]^d$, whenever~$Y$ has distribution~$\nu$. Thus, when~$\nu$ is absolutely continuous, the map~$Q_\nu^{-1}$ is a cyclically monotone map that transports~$\nu$ into the uniform~$\mu$ on~$[0,1]^d$. Hence, it can be construed as a cardinal to ordinal transformation: It transforms a random vector~$Y$ with distribution~$\nu$ into a uniformly distributed random vector~$U=Q_\nu^{-1}(Y)$ while minimizing distortion in the transformation (because of cyclical monotonicity). This motivates the definition of multivariate ranks in \cite{chernozhukov2017monge}: The multivariate rank of vector~$Y$ is~$U=R_\nu(Y):=Q_\nu^{-1}(Y)$. This also motivates the extension of definition~\ref{def:comonotone-1} to random vectors:
\begin{definition}\label{def:comonotone-d}
    The random vectors~$Y\sim\nu$ and~$Y^\prime\sim\nu^\prime$ are called vector comonotone if they have the same ranks, i.e., if there is a uniformly distributed random vector~$U$ on~$[0,1]^d$ such that~$Y=Q_\nu(U)$ and~$Y^\prime=Q_{\nu^\prime}(U)$.
\end{definition}
Definition~\ref{def:comonotone-d} was originally proposed by \cite{ekeland2012comonotonic}. \cite{puccetti2010multivariate} discuss alternative notions of multivariate comonotonicity, and \cite{torous2024optimal} call it {\em cyclical comonotonicity} and use it to define an analogue of parallel trends for distributional difference-in-differences.

The vector quantile of a random vector with distribution~$\nu$ can be estimated using the semi-discrete algorithm of section~\ref{sec:power} to compute the optimal transport map between the uniform distribution on~$[0,1]^d$ and the empirical distribution~$\nu_n$ of a sample~$(Y_1,\ldots,Y_n)$. The vector rank map can be estimated with the optimal transport map from the empirical distribution~$\nu_n$ to the empirical distribution~$\mu_n$ of a Halton low discrepancy sequence of size~$n$ on~$[0,1]^d$. See section~\ref{sec:sink} for the definition of the discrete optimal transport map and the simplex algorithm~\ref{sec:simplex} for its computation.

In addition to distributional difference-in-difference already mentioned, vector quantiles and ranks are applied to quantile regression in \cite{carlier2016vector}, multivariate stochastic orders in \cite{ekeland2012comonotonic}, \cite{galichon2012dual}, \cite{charpentier2016local} and \cite{fan2022lorenz}, the measurement of risk and inequality in \cite{ekeland2012comonotonic}, \cite{fan2022lorenz} and \cite{hallin2025multiple}, the characterization of dependence between random vectors in \cite{fan2023vector}, distribution-free multivariate inference in \cite{hallin2021distribution} and \cite{deb2023multivariate}, the identification of nonlinear simultaneous equations in \cite{chernozhukov2021identification} and nonlinear principal component analysis in \cite{gunsilius2023independent}. We examine each in turn.

    \subsubsection{Vector quantile regression}

    As noted before, in econometric applications, we are often interested in the conditional version of the optimal transport problem. As vector quantiles are defined based on theorem~\ref{thm:BMcC}, conditional vector quantiles can be defined based on a conditional version of theorem~\ref{thm:BMcC}. We call conditional vector quantile of random vector~$Y$ on~$\mathbb R^d$ conditional on~$X=x$ the almost surely unique cyclically monotone map~$Q_{Y\vert X=x}$ such that~$Q_{Y\vert X=x}(U)$ has the same distribution as~$Y\vert X=x$ for any uniformly distributed~$U$ on~$[0,1]^d$. \cite{carlier2016vector} propose a parametric version of this conditional vector quantile, which extends traditional quantile regression to the multivariate outcome case. The parametric vector quantile takes the form
    \begin{eqnarray}\label{eq:VQR}
        Q_{Y\vert X=x}(u\vert X=x) & = & \beta(u)^\top x,
    \end{eqnarray}
    where~$x$ is a~$p$-valued vector of covariates (including an intercept), and~$\beta(u)$ is the~$p\times d$ valued matrix of regression coefficients and~$\beta(u)^\top x$ is constrained to be cyclically monotone in~$u$ for each~$x$. When~$d=1$, (\ref{eq:VQR}) reduces to the traditional quantile regression. We now explain how the optimal transport formulation allows efficient computation of the regression weights~$\beta(u)$. By definition of the vector quantile, we know from theorem~\ref{thm:BMcC} that~$Q_{Y\vert X=x}(u)=\nabla\varphi(u,x)$,
    where~$\varphi$ solves the semi-dual optimization program
    \begin{eqnarray}\label{eq:VQR semi-dual}
        \inf_{\varphi} \int_{[0,1]^d}\varphi(u,x) du + \int_{\mathcal Y}\varphi^\ast(y,x) \;dF_{Y\vert X=x}(y\vert X=x),
    \end{eqnarray}
    where the infimum is taken over all convex functions, and where $\varphi^\ast(y,x)$ is understood as the convex conjugate at $y$ of $u \mapsto \varphi(u,x)$ for fixed $x$. Under the assumption that~$Q_{Y\vert X=x}(u)=\beta(u)^\top x$, we have~$\nabla\varphi(u,x)=\beta(u)^\top x$, hence the solution to the semi-dual program~(\ref{eq:VQR semi-dual}) is~$\varphi(u,x):=B(u)^\top x$, where~$B$ solves the following program among functions~$B$ such that~$B(u)^\top x$ is a convex function of~$u$.
    \begin{eqnarray*}
        \inf_{B} \int_{[0,1]^d}B(u)^\top x \;du + \int_{\mathcal Y}\sup_u\{u^\top y - B(u)^\top x\} \;dF_{Y\vert X=x}(y\vert X=x).
    \end{eqnarray*}
    After integration over~$X$, the function~$B$ solves
    \begin{eqnarray*}
        \inf_{B} \int_{[0,1]^d}\left(B(u)^\top \mathbb E[X]\right) du + \int_{\mathcal Y\times\mathcal X}\sup_u\left\{u^\top y - B(u)^\top x\right\} \;dF_{Y,X}(y,x).
    \end{eqnarray*}
    Finally, the solution is~$\beta(u)=D B(u)$, the Jacobian of $B$. 
    
    
    \subsubsection{Multivariate stochastic orders and the measurement of risk and inequality}

    Multivariate quantiles and ranks are natural building blocks of multivariate risk and inequality measures. Consider a random vector~$Y$ on~$\mathbb R^d$ with probability distribution~$\nu$ and vector quantile function~$Q_\nu$. In risk analysis, vector~$Y$ is interpreted as a vector of uncertain prospects or financial exposures. In inequality analysis, the vector~$Y$ is interpreted as the vector of attributes (wealth, health, education level) of an individual randomly drawn from the population of interest. A risk or inequality measure is a functional~$\varrho$ that takes random vector~$Y$ and returns a real number. We consider only {\em law invariant} functionals, i.e., functionals that only depend on the distribution of~$Y$ or, equivalently, on its vector quantile function. \cite{galichon2012dual} show that functionals~$\varrho$ are comonotonic additive, i.e., $\varrho(Y+Y^\prime)=\varrho(Y)+\varrho(Y^\prime)$ for comonotonic~$Y$ and~$Y^\prime$ if and only if they are of the form~$\varrho(Y)={\textstyle\int_{[0,1]^d}}\left(\phi(u)^\top Q_Y(u)\right)du$, for some vector function~$\phi:\mathbb R^d\rightarrow\mathbb R^d$. Such a functional~$\varrho$ is called a rank dependent functional, since it is a weighted average of vector ranks only. Rank dependent risk and inequality measures are of this form.

    \cite{charpentier2016local} extend the Bickel-Lehmann dispersive order and its characterization in \cite{landsberger1994co}. A distribution~$\nu$ is more dispersed than a distribution~$\nu^\prime$ if there are random vectors~$Y\sim\nu$, $Y^\prime\sim\nu^\prime$ and~$Z$ vector comonotonic with~$Y^\prime$ such that~$Y=Y^\prime+Z$. They show that it is equivalent to cyclical monotonicity of the difference in quantile maps~$Q_\nu-Q_{\nu^\prime}$. \cite{fan2022lorenz} extend the Lorenz order of increasing inequality to multi-attribute inequality. A population with resource distribution~$\nu$ is more unequal than a population with resource distribution~$\nu^\prime$ if 
    \begin{eqnarray*}
        \int_0^{r_1}\!\!\!\!\cdots\!\int_0^{r_d} Q_\nu(u_1,\ldots,u_d)\;du_1\ldots du_d & \leq & \int_0^{r_1}\!\!\!\!\cdots\!\int_0^{r_d} Q_{\nu^\prime}(u_1,\ldots,u_d)\;du_1\ldots du_d,
    \end{eqnarray*}
    for all~$r=(r_1,\ldots,r_d)\in[0,1]^d$. Similarly to the traditional scalar case, this has the following simple interpretation. For any rank in the population, the cumulative resource in each attribute held by all individuals below that rank is larger in the less unequal population. \cite{fan2022lorenz} show that the increasing inequality ordering shares the traditional interpretation in terms of inequality increasing transfers and preference by any rank dependent inequality measure.

    \subsubsection{Dependence between random vectors}
    
    \cite{fan2023vector} propose an extension of the Sklar theorem and a definition of copula to characterize dependence between two or more random vectors. Let~$Y=(Y_1^\top,\ldots,Y_K^\top)^\top$ be a random vector with distribution~$\nu$ on~$\mathbb R^{d_1}\times \ldots\times \mathbb R^{d_K}$. Each component~$Y_k$ is a random vector on~$\mathbb R^{d_k}$ with vector rank map~$R_k$. Then there is a unique probability distribution~$\nu_C$ such that for each collection of measurable subsets~$(A_1,\ldots,A_K)$
    \begin{eqnarray*}
        \nu(A_1\times \cdots A_K)=\nu_C(R_1(A_1) \times \ldots \times R_K(A_K)).
    \end{eqnarray*}
    The probability distribution~$\nu_C$ on~$[0,1]^{d_1}\times \ldots\times[0,1]^{d_K}$ characterizes the dependence between random vectors~$Y_1,\ldots,Y_K$ independently of each of their multivariate marginals. 
    
    \subsubsection{Multivariate distribution-free testing}
 
    \cite{deb2023multivariate} propose rank based inference for mutual independence and for equality of two distributions. Tests are exactly distribution-free, which means that the null distribution of the test statistics are free of the unknown data generating process at all sample sizes. Let~$(Y_1,\ldots,Y_n)$ be a sample of independently and identically distributed random vectors with distribution~$\nu$ on~$\mathbb R^d$. The vector rank map~$R_\nu$ is the optimal transport map from~$\nu$ to the uniform distribution~$\mu$ on~$[0,1]^d$. Empirical ranks are defined in the following way. Let~$\mathcal V_n:=(v_1,\ldots,v_n)$ be a low discrepancy sequence, i.e., a deterministic sequence of points in~$[0,1]^d$ such that the empirical distribution~$\mu_n:=\Sigma_i\delta_{v_i}/n$ approximates~$\mu$ and converges in distribution faster than the empirical distribution of a random sample. A popular choice is a Halton sequence. The empirical rank map~$R_{\nu,n}$ is the optimal transport map from~$\nu_n$ to~$\mu_n$, i.e., it solves~(\ref{eq:discrete Monge}) in section~\ref{sec:sink}. Finally, define order statistics~$(Y_{(1)},\ldots,Y_{(n)})$ as any arbitrary ordering of the sample (for instance ordering with respect to the first coordinate). Then, the following theorem (proposition~2.2 in \cite{deb2023multivariate} and proposition~1.6.1 of \cite{hallin2021distribution}) shows that multivariate ranks inherit the distribution-free property of univariate ranks.
    \begin{theorem}\label{thm:DFP} Let~$(Y_1,\ldots,Y_n)$ be i.i.d. with absolutely continuous distribution~$\nu$ on~$\mathbb R^d$. Then the following hold.
        \begin{enumerate}
            \item $(R_{\nu,n}(Y_1),\ldots,R_{\nu,n}(Y_n))$ is uniformly distributed on the~$n!$ permutations of~$\mathcal V_n$;
            \item $(R_{\nu,n}(Y_1),\ldots,R_{\nu,n}(Y_n))$ and~$(Y_{(1)},\ldots,Y_{(n)})$ are mutually independent.
        \end{enumerate}
    \end{theorem}
    Using theorem~\ref{thm:DFP} and the combinatorial central limit theorem of \cite{hoeffding1951combinatorial}, \cite{deb2023multivariate} propose a test of independence of two random vectors, and a test of the hypothesis that two random vectors have the same distribution. In both cases, the test statistic is a function of empirical vector ranks only, and the limiting distribution is of the form~$\Sigma_{j=1}^\infty\eta_jZ_j^2$, where the~$Z_j$'s are standard normals and the weights~$\eta_j$ do not depend on the underlying distributions, or the choice of low discrepancy sequence~$\mathcal V_n$. 
    
    \subsubsection{Identification of nonlinear simultaneous equations}

    In traditional systems of~$d$ linear simultaneous equations, the outcome variable~$Y$ is a random vector in~$\mathbb R^{d_y}$ that satisfies~$Y=AX+\varepsilon$, where~$A$ is a matrix of parameters,~$X$ is a vector of covariates and~$\varepsilon$ is a random vector in~$\mathbb R^d_{\varepsilon}$ satisfying~$\mathbb E[\varepsilon\vert X]=0$. We consider identification of the nonparametric non additively separable extension~$Y=F(X,\varepsilon)$. Without further assumption, the distribution of~$\varepsilon$ and the function~$F$ cannot be separately identified. More precisely, for any distribution~$P_{Y\vert X=x}$, and any pair of absolutely continuous distributions~$(P_{\varepsilon\vert X=x},\tilde P_{\tilde\varepsilon\vert X=x})$, we know by \cite{mccann1995existence} that there exists an invertible map~$T$ such that~$F(x,\varepsilon)$ and~$\tilde F(x,\tilde\varepsilon):=F(x,T(\tilde\varepsilon))$ have the same distribution~$P_{Y\vert X=x}$. Therefore, $(F,P_{\varepsilon\vert X=x})$ and~$(\tilde F,\tilde P_{\tilde\varepsilon\vert X=x})$ are observationally equivalent. Hence, we normalize the conditional distribution of~$\varepsilon$ to be a fixed known distribution~$P_{\varepsilon\vert X=x}$. In case~$d_y=d_\varepsilon$, function~$F$ is identified as the unique cyclically monotone map that transports probability distribution~$P_{\varepsilon\vert X=x}$ to~$P_{Y\vert X=x}$. This extends the nonparametric identification result in \cite{matzkin2003nonparametric} to~$d_y=d_\varepsilon>1$. Note that the distribution of~$Y$ need not be absolutely continuous. More generally, \cite{chernozhukov2021identification} also show nonparametric identification of the model~$\varepsilon=H(Y,X)$ in case~$d_Y\geq d_\varepsilon\geq1$. The condition for identification of~$H(Y,X)$ is that~$Y$ is obtained as the hedonic demand of a consumer with type~$(x,\varepsilon)$ and the identification is based on theorem~\ref{thm:T}.
        
    \subsubsection{Principal component analysis} 
        
        \cite{gunsilius2023independent} propose an alternative to principal component analysis based on optimal transport. Consider a random vector~$Y$ in~$\mathbb R^d$ with absolutely continuous distribution~$\nu$ and a predetermined desired number~$k<d$ of principal factors. The proposed procedure is as follows: 
        \begin{enumerate}
            \item Call~$T_\nu$ be the optimal transport map with quadratic cost that transports the standard multivariate normal distribution to~$\nu$, and let~$Z$ be a standard multivariate normal random vector such that~$Y=T_\nu(Z)$. 
            \item Express~$Z$ in terms of an orthonormal basis~$(v^1,\ldots,v^d)$ of~$\mathbb R^d$.
            \item Decompose the relative entropy between~$\nu$ and the standard multivariate normal as a sum of components attributable to each element of the basis.
            \item Optimize the orthonormal basis such that the first~$k$ elements of the basis~$(v^1,\ldots,v^d)$ yield the largest contribution to the relative entropy.
        \end{enumerate}
        The basis vectors~$(v^1,\ldots,v^k)$ are the principal directions. The vector~$Y^{*k}$ is the~$k$-factor approximation of~$Y$. The linear transformation in PCA is replaced with the optimal transport map~$T_\nu$ (if~$Y$ is normal, they coincide) and entropy replaces variance as a way to quantify the relevance of components.

\subsection{Optimal transport distance between distributions}

The Wasserstein distance induces a geometry on the space of probability distributions. This geometry inherits features from the geometry of the base space, typically~$\mathbb R^d$. Heuristically, this is because it is more costly to transport mass far away than to transport it close by. As a result, the Wasserstein distance is useful to model robustness concerns to mis-specification of baseline distributions, in \cite{blanchet2019quantifying}, \cite{adjaho2022externally}, \cite{kido2022distributionally}, \cite{gu2023dual}, and \cite{fan2025quantifying}, and measurement error in \cite{schennach2022optimally} and \cite{forneron2024fitting}. The Wasserstein distance is based on shifts of mass around the baseline space, so it naturally allows the comparison of probability distributions with different supports. This is particularly useful in applications to treatment effect problems with limited overlap between the supports of covariate distributions for the treated and control populations.\footnote{The two properties of Wasserstein distance mentioned above also make it a good candidate to define indices of dependence between random vectors or more abstract random elements. See \cite{catalano2025measures} for a survey on indices based on the Wasserstein distance between the joint distribution and the independent coupling, or between the conditional and unconditional distributions.} The Wasserstein barycenter is used in \cite{gunsilius2023distributional} to define distributional synthetic controls and applied to distributional regression by \cite{zhu2023autoregressive}. The Wasserstein distance has computational advantages that make it useful to simulate from distributions, as shown in \cite{arellano2023recovering} and \cite{athey2024using}, and to estimate parametric models in \cite{kaji2023adversarial} and \cite{fan2024minimum}.

    \subsubsection{Distributional robustness}

    The prototypical distributional robust optimization framework features a reference probability distribution~$\mu$, which is known (possibly only up to a finite dimensional parameter vector), and an adversarial view of nature, which is supposed to choose the true data generating process~$\nu$ within a Wasserstein neighborhood. In this section, we will call Wasserstein neighborhood of a probability distribution~$\mu$ a set of probability distributions~$\nu$ such that
    \begin{eqnarray*}
        \begin{array}{lllll}
           W_\delta(\nu,\mu) & :=  &  \inf_{\pi\in\mathcal M(\nu,\mu)}\int \delta(y,\tilde y) \; d\pi(y,\tilde y) & \leq & \rho,
        \end{array}
    \end{eqnarray*}
    for some~$\rho>0$ and some cost function~$\delta$, in most cases a metric on~$\mathcal Y\subseteq\mathbb R^d$. 
    
    In \cite{blanchet2019quantifying}, the parameter of interest is the expectation of a function~$f$, so that the distributional robust optimization problem is specified as
    \begin{eqnarray}\label{eq:Blanchet1}
        \sup_{\nu} \int f(y)\;d\nu(y), \mbox{ s.t. }W_\delta(\nu,\mu)\leq\rho.
    \end{eqnarray}
    Letting~$\lambda$ denote the Lagrange multiplier associated with the Wasserstein constraint, the Lagrangian formulation of the constrained optimization problem is
    \begin{eqnarray*}
        &&\inf_{\lambda\geq0} \left( \sup_{\pi\in\mathcal M(\nu,\mu)}\int \left(f(y)+\lambda\left(\rho -  \delta(y,\tilde y) \right)\right) \; d\pi(y,\tilde y) \right)\\
        &=&\inf_{\lambda\geq0} \left( \lambda\rho +\sup_{\pi\in\mathcal M(\nu,\mu)}\int \left(f(y)-  \lambda\delta(y,\tilde y) \right) \; d\pi(y,\tilde y) \right).
    \end{eqnarray*}
    The inner optimal transport problem has Kantorovich dual
    \begin{eqnarray*}
        \inf_{\varphi,\psi}\left(\int \varphi(y) \; d\nu(y)+\int \psi(\tilde y) \; d\mu(\tilde y) \right) \mbox{ s.t. }\varphi(y)+\psi(\tilde y)\geq f(y)-  \lambda\delta(y,\tilde y),
    \end{eqnarray*}
    and the semi-dual is achieved by taking~$\varphi(y)=0$ and~$\psi(\tilde y)=\sup_y(f(y)-\lambda\delta(y,\tilde y))$. Hence, the original problem is equal to
    \begin{eqnarray}\label{eq:Blanchet2}
        \inf_{\lambda\geq0} \left\{ \lambda\rho + \int \sup_y \left\{ f(y)-\lambda \delta(y,\tilde y) \right\} \; d\mu(\tilde y) \right\}.
    \end{eqnarray}
    \cite{blanchet2019quantifying} apply this duality result to the computation of worst-case probabilities, i.e., probability distributions~$\nu_{\tiny\mbox{max}}$ and~$\nu_{\tiny\mbox{min}}$ in the Wasserstein neighborhood~$W_\delta(\nu,\mu)\leq\rho$ that achieve the maximum and the minimum of~${\textstyle\int} fd\nu$ respectively. 
    
    In econometric applications of distributionally robust optimization, the problem of conditioning on exogenous covariates is addressed
    in a variety of ways. \cite{adjaho2022externally} consider the problem of choosing treatment assignment~$\tau(x)\in\{0,1\}$ to maximize the utilitarian welfare under adversarial shifts in the joint probability distribution~$P_{Y_0,Y_1,X}$ for the scalar potential outcomes~$Y_0$, $Y_1$, and the vector of covariates~$X$. They assume the planner considers a potentially mis-specified reference joint probability distribution~$P_{Y_0,Y_1,X}$. This can be obtained for instance with access to a pilot treatment data set with selection on observables and an assumption of either perfect correlation of~$Y_0$ and~$Y_1$ conditional on~$X$, or independence of~$Y_0$ and~$Y_1$ conditional on~$X$. The planner then seeks robustness to the misspecification of the joint probability (in particular robustness to the assumption of comonotone or independent coupling) and \cite{adjaho2022externally} define the robust welfare criterion as the maximin response to an adversarial nature. Nature is assumed to choose the worse case data generating process~$\nu$ in the Wasserstein neighborhood~$W_\delta(\nu,P)$ with 
    \begin{eqnarray*}
        \delta((y_0,y_1,x),(\tilde y_0,\tilde y_1,\tilde x)) & = & |y_0-\tilde y_0|+|y_1-\tilde y_1|+\|x-\tilde x\|.
    \end{eqnarray*}
    The robust welfare criterion is therefore defined as
    \begin{eqnarray*}
        \mbox{RW}(\tau) & := & \inf_{W_\delta(\nu,P)\leq\rho}\int \left(y_0(1-\tau(x))+y_1\tau(x)\right) d\nu(y_0,y_1,x).
    \end{eqnarray*}
    Using the equality between~(\ref{eq:Blanchet1}) and~(\ref{eq:Blanchet2}) above, and the fact that~$y+\lambda|y-\tilde y|$ is minimized by setting~$y=-\infty$ if~$\lambda\in[0,1)$ and~$y=\tilde y$ if~$\lambda\geq1$, the robust welfare criterion is equal to
    \begin{eqnarray*}
        \sup_{\lambda\geq1}\left(\int\min\left\{y_0+\lambda\inf_{\tau(\tilde x)=0}\|x-\tilde x\|,y_1+\lambda\inf_{\tau(\tilde x)=1}\|x-\tilde x\|\right\}dP(y_0,y_1,x)-\lambda\rho\right).
    \end{eqnarray*}
    \cite{fan2025quantifying} consider a similar robustness concern in the allocation of treatment. However, they do not assume that the planner has a reference joint probability distribution fo~$(Y_0,Y_1,X)$. Instead, there is a reference probability distribution for~$(Y_0,X)$ and~$(Y_1,X)$, justified for instance with access to a pilot treatment data set with selection on observables, but no additional information on the joint probability distribution. \cite{fan2025quantifying} derive bounds for the expectation of a function of~$Y_0,Y_1,$ and~$X$. They combine the method of conditioning of \cite{ruschendorf1991bounds} to deal with the overlapping marginals with the distributional robust optimization approach of~\cite{blanchet2019quantifying} to account for the uncertainty about the marginals. The method applies to a large variety of settings, including ones with multivariate potential outcomes. However, we choose to illustrate it on a special case with scalar potential outcomes. We compare the result in \cite{fan2025quantifying} to \cite{adjaho2022externally} in the special case of robust treatment assignment. For~$l\in\{0,1\}$, define the following quantities. Let~$\mu_l$ be the reference distribution for~$(Y_l,X)$. Let the Wasserstein neighborhood~$W_\delta(\cdot,\mu_l)\leq \rho_l$ of~$\mu_l$ be defined with radius~$\rho_l>0$ and
    \begin{eqnarray*}
        \delta((y_l,x),(\tilde y_l,\tilde x)) & = & |y_l-\tilde y_l|+\|x-\tilde x\|.
    \end{eqnarray*} 
    The robust treatment criterion is
    \begin{eqnarray*}
        \mbox{RW}^\prime(\tau) & := & \inf_\pi\int \left(y_0(1-\tau(x))+y_1\tau(x)\right) d\pi(y_0,y_1,x),
    \end{eqnarray*}
    where the infimum is now taken with respect to probability distribution~$\pi\in\mathcal M(\nu_0,\nu_1)$ for~$(Y_0,Y_1,X)$ with overlapping marginals~$\nu_0$ for~$(Y_0,X)$ and~$\nu_1$ for~$(Y_1,X)$, and each~$\nu_l$ lies in the Wasserstein neighborhood~$W_\delta(\nu_l,\mu_l)$. Using the same technique, the robust welfare criterion is shown to equal
    \begin{eqnarray*}
        \sup_{\lambda\geq1}\left(\inf_{\pi\in\mathcal M(\mu_0,\mu_1)}\int\min\left\{y_0+\phi_\lambda^0(x_0,x_1),y_1+\phi_\lambda^1(x_0,x_1)\right\}d\pi(y_0,x_0,y_1,x_1)-\lambda^\top\rho\right),
    \end{eqnarray*}
    with~$\phi_\lambda^l:=\inf_{\tau(x)=l}\{\lambda_0\|x_0-x\|+\lambda_1\|x_1-x\|\}$. The difference with the \cite{adjaho2022externally} case is the data combination problem and the duplication of covariates to deal with overlapping marginals.
    
    \cite{gu2023dual} add distributional robustness concerns to the class of incomplete models described in section~\ref{sec:incomplete}.\footnote{The sensitivity analysis in \cite{gu2023dual} is related to prior work by \cite{christensen2023counterfactual}, where the distributional robustness neighborhood are characterized by~$f$-divergences instead of the Wasserstein distance.} Observed variables are collected in~$(Y,X)$. In the identification analysis below, we assume the probability distribution~$P_{Y,X}$ generating them is known. There is also a vector~$U$ of latent variables with distribution~$Q_U$. 
    The model structure is characterized by the support restriction~$(Y,U,X)\in\Gamma(\theta)$. We are interested in a functional~$\mathbb E[\varphi(Y^\ast,U,X;\theta)]$ of counterfactual outcomes~$Y^\ast$, which satisfy a counterfactual variant of the structural model, namely~$(Y^\ast,U,X)\in\Gamma^\ast(\theta)$. Both actual structure~$\Gamma(\theta)$ and counterfactual structure~$\Gamma^\ast(\theta)$ are known up to the finite dimensional unknown parameter vector~$\theta$. Because of distributional robustness concerns, the distribution~$Q_U$ of the latent vector~$U$ is only assumed to lie in a Wasserstein neighborhood of a reference distribution~$Q_V$: $W_1(Q_U,Q_V)\leq\rho$. By definition of the Wasserstein distance~$W_1$, there is a coupling~$(U,V)$ of the vectors~$U$ and~$V$ with respective distributions~$Q_U$ and~$Q_V$ such that~$\mathbb E\|U-V\|\leq\rho.$
    Hence, the lower bound for the parameter of interest is given by
    \begin{eqnarray}\label{eq:Russell}
        \inf_\theta\inf_\pi \int\varphi(y^\ast,u,x;\theta)\;
        d\pi(y^\ast,y,u,v,x),
    \end{eqnarray}
    where the inner infimum is over all probability distributions~$\pi$ for~$(Y^\ast,Y,U,V,X)$ satisfying the following constraints:
    \begin{enumerate}
        \item The support of~$\pi$ is constrained by~$(Y,U,X)\in\Gamma(\theta)$ and~$(Y^\ast,U,X)\in\Gamma^\ast(\theta)$, $\pi$ almost surely;
        \item The marginal of~$\pi$ relative to~$(Y,X)$ is~$P_{Y,X}$;
        \item The marginal of~$\pi$ relative to~$V$ is~$Q_V$; 
        \item $\mathbb E\|U-V\|\leq\rho$.
    \end{enumerate}
    The inner optimization program in~(\ref{eq:Russell}) has Lagrangian formulation:
    \begin{eqnarray}\label{eq:inner}
        \sup_{\lambda\geq0}\inf_\pi
        \int\left[\varphi(y^\ast,u,x;\theta)+\lambda(\|u-v\|-\rho)\right]\,d\pi(y^\ast,y,u,v,x),
    \end{eqnarray}
    subject to~(1), (2) and~(3) above. Call
    \begin{eqnarray*}
        \mathcal G(y,x;\theta) & := & \{(y^\ast,u): (y,u,x)\in\Gamma(\theta)\mbox{ and }(y^\ast,u,x)\in\Gamma^\ast(\theta)\}
    \end{eqnarray*}
    the collection of values of the latent variables~$(y^\ast,u)$ compatible with observations~$(y,x)$. Define the random set
    \begin{eqnarray*}
        (y,v,x) & \mapsto & \Phi(y,v,x;\theta,\lambda) \\
        & := &\{\varphi(y^\ast,u,x;\theta)+\lambda(\|u-v\|-\rho):\;(y^\ast,u)\in\mathcal G(y,x;\theta)\},
    \end{eqnarray*}
    as the set of values that the function~$\varphi(y^\ast,u,x;\theta)+\lambda(\|u-v\|-\rho)$ can take when latent variables vary over~$\mathcal G(y,x;\theta)$ and~$v$ is free. The inner optimization problem in~(\ref{eq:inner}) is equal to the infimum over the Aumann expectation of the random set~$\Phi$, i.e., over the collection of integrals~${\textstyle\int_\pi}\left[\varphi(y^\ast,u,x;\theta)+\lambda(\|u-v\|-\rho)\right]\,d\pi(y,v,x)$ subject to~(2) and~(3) when latent variables vary over~$\mathcal G(y,x;\theta)$ and~$v$ is free. Under the assumptions of theorem~2.1.20 page~236 of \cite{molchanov2005theory}, the infimum of the Aumann expectation is the expectation of the infimum over~$\mathcal G(y,x;\theta)$. Hence, calling
    \begin{eqnarray*}
        c((y,x),v;\theta,\lambda) & := & \inf_{(y^\ast,u)\in\mathcal G(y,x;\theta)} \left( \varphi(y^\ast,u,x;\theta)+\lambda(\|u-v\|-\rho) \right),
    \end{eqnarray*}
    the inner optimization problem in~(\ref{eq:inner}) is equal to the optimal transport problem with cost~$c((y,x),v;\theta,\lambda)$:
    \begin{eqnarray*}
        \mathcal C(\theta,\lambda) & := & \inf_{\pi\in\mathcal M(P_{Y,X},Q_V)} \int c((y,x),v;\theta,\lambda)\;d\pi((y,x),v).
    \end{eqnarray*}
    Finally, the sharp lower bound for the parameter of interest~$\mathbb E[ \varphi(Y^\ast,U,X;\theta)]$ is therefore equal to~$\inf_\theta \sup_{\lambda\geq0}\mathcal C(\theta,\lambda)$.

    \cite{Xu2025distributionally} study treatment effect prediction under population shift by defining the target object through a distributionally robust optimization problem. Let~$P$ denote the source distribution of potential outcomes and let $\mathcal P_\delta = \{ Q : W_p(Q,P) \le \delta \}$ be a Wasserstein neighborhood. The robust treatment effect is defined as the minimax solution
    \[
        \tau_\delta^\ast
        = \mbox{arg}\min_{\tau} \sup_{Q \in \mathcal P_\delta}
        \mathbb E_Q \bigl[(Y(1)-Y(0)-\tau)^2\bigr].
    \]
    Because the joint distribution of $(Y(1),Y(0))$ is not identified, the problem is coupled with partial identification over the set of copulas consistent with the marginal potential outcome distributions, yielding sharp optimistic and pessimistic minimax solutions. The resulting robust predictor coincides with the source average treatment effect when $\delta=0$ and exhibits shrinkage toward zero as $\delta$ increases.

    \subsubsection{Measurement error}

    \cite{schennach2022optimally} propose to model robustness to measurement error using the Wasserstein distance. Suppose the data~$(Y_1,\ldots,Y_n)$ is a sample of possibly mismeasured version of a latent variable~$Z$, which satisfies moment conditions~$\mathbb Eg(Z,\theta)=0$. The quantity of interest is the finite dimensional parameter~$\theta$. \cite{schennach2022optimally} propose the optimally transported generalized method of moments (OTGMM) estimator for~$\theta$ which minimizes the solution of the following program:
    \begin{eqnarray*}
        \min_{\tilde Y=(\tilde Y_1,\ldots,\tilde Y_n)}\sum_{i=1}^n\|Y_i-\tilde Y_i\|^2, \mbox{ s.t. }\sum_{i=1}^ng(\tilde Y_i;\theta)=0.
    \end{eqnarray*}
    For each value of~$\theta$, the program finds a set of points~$(\tilde Y_1,\ldots,\tilde Y_n)$ that satisfy the empirical moments, while minimizing the 2-Wasserstein distance between the empirical distribution~$\Sigma_{i=1}^n\delta_{\tilde Y_i}/n$ relative to~$(\tilde Y_1,\ldots,\tilde Y_n)$ and the empirical distribution~$\Sigma_{i=1}^n\delta_{Y_i}/n$ relative to the actual sample. Where generalized empirical likelihood estimators re-weight observations to minimize KL discrepancy to account for sampling bias, OTGMM adjusts sample points to minimize 2-Wasserstein distance to account for measurement error. \cite{forneron2024fitting} apply related ideas to robustness in dynamic state space models. \cite{daljord2021black} also use an optimal transport approach to the challenge of measuring illicit trade volume.
    
        
    \subsubsection{Barycenters, synthetic controls and distributional regression}

    The geometry induced by the Wasserstein distance in the space of probability measures makes it possible to work with averages of distributions for distributional regression and synthetic controls in particular. \cite{gunsilius2023distributional} extends the synthetic control methodology to estimate distributional causal effects. In the traditional setting, the analyst observes an aggregate outcome~$Y_{jt}$ for each unit~$j=1,\ldots,J$ and time period~$t=1,2$. Unit~$J$ receives a treatment at period~$2$ only. All other units are never treated. This is easily generalizable to more treatment units and more periods. In case we have access to disaggregated data within each unit, we can identify and estimate the probability distribution~$P_{jt}$ that generates~$Y_{jt}$ for each unit and time period. The synthetic control distribution~$P_{J2}^N$ is a weighted average of never treated units. The averaging is with respect to Wasserstein distance: It is a Wasserstein barycenter as defined in~(\ref{eq:barycenter}). The weights are chosen so that the synthetic control distribution in period~$1$ is closest in Wasserstein distance to the distribution for unit~$J$ in period~$1$. 
    
    
    \cite{zhu2023autoregressive} use the Wasserstein geometry to define autoregressive processes for distribution regression.
    
        
    \subsubsection{Simulation}

    The Wasserstein distance can be used to simulate data. One example is the filtering of the latent variable~$U_t$ in \cite{forneron2024fitting} described above. In \cite{arellano2023recovering}, observed outcomes~$Y$ are assumed to be the sum~$Y=\eta+\eta^\prime$ of two latent factors of interest~$\eta$ and~$\eta^\prime$. The latent factors are the objects of interest and we wish to simulate samples of the latter. A sample of observable outcomes~$(Y_1,\ldots,Y_n)$ is available. Let~$\sigma_\eta$ and~$\sigma_\eta^\prime$ be independent permutations. We want to generate samples~$(\hat \eta_1,\ldots,\hat\eta_n)$ and~$(\hat \eta_1^\prime,\ldots,\hat\eta_n^\prime)$ to minimize the 2-Wasserstein distance
    \begin{eqnarray*}
        \min_{\hat\eta,\hat\eta^\prime\in\mathcal E}
        \min_\sigma \sum_{i=1}^n \left( Y_{\sigma(i)}-(\hat\eta_{\sigma_\eta(i)}+\hat\eta^\prime_{\sigma_\eta^\prime(i)})\right)^2,
    \end{eqnarray*}
    where the regularization (nonparametric) set~$\mathcal E$ is defined by~$|\eta| \leq\bar C_n$ and
    \begin{eqnarray*}
        \begin{array}{lllll}
            \underline C_n  & \leq & (n+1)(\eta_{i+1}-\eta_i) & \leq & \bar C_n.
        \end{array}
    \end{eqnarray*}

    \cite{athey2024using} propose to use Wasserstein GANs (described in section~\ref{sec:WGAN}) to produce synthetic data that closely mimic a given real data set. 
    
    \subsubsection{Estimation}
        
    Wasserstein GANs can also be used in estimation. \cite{kaji2023adversarial} propose adversarial estimation\footnote{See also \cite{kaji2021adversarial}.}. They use traditional GANs, but their method can be trivially adapted to Wasserstein GANs. Let~$P_\theta$ be the parametric model entertained, such that data can be simulated via~$g_\theta(Z_i)$, where~$Z_i$ follows a given reference distribution (as explained in section~\ref{sec:WGAN}). Let~$P_{\theta,m}$ be the empirical distribution relative to a simulated sample~$(g_\theta(Z_1),\ldots,g_\theta(Z_m))$. The adversarial estimator for~$\theta$ is the minimizer~$\hat\theta$ of the 1-Wasserstein distance between~$P_{\theta,m}$ and the empirical distribution~$P_n$ of the data sample~$(X_1,\ldots,X_n)$: 
    \begin{eqnarray*}
        \hat\theta & := & \mbox{arg}\min_\theta W_1(P_n,P_{\theta,m}) \\
        & = & \mbox{arg}\min_\theta \sup_{\varphi\in\mbox{\tiny Lip}_1}
        \left( \frac{1}{n}\sum_{i=1}^n\varphi(X_i) - \frac{1}{m}\sum_{j=1}^m\varphi(g_\theta(Z_j)) \right). 
    \end{eqnarray*}
    In \cite{kaji2023adversarial}, $n/m\rightarrow0$, to avoid needlessly suffering from simulation noise when~$n$ is small.
    \cite{fan2024minimum} also use a minimum Wasserstein distance estimation principle. However, to circumvent the curse of dimensionality, they propose to minimize the sliced Wasserstein distance defined in~\ref{sec:W estimation}. With the notation of section~\ref{sec:W estimation}, the sliced Wasserstein distance between the empirical distribution~$\mu_n$ of a sample of observations and the parametric distribution (or the empirical distribution of a simulation sample thereof) is
    \begin{eqnarray*}
        \widehat{\mbox{SW}}_p(\theta) & := & \left( \int\int_0^1\left|G_{\mu_n}^{-1}(u;\alpha)-G_{\nu_\theta}^{-1}(u;\alpha)\right|^pdu\;d\rho(\alpha)\right)^\frac{1}{p}.
    \end{eqnarray*}
    The MSWD (Minimum Sliced Wasserstein Distance) estimator~$\hat\theta$ is defined by
    \begin{eqnarray*}
        \widehat{\mbox{SW}}(\hat\theta) & := & \inf_\theta\widehat{\mbox{SW}}(\theta) + o_p\left(\frac{1}{n}\right).
    \end{eqnarray*}
    One notable feature of the minimum sliced Wasserstein estimator is that it is asymptotically normal even when the support of the data generating process depends on the parameter. In a similar spirit, \cite{qu2024distributionally} propose a distributionally robust instrumental variable estimation, which solves
    \begin{eqnarray*}
        \min_\beta\sup_{Q: W_2(Q,\tilde P_n)<\rho}\mathbb E_Q(Y-\beta X)^2,
    \end{eqnarray*}
    where~$\tilde P_n$ is the empirical distribution of~$(\Pi_ZX_i,\Pi_ZY_i)_i$.

    A final application of the Wasserstein distance to inference problems can be adapted from \cite{galichon2013dilation}. The latter use a different metric on the space of probability measures, but the problem can be adapted in the following way. In the framework of section~\ref{sec:incomplete}, we now consider inference on the basis of an empirical distribution~$P_{Y,X;n}$. Hence the true distribution~$P_{Y,X}$ is unknown. A confidence region for the parameter vector~$\theta$ can be characterized by
    \begin{eqnarray*}
        &&\inf_{\pi} \int \delta((u,x),(y,x^\prime);\theta) \;d\pi((u,x),(y,x^\prime)) \; = \; 0,\\
        && \hskip50pt \mbox{ subject to } \pi\in\mathcal M(Q_U\otimes P_X,P_{Y,X}) \mbox{ and }W_p(P_{Y,X},P_{Y,X;n})\leq \rho_n.
    \end{eqnarray*}
    In the expression above,~$\rho_n$ is calibrated to the desired confidence level.

\newpage
\section{Optimal transport as a model: The econometrics of matching}
\label{sec:IOT}


While this article is dedicated to the use of optimal transport {in} econometrics, it is particularly relevant to discuss the {econometrics of optimal transport} itself, or, equivalently, the {econometrics of matching models with transferable utility}. In the transferable-utility (TU) framework, the total utility generated by a match between types $x$ and $y$ can be summarized by a {joint surplus} $\Phi_{xy}$, which is split between the two agents through transfers. As noted  by \cite{Graham2011AssignmentProblems}, the econometric literature on matching and assignment problems is fundamentally organized around whether match output $\Phi$ is observed. When match output is observed, the analyst can relate realized surplus or performance directly to the characteristics of the matched agents. When it is not observed, identification and estimation must instead proceed from the structure of the assignment itself, together with the maintained assumptions on preferences, complementarities, and equilibrium behavior. This chapter focuses on this second, more indirect case, in which match output is unobserved.

Empirically, the econometrician often observes the {equilibrium matching pattern} between two finite populations---firms and workers, men and women, origins and destinations in migration flows---but the underlying systematic surplus $\Phi$ (or the corresponding transportation cost $c(x,y)=-\Phi_{xy}$) that rationalizes this allocation is not directly observed. Recovering this economic primitive from the observed allocation is the object of {Inverse Optimal Transport} (IOT). Because structural matching models with transferable utility are fundamentally optimal-transport problems, and equilibrium allocations in these models are the solutions to such problems, IOT provides the natural framework for structural identification and estimation.

This perspective has proved empirically powerful in a wide range of applications. In the marriage market, \cite{choo2006marries} use a TU matching model with unobserved heterogeneity to measure the gains to marriage over time and to assess how the U.S.\ Supreme Court’s ruling in {Roe v.\ Wade} affected the ``value of marriage'' implicit in observed sorting patterns.  \cite{dupuy2014personality} construct {indices of mutual attractiveness} that summarize, in a low-dimensional way, how a rich set of characteristics (education, height, BMI, health, risk attitudes, personality traits) contributes to sorting, by estimating the surplus function in a TU matching model.  In a related vein, \cite{chiappori2012fatter} investigate how physical appearance and body shape affect the structure of matching gains, showing that the valuation of body size across genders interacts with that of education and income to shape systematic sorting in marriage markets.

\citet{CiscatoGalichonGousse2020} structurally compare homogamy across same-sex and different-sex couples, quantifying how assortative mating on education, age, race, and wages differs across these groups. \citet{ChiapporiSalanieWeiss2017} study the {marital college premium}, showing how changes in education levels and educational assortative matching translate into changes in the returns to college in the marriage market. More recently, \citet{ChiapporiFiorioGalichonVerzillo2025} exploit an extremely rich administrative income dataset to measure assortative matching on income, using a flexible TU matching framework that can accommodate complex sorting patterns and their implications for household-income inequality. Beyond household and labor applications, TU matching methods have also been applied to corporate finance: \citet{GuadalupeRappoportSalanieThomas2023} use a structural assortative-matching framework to study mergers and acquisitions, showing how the matching between targets and acquirers reflects systematic complementarities in firm characteristics.
Related TU matching models have also been applied to industrial organization contexts, for instance in the estimation of matching games with transfers between firms and their trading partners or suppliers, e.g., \cite{Fox2018MatchingTransfers}, underscoring the ubiquity of IOT-type questions whenever economic agents sort into relationships on the basis of underlying complementarities.

\subsection{Definition of IOT in the unregularized discrete setting}

Throughout this section we place ourselves in the {discrete} setting with the notation of section~\ref{sec:sink}. In economic applications, it is often more natural to work not with the cost but with the {systematic surplus}~$\Phi_{xy} = -c(x,y),$ since \(\Phi_{xy}\) represents the deterministic component of the joint utility that two agents of types~\(x\) and~\(y\) derive from matching. Given a surplus matrix \(\Phi=(\Phi_{xy})\), the classical (unregularized) optimal transport problem~(\ref{eq:discrete primal}) is repeated here for convenience:
\begin{equation}
\label{eq:unregOT}
\max_{\pi\in\Pi(\mu,\nu)}
\sum_{x,y} \pi_{xy}\Phi_{xy},
\end{equation}
where \(\Pi(\mu,\nu)\) denotes the \emph{transportation polytope}, which is the set of feasible couplings between \(\mu\) and~\(\nu\), that is, the set of $\pi_{xy} \geq 0$ such that $\sum_y \pi_{xy}=\mu_x$ and $\sum_x \pi_{xy} = \nu_y$. In the inverse optimal transport, the econometrician observes the matching pattern~\(\widehat{\pi}\) and seeks to infer the surplus matrix~\(\Phi\) that rationalizes it.

\begin{definition}[Inverse Optimal Transport (unregularized)]
Given observed marginals \((\mu,\nu)\) and an observed transport plan \(\widehat{\pi}\in\Pi(\mu,\nu)\), the (unregularized) IOT problem consists of recovering a cost function \(c=-\Phi\) such that \(\widehat{\pi}\) is a solution of the optimal transport problem \eqref{eq:unregOT}.
\end{definition}

\medskip

IOT is a notoriously hard problem in its unregularized form. A detailed account of the difficulties associated with unregularized assignment and transportation problems can be found in section~6.7 of the monograph of \cite{burkard2012assignment}, which documents the combinatorial structure and degeneracies of assignment polytopes.

The fundamental difficulty is simple to state: {unregularized optimal transport typically produces extremely sparse optimal solutions}. In the fully balanced assignment case ($\mu_x=\nu_y=\tfrac{1}{N}$), the optimal solution is always a {vertex} of the Birkhoff polytope of doubly stochastic matrices and therefore corresponds to a {permutation matrix}. More generally, 
the optimal solution is an {extreme point} of the transportation polytope (see section~\ref{sec:simplex}), and extreme points have support size at most $M+N-1$. This means the mass of~$\pi$ concentrates on a sparse set of pairs.

In many important cases in the continuous setting too --- for instance, when the cost function~$c$ is quadratic --- the optimizer is a {Monge map}: a deterministic matching function $y=T(x)$. Thus the mapping from~$c$ to~$\pi$ is highly discontinuous and set-valued. Inverting it is extremely challenging, and in most cases impossible without strong structural restrictions.

A classical example from economics makes the identification problem vivid. In Becker’s model of marriage markets, suppose types are ordered scalars and the matching is positively assortative. Then any {supermodular} surplus function~$\Phi$ generates positive assortative matching as the unique stable (and optimal transport) outcome. Because the class of supermodular functions is extremely large, the observed matching pattern carries very little identifying information: the same allocation can be rationalized by an infinite-dimensional family of surplus functions. Thus, in the absence of regularization, the IOT problem is fundamentally {under-identified}. Many surplus functions~$\Phi$ (or cost functions~$c=-\Phi$) produce the same optimal transport plan. This motivates the need for additional structure to restore identifiability and to regularize the inverse problem. In practice, this structure may come from parametric restrictions on the surplus $\Phi$ (for example, bilinear or low-rank specifications), from sparsity-promoting priors on its parameters, or---most powerfully—from {entropic regularization}. Entropic regularization plays a dual role: it introduces a smooth, strictly convex component into the forward optimal transport problem, which makes the mapping $c\mapsto\pi$ single-valued and differentiable; and it has the natural economic interpretation of capturing the effect of unobserved heterogeneity in preferences. In the next subsection we formalize this regularized formulation of IOT, and we show how it yields a well-posed econometric problem whose structure is closely connected to Poisson pseudo–maximum likelihood (hereafter PPML) and to gravity models of bilateral trade flows.

\subsection{Entropic regularization and tractable IOT}
\label{sec:entropic-IOT}

We begin by formally defining the regularized IOT problem in the discrete setting. Let~$\Phi_{xy}=-c(x,y)$ be the surplus matrix, and let~$\pi^\sigma(\Phi)$ denote the unique solution of the entropically regularized optimal transport problem
\begin{eqnarray*}
    \pi^\sigma(\Phi)
& = & 
\mbox{arg}\max_{\pi\in\Pi(\mu,\nu)}
\left\{
\sum_{x,y} \Phi_{xy}\,\pi_{xy}
-
2\sigma \sum_{x,y}\pi_{xy}(\log\pi_{xy}-1)
\right\}
\end{eqnarray*}
with marginals $(\mu,\nu)$.  

\begin{definition}[Regularized Inverse Optimal Transport]
Given marginals $(\mu,\nu)$ and an observed coupling $\widehat{\pi}\in\Pi(\mu,\nu)$, the {regularized IOT} problem consists of finding a cost matrix $c$ such that~$\pi^\sigma(\Phi)=\widehat{\pi}$.
\end{definition}

The entropic term makes the forward map $\Phi \mapsto\pi^\sigma(\Phi)$ smooth, strictly convex, and single-valued, turning the inverse problem into a well-posed estimation problem. Crucially, entropic OT always yields a coupling with {full support}, i.e.,~$\pi^\sigma_{xy}>0$ for all~$(x,y)$, which eliminates the sparsity and non-invertibility issues inherent in classical (unregularized) IOT.

\subsubsection{Discrete logit unbalanced matching}
The model of \citet{choo2006marries} provides a prototypical empirical setting in which entropic regularization arises naturally. There are finite type sets $\mathcal{X}$ and $\mathcal{Y}$, and agents on both sides may remain unmatched. The systematic surplus shared by types $(x,y)$ is
\[
\Phi_{xy}=U_{xy}+V_{xy},
\]
where $U_{xy}$ and $V_{xy}$ are deterministic components of utility for each side. The systematic utilities of unmatched individuals are normalized to zero, so $U_{x0}=V_{0y}=0$. Additive i.i.d. Gumbel shocks generate a multinomial-logit structure. 

Let~$m_x$ and~$m_y$ denote the masses of women and men of each type. The equilibrium matching flows~$\pi_{xy}$ (i.e., mass of~$(x,y)$ pairs, including~$x=0$ and~$y=0$) satisfy
\begin{eqnarray*}
\label{eq:CS-probabilities}
\begin{array}{lllll}
    \pi_{xy} 
    & = & 
    \mu_x \frac{\exp(U_{xy})}{
    1 + \sum_{y'} \exp(U_{xy'})}
    & = &
    \nu_y \frac{\exp(V_{xy})}{
    1  + \sum_{x'} \exp(V_{x'y})},  
\end{array}
\end{eqnarray*}
where the utilities of unmatched individuals are normalized to zero. Eliminating type-specific utilities yields the well-known \cite{choo2006marries} identification formula:
\begin{eqnarray*}
\label{eq:CS-identification}
\Phi_{xy}
& = &
\log \pi_{xy}
- \log \pi_{x0}
- \log \pi_{0y},
\end{eqnarray*}which allows to identify the entire $\Phi$ directly from matching flows. Moreover, as shown by~\citet{galichonSalanie2009,galichon2022cupid}, the equilibrium is the solution of the {unbalanced} entropically regularized optimal transport problem, which is a convex optimization problem:
\begin{eqnarray}
&& \max_{\pi \in \Pi(\mu ,\nu)}
\left\{
\sum_{x,y} \pi_{xy}\Phi_{xy}
-
2\sigma\sum_{x,y} \pi_{xy}(\log \pi_{xy}-1) 
\right.\nonumber\\
&& \hskip100pt - \left.
\sigma\sum_{x} \pi_{x0}(\log \pi_{x0}-1)
-
\sigma\sum_{y} \pi_{0y}(\log \pi_{0y}-1)
\right\}.
\label{eq:CS-entropic-OT}
\end{eqnarray}
\cite{decker2013unique} study the uniqueness of the equilibrium and substitution effects in this model.

\subsubsection{Continuous logit balanced matching and fixed effects}
\cite{dupuy2014personality} develop a continuous-logit matching model without singlehood, where every agent is matched. Let~$\mathcal X$ and~$\mathcal Y$ be continuous type spaces with densities~$\mu(x)$ and~$\nu(y)$. 
With i.i.d. Gumbel shocks, the equilibrium matching density is:
\begin{eqnarray*}
\label{eq:DG-Gibbs}
\pi(x,y)
& = &
\exp\!\left(\frac{\Phi(x,y)-a(x)-b(y)}{2\sigma}\right),
\end{eqnarray*}
where $a(x)$ and $b(y)$ are endogenous potentials ensuring the correct marginals.
Taking logs yields the identification formula
\begin{eqnarray*}
\label{eq:DG-identification}
\Phi(x,y)
& = &
2\sigma \log \pi(x,y) + a(x) + b(y),
\end{eqnarray*}
so~$\Phi$ is identified up to the additive fixed effects~$a(x)$ and~$b(y)$. 

\subsubsection{Beyond the logit case: General heterogeneity distributions.}
A central contribution of \cite{galichon2022cupid} is to show that the convex-analytic formulation above extends far beyond the Gumbel (logit) specification. Let $\varepsilon=(\varepsilon_{y})_{y\in\mathcal{Y}\cup\{ 0\} } $ denote the vector of taste shocks for a type-$x$ individual, with some arbitrary joint distribution, and let $z_x=(z_{xy})_{y\in\mathcal{Y}\cup\{ 0\} }$ be a vector of {systematic utilities} associated with the various matching options (including the option of being unmatched). We classically define the social surplus function as the expected indirect utility of each type
\begin{equation}
\label{eq:GS-Gx}
G_x(z_x)
=
\mathbb{E}\Big[ \max_{y\in\mathcal{Y}} \{ z_{xy} + \varepsilon_{y},\varepsilon_{0} \} \Big],
\end{equation}
which is finite and convex in $z_x$. Galichon and Salanié define the \emph{entropy of choice} as the convex conjugate of the former, that is
\begin{equation}
\label{eq:GS-Gx-star}
G_x^\star(p_x)
=
\sup_{z_x}\Big\{
\sum_{y} p_{xy} z_{xy} - G_x(z_x)
\Big\}
\end{equation}
is finite when $p_x=(p_{xy})_{y}$ lies in the simplex (interpreted as the vector of choice probabilities of a type-$x$ woman across all options). An analogous pair of functions $H_y(\cdot)$ and $H_y^\star(\cdot)$ can be defined for type-$y$ men.

At the aggregate level, if $\pi_{xy}$ is the matching measure and~$m_x$ (resp.\ $m_y$) the mass of type-$x$ women (resp. type-$y$ men), \cite{galichon2022cupid} define the \emph{entropy of matching} as the weighted sums of the entropy of choice, that is
\begin{equation}
\label{eq:GS-E-pi}
\mathcal{E}(\pi)
=
\sum_{x} m_x \, G_x^\star\!\left( \frac{\pi_{x\cdot}}{m_x} \right)
+
\sum_{y} m_y \, H_y^\star\!\left( \frac{\pi_{\cdot y}}{m_y} \right),
\end{equation}
where $\pi_{x\cdot}=(\pi_{xy})_{y}$ and $\pi_{\cdot y}=(\pi_{xy})_{x}$ denote the row and column profiles of $\pi$. Equilibrium matching is then characterized as the solution of the convex program, which characterizes the social welfare as the convex conjugate of the entropy of matching, which is given by 
\begin{equation}
\label{eq:GS-general-RUM}
\mathcal{E}^\star (\Phi) = \max_{\pi\in\Pi(m,n)}
\Big\{
\langle \Phi, \pi \rangle - \mathcal{E}(\pi)
\Big\}.
\end{equation}
In the Gumbel (logit) case, $G_x$ takes the log-sum-exp form, and $G_x^\star$ reduces to an entropy term so that $\mathcal{E}(\pi)$ is (a multiple of) the Kullback–Leibler divergence; this recovers the entropic OT formulation \eqref{eq:CS-identification}. For general random-utility models, the distribution of the shocks is encoded in the convex functionals $G_x^\star$ and $H_y^\star$, yielding a large class of entropy-like regularizers. In particular, \cite{galichon2022cupid} show that for any fixed heterogeneity distribution satisfying mild convexity conditions, the mean utilities and the matching measure are identified from observed matching patterns via the convex program \eqref{eq:GS-general-RUM}. This firmly embeds regularized IOT as a general econometric framework, not restricted to the logit case.

\subsection{Parametric IOT}
\label{sec:parametric-IOT}

The nonparametric model of \cite{choo2006marries} identifies the entire surplus matrix $\Phi$ up to a normalization when matching flows are observed, but in many empirical settings a {parametric} structure on $\Phi$ is both substantively meaningful and statistically advantageous. Parametric restrictions impose economic structure, reduce dimensionality, and enable the use of rich covariates on both sides of the market. They also allow the econometrician to examine how specific characteristics contribute to the joint surplus, much as parametric discrete-choice models allow one to interpret the role of covariates in choice probabilities.

\subsubsection{Parameterizing the surplus}

\cite{galichon2022cupid} suggest to parameterize the matching surplus using
\begin{equation}
    \Phi_{xy} (\lambda)= \sum_{k} \phi_{xyk} \lambda_k 
\end{equation}
where $\lambda \in \mathbb R^K$ is the parameter to be estimated, and  $(\phi_{xyk})_{xy,k}$ are basis functions that parameterize the surplus.  For instance, the various elements $(\phi_{xyk})_{xy,k}$ associated with various indices $k$ may be distances in various socio-demographic dimensions, such as age, education, or socioeconomic status, and so on. By an application of the envelope theorem in expression~\eqref{eq:GS-E-pi}, the derivative of the social welfare with respect to $\lambda_k$ is seen to be
\begin{equation}
\label{eq:env-thm-matching}
    \frac { \partial\mathcal{E}^\star } {\partial \lambda_k}(\Phi(\lambda)) = \sum_{xy} \pi^\lambda _{xy} \phi_{xyk}
\end{equation}
where $\pi^\lambda $ is the maximizer of~\eqref{eq:GS-E-pi}. \cite{galichon2022cupid} define the \emph{moment matching estimator} $\hat{\lambda}$ as the value of the parameter vector $\lambda$ for which the predicted moments $\sum_{xy} \pi^\lambda _{xy} \phi_{xyk}$ coincide with the observed moments $\sum_{xy} \hat{\pi} _{xy} \phi_{xyk}$, and, using equation~\eqref{eq:env-thm-matching}, they note that the moment matching estimator is obtained as the unique solution to the convex optimization problem 
\begin{equation}
\label{mm-as-convex-optim}
    \min_{\lambda} \{ \mathcal{E}^\star (\Phi(\lambda)) - \sum_{xyk}\hat{\pi}_{xy} \phi_{xyk} \lambda_k \}.
\end{equation}

\cite{dupuy2014personality} consider an extension of the previous framework in the continuous case, with the surplus being specified as
\begin{eqnarray*}
\Phi_{xy} (\lambda) & = & W_x^\top A\, W_y,
\end{eqnarray*}
where~$W_x$ and~$W_y$ are vectors of observable characteristics of types~$x$ and~$y$ respectively, and the parameter $\lambda$ is $A$, a matrix of parameters capturing complementarities, so that the vector of entries of $A$ is the parameter vector $\lambda$. 
Such models allow the econometrician to estimate how education, income, health, age, personality traits, or other observables contribute to sorting patterns through the entries of $A$. Imposing a bilinear structure improves statistical efficiency, supports interpretation, and, when supplemented with low-rank or sparsity restrictions, promotes parsimony.

\subsubsection{Poisson regression} 
In the case when the random utilities are Gumbel, one can show that the dual expression of  $\mathcal{E}^\star (\Phi)$ in~\eqref{eq:GS-general-RUM} boils down to
\begin{equation}
\label{eq:dual-GS-logit}
\mathcal{E}^\star (\Phi) = \min_{a,b} \left\{ \sum_{x} \mu_x a_x + \sum_y \nu_y b_y +2  \sigma \sum_{xy} e^{\frac { \Phi_{xy}-a_x - b_y} {2 \sigma } } + \sigma \sum_x e^{-\frac {a_x} \sigma } + \sigma \sum_y e^{-\frac{ b_y} \sigma  }  \right\},
\end{equation}

\cite{galichon2022cupid} propose several estimation methods for parametric IOT. Taking the normalization~$\sigma=1/2$, the entropically regularized optimal matching induced by $\Phi_{xy}(\lambda)=\sum_{k} \phi_{xyk} \lambda_k $ satisfies
\begin{eqnarray}
\label{eq:pi(A)}
    \pi_{xy} & = & \exp{\left( \sum_{k} \phi_{xyk} \lambda_k -a_x - b_y\right)},
\end{eqnarray}
and the moment matching estimation problem~\eqref{mm-as-convex-optim} becomes
\begin{equation}\label{mm-estimation-logit}
    \min_{\lambda ,a,b}\left\{ 
\begin{array}{c}
\sum_{xy}\hat{\pi}_{xy}\left( a_{x}+b_{y}-\Phi _{xy}\left( \lambda \right)
\right)  \\ 
+\sum_{xy}e^{\Phi _{xy}\left( \lambda \right) -a_{x}-b_{y}}+\frac{1}{2}%
\sum_{x}e^{-2a_{x}}+\frac{1}{2}\sum_{y}e^{-2b_{y}}%
\end{array}%
\right\} 
\end{equation}

Letting $\theta = (\lambda^\top,a^\top,b^\top)^\top$, this suggests the following Poisson regression with two-way fixed effects for the observed matches~$\hat\pi_{xy}$:
\begin{eqnarray*}
    \label{eq:Poisson}
\mathbb{E}[\hat{\pi}_{xy}\,|\,xy] &=&\exp \left( \Phi _{xy}\left( \lambda
\right) -a_{x}-b_{y}\right) , \\
\mathbb{E}[\hat{\pi}_{x0}\,|\,x0] &=&\exp \left( -2a_{x}\right) , \\
\mathbb{E}[\hat{\pi}_{0y}\,|\,0y] &=&\exp \left( -2b_{y}\right) .
\end{eqnarray*}
If the observations~$\hat\pi_{xy}$ were drawn from a Poisson distribution with intensity equal to these expectations, and if one assigns a weight of 1 to matched pairs, and a weight 1/2 to unmatched elements, then the weighted log-likelihood would be equal to
\begin{equation*}
    l(\theta) = \sum_{xy}\hat{\pi}_{xy}\left( \Phi _{xy}\left( \lambda \right)
-a_{x}-b_{y}\right) -\sum_{xy}e^{\Phi _{xy}\left( \lambda \right)
-a_{x}-b_{y}}-\frac{1}{2}\sum_{x}e^{-2a_{x}}-\frac{1}{2}\sum_{y}e^{-2b_{y}}.
    \label{eq:Poisson log likelihood}
\end{equation*}
Therefore $\theta$ is estimated by a weighted Poisson regression. Yet since Poisson sampling is not assumed, inference for the estimator of~$\theta $ that maximizes~$l(\theta)$ is a PPML. 

 Without unmatched agents, the entropy regularized optimal transport solution~(\ref{eq:pi(A)}) can be reinterpreted as a structural gravity model of bilateral trade flows, where~$\pi_{xy}$ are the flows from exporter~$x$ to importer~$y$, and $a_x$ and~$b_y$ are exporter and importer fixed effects respectively. As is well-known in the trade literature since the paper by \citet{silva2006log}, the gravity equation can be obtained by an unweighted Poisson regression, where the uniform weights stems from the fact that there are no unmatched agents to account for.  

\subsubsection{LASSO regularization and the SISTA algorithm.}
While entropic regularization smooths the matching pattern and guarantees full support, empirical researchers are often interested in {sparse} or {structured} specifications of the surplus function, especially when $\Phi_{xy}$ is parameterized by a high-dimensional matrix $A$. Sparsity can be promoted by adding an $\ell_1$-penalty to the parametric IOT likelihood:
\[
\max_{A}
\left\{
\sum_{x,y} \widehat{\pi}_{xy} \log \pi_{xy}(A)
-
\sum_{x,y} \pi_{xy}(A)
-
\lambda \|A\|_{1}
\right\},
\]
which drives many entries of $A$ exactly to zero.  
This yields a composite optimization problem combining a smooth (entropic-OT) component and a nonsmooth ($\ell_1$) penalty.

A computationally efficient solution to this problem is the SISTA algorithm ({Sinkhorn Iterative Soft-Thresholding Algorithm}) of \citet{carlier2023sista}.  
SISTA alternates between two steps:
(i) a {Sinkhorn/Bregman iteration} that updates the dual potentials of the entropic OT problem and computes the gradient of the smooth part with respect to $A$, and
(ii) a {soft-thresholding proximal step}
\[
A \leftarrow \operatorname{Soft}_{\lambda \tau}
\big( A + \tau\, \nabla_A \log \pi(A) \big),
\]
where $\tau$ is a step size.  
The first step exploits the structure of entropic OT to compute $\nabla_A \pi(A)$ efficiently via the dual potentials, and the second step implements the proximal operator for the $\ell_1$ penalty.  
SISTA is monotone, scalable, and tailored to situations where the parameter matrix is large but the OT structure provides efficient low-dimensional updates.

Conceptually, LASSO regularization plays a role opposite to that of entropy: the entropic term {diffuses} mass and generates dense matching patterns, while the $\ell_1$ penalty {concentrates} the parameters by shrinking many coefficients to zero.  
Together, the two regularizers offer complementary control of smoothness, sparsity, interpretability, and generalization in parametric IOT.

\subsection{IOT and metric learning}
\label{sec:metric-learning}

The IOT problem is closely related to---and in some ways a special case of---the broader field of {metric learning}.  
In metric learning, one seeks to infer a distance or similarity function from observed pairwise relationships; in IOT, one seeks to infer a {transport cost} (or surplus) function from observed matching flows.  
Both problems aim to recover the geometry of a space from behavioral or relational data.  
However, the constraints imposed by transferable-utility equilibrium and the convexity induced by entropic regularization give IOT several structural advantages over classical metric learning, which is typically nonconvex and statistically more delicate.

\subsubsection{Classical metric learning}
In its conventional form, metric learning seeks a matrix $M\succeq 0$ such that the Mahalanobis distance
\[
d_M(x,y) = \|x - y\|_M^2 = (x-y)^\top M (x-y)
\]
reflects observed similarities or dissimilarities between data points.  
To ensure $M\succeq 0$, one writes $M = L^\top L$ and optimizes over $L$, which makes the problem inherently nonconvex.  
Triplet-loss objectives, hinge losses, and neighborhood constraints, as in \cite{weinberger2009distance}, \cite{hoffer2015deep}, or \cite{bellet2013supervised}, all involve nonconvex formulations.  
As a result, solutions may be sensitive to initialization, prone to local minima, and difficult to interpret.
In addition, classical metric learning does not enforce consistency with marginal distributions or equilibrium conditions: similarity constraints are typically local (e.g.~triplets) rather than arising from a global balance condition such as feasibility of a transport plan.

\subsubsection{Metric learning through IOT: An intermediate perspective}
A natural bridge between metric learning and IOT emerged with the use of differentiable entropy-regularized OT as a loss function, as in \cite{cuturi2014fast} and \cite{courty2016optimal}.  
By learning the cost matrix $c_\theta$ parameterized by $\theta$ so as to match observed transport plans (or distributions, in domain adaptation), these approaches cast metric learning as the minimization of
\[
\mathcal{L}(\theta)
=
D\big( \pi^\sigma(c_\theta), \widehat{\pi} \big),
\]
where~$\pi^\sigma(c_\theta)$ is the solution to the entropy regularized optimal transport with cost~$c_\theta$, $\hat\pi$ is the observed matching pattern, and~$D$ is a divergence between probability distributions.  
Differentiability of $\pi^\sigma$ (thanks to entropy) gives access to stochastic-gradient methods, moving metric learning closer to convex-analytic optimal transport.

However, these methods still (i) lack equilibrium structure, (ii) do not incorporate marginal constraints as equilibrium conditions, and (iii) often rely on heuristic or local triplet-loss-like objectives.

\subsubsection{IOT as equilibrium-based metric learning}
The IOT framework imposes the full set of constraints from transferable-utility equilibrium.  
Instead of comparing distances for isolated triplets or neighborhoods, IOT uses the {entire joint distribution} of matches to recover the structural surplus.
Metric learning becomes:
\[
\text{find } \theta \text{ such that } 
\pi^\sigma\!\left( c_\theta \right)
\approx
\widehat{\pi},
\]
where the matching pattern is an equilibrium object satisfying marginal constraints and dual optimality conditions.  
This transforms metric learning from a nonconvex factorization problem into a convex estimation problem. In particular, if the parameter~$\theta$ enters linearly in the cost function (as in the bilinear surplus case studied above), 
then, the entropically regularized OT mapping is {smooth and globally convex},  
convex duality provides explicit gradients and fixed-point characterizations,  
and the equilibrium constraints ensure global rather than local consistency.
In economic terms, IOT learns the affinity between types in a way that respects a TU equilibrium: the recovered geometry is the one generating the observed sorting pattern under optimality and feasibility.

\subsubsection{Comparison of difficulty: IOT vs.~metric learning}
Classical metric learning is notoriously difficult because the Mahalanobis parametrization $M=L^\top L$ renders the optimization problem nonconvex, because similarity constraints are typically local (for example, triplets) rather than global equilibrium conditions, because distances are identifiable only up to arbitrary additive and multiplicative normalizations, and because gradient-based methods can be numerically unstable in the absence of entropy or other smoothing devices. In contrast, IOT with entropic regularization is considerably more tractable: when~$\theta$ enters linearly in the cost function~$c_\theta$, the objective in~$\theta$ is globally concave (as in log-likelihood or KL-fitting formulations), the transport plan $\pi^\sigma(c_\theta)$ is smooth, unique, and has full support, the dual potentials yield stable gradients through convex duality, and the equilibrium conditions impose a global structure that sharpens identification. Seen from this angle, IOT can be viewed as {metric learning under convexity and equilibrium constraints}, which explains why entropically regularized IOT enjoys strong computational and statistical advantages over classical metric-learning formulations.

\subsubsection{Economic relevance of the metric-learning perspective}
Viewing IOT as metric learning reveals why bilinear or low-rank surplus models are so powerful in empirical matching: they learn a latent metric or affinity space in which assortative patterns become linear (or low-dimensional).  
Structured sparsity (group Lasso, nuclear norm, SISTA) further enhances interpretability by identifying which characteristics of agents matter for sorting and which complementarities drive equilibrium behavior.
This interpretation also emphasizes the conceptual contrast with classical discrete-choice estimation: discrete-choice models learn individual preferences, whereas IOT learns {pairwise complementarities} governing equilibrium matches. Both are metric-learning problems, but only IOT uses equilibrium OT geometry.

\subsection{Other developments in IOT} Fox’s work provides an important alternative to  the inverse optimal transport perspective, by relying on maximum score. In particular, \cite{Fox2010IdentificationMatchingGames} studies what features of the match surplus can be identified from equilibrium matching patterns, while \cite{Fox2018MatchingTransfers} develops a tractable matching maximum score approach for estimating transferable-utility matching games from observed assignments. Outside the economics literature, the paper by \citet{StuartWolfram2020} formulates IOT as a fully fledged Bayesian inverse problem, studying well-posedness and posterior contraction for cost learning from noisy observations of optimal transport plans. More recently, \citet{AndradePeyrePoon2024} analyze $\ell_1$-regularized IOT in an entropic setting and establish \emph{sparsistency} of the resulting estimators, that is, consistent recovery of the support of the true cost function. They show that, as the entropic penalty varies, IOT interpolates between classical Lasso and graphical Lasso, thereby connecting sparse IOT to sparse precision-matrix and graph estimation problems that are central in modern statistics and machine learning.

\section{Concluding remarks}

Optimal transport has become an exceptionally powerful tool in econometrics because it unifies, extends, and operationalizes several fundamental ideas in a way that is both conceptually elegant and empirically tractable. First, OT generalizes the notion of distance from points to probability distributions, yielding metrics---such as Wasserstein distances---that metrize weak convergence and underlie a wide range of applications in distributional comparisons, semiparametric inference, and distributionally robust methods. Second, OT extends the classical univariate theory of monotone rearrangements to the multivariate setting, thereby providing economically meaningful notions of multivariate quantiles, ranks, copula-based dependence, and multidimensional measures of inequality and risk, all of which are central in modern empirical work with rich heterogeneity. Third, entropically regularized OT reveals a deep algebraic connection with generalized linear models featuring two-way fixed effects, placing OT squarely within the econometric panel-data tradition and allowing researchers to draw on a mature body of identification, estimation, and inference tools. Finally, OT enjoys outstanding computational properties: exact solutions reduce to linear programming in finite dimensions, while regularized variants admit scalable Sinkhorn-type algorithms in high dimensions, enabling applications to large datasets and machine-learning environments. These combined features---a rich geometric structure, powerful generalizations of classical econometric concepts, tight connections to familiar statistical models, and remarkable computational efficiency---explain why optimal transport has rapidly become a central instrument in modern econometric analysis.



















\newpage

\bibliographystyle{abbrvnat}
\bibliography{OT-ER}

\end{document}